\documentclass[fleqn,usenatbib]{aastex62}

\usepackage{graphicx}	% Including figure files
\usepackage{amsmath}	% Advanced maths commands
\usepackage{amssymb}	% Extra maths symbols

\newcommand{\pointless}[1]{#1}
\definecolor{scc}{rgb}{1.0, 0.49, 0.0}
\graphicspath{{./}{figures/}}

\received{May 23, 2018}
\revised{\today}
\accepted{\today}
\submitjournal{PASP}

\shorttitle{Photoz for radio surveys}
\shortauthors{Norris et al.}
\begin{document}

\title{A Comparison of Photometric Redshift Techniques  for Large Radio Surveys}

\correspondingauthor{Ray Norris}
\email{raypnorris@gmail.com}

\author[0000-0002-4597-1906]{Ray P. Norris}
\affiliation{Western Sydney University, Locked Bag 1797, Penrith South, NSW 1797, Australia}
\affiliation{CSIRO Astronomy \& Space Science, PO Box 76, Epping, NSW 1710, Australia}

\author{M. Salvato}
\affiliation{MPE, Giessenbach Strasse 1, D-85748, Garching, Germany}
\affiliation{Excellence Cluster, Boltzmanstrasse 2, D-85748, Germany}

\author[0000-0002-9182-8414]{G. Longo}
\affiliation{Dept. of Physics Ettore Pancini, University Federico II, via Cintia, Napoli, Italy}
\affiliation{California Institute of Technology, Pasadena, CA, USA}

\author[0000-0001-9506-5680]{M. Brescia}
\affiliation{INAF - Astronomical Observatory of Capodimonte, via Moiariello 16, 80131, Napoli, Italy}

\author{T. Budavari}
\affiliation{Johns Hopkins University, Baltimore, MD, USA}

\author{S. Carliles}
\affiliation{Johns Hopkins University, Baltimore, MD, USA}

\author[0000-0002-3787-4196]{S. Cavuoti}
\affiliation{Osservatorio Astronomico di Capodimonte, INAF, via Moiariello 16, 80131, Napoli, Italy}
\affiliation{Dept. of Physics Ettore Pancini, University Federico II, via Cintia, Napoli, Italy}
\affiliation{INFN Naples section, via Cinthia 6, I-80126, Napoli, Italy}

\author{D. Farrah}
\affiliation{University of Sussex, Sussex, UK}

\author{J. Geach}
\affiliation{University of Hertfordshire, Hatfield, Herts., UK}

\author{K. Luken}
\affiliation{Western Sydney University, Locked Bag 1797, Penrith South, NSW 1797, Australia}
\affiliation{CSIRO Astronomy \& Space Science, PO Box 76, Epping, NSW 1710, Australia}

\author{A. Musaeva}
\affiliation{Sydney Institute for Astronomy, School of Physics, The University of Sydney, NSW 2006, Australia}
\affiliation{Centre of Excellence for All-sky Astrophysics (CAASTRO), The University of Sydney, NSW 2006, Australia}

\author{K. Polsterer}
\affiliation{HITS, Schloss-Wolfsbrunnenweg 35 69118, Heidelberg, Germany}

\author[0000-0001-7020-1172]{G. Riccio}
\affiliation{Osservatorio Astronomico di Capodimonte, INAF, via Moiariello 16, 80131, Napoli, Italy}

\author{N. Seymour}
\affiliation{Curtin University, Perth, WA, Australia}

\author{V. Smol\v{c}i\'{c}}
\affiliation{Department of Physics, Faculty of Science, University of Zagreb,  Bijeni\v{c}ka cesta 32, 10000  Zagreb, Croatia}

\author[0000-0002-6748-0577]{M. Vaccari}
\affiliation{Department of Physics and Astronomy, University of the Western Cape, Robert Sobukwe Road, 7535 Bellville, Cape Town, South Africa}
\affiliation{INAF - Istituto di Radioastronomia, via Gobetti 101, 40129 Bologna, Italy}

\author{P. Zinn}
\affiliation{Astronomisches Institut, Ruhr-UniversitaÌt Bochum, Universitatstrasse 150, 44801 Bochum, Germany}



\begin{abstract}

Future radio surveys will generate catalogues of tens of millions of radio sources, for which redshift estimates will be essential to achieve many of the science goals.  However, spectroscopic data will be available for only a small fraction of these sources, and in most cases even the optical and infrared photometry will be of limited quality. Furthermore,  radio sources tend to be at higher redshift than most optical sources (most radio surveys have a median redshift greater than $1$) and so a significant fraction of radio sources hosts  differ from those for which most photometric redshift templates are designed. We therefore need to develop new techniques for estimating the redshifts of radio sources. As a starting point in this process, we evaluate a number of machine-learning techniques for estimating redshift, together with a conventional template-fitting technique. We pay special attention to how the performance is affected by 
the incompleteness of the training sample and by sparseness of the parameter space or by limited 
availability of ancillary multi-wavelength data. As expected, we find that the quality of the photometric-redshift degrades as the quality of the photometry decreases, but that even with the limited quality of photometry available for all sky-surveys, useful redshift information is available for the majority of sources, particularly at low redshift. 
%We find that while template-fitting techniques perform best with very high quality data,  machine-learning techniques perform almost as well on lower-quality data. 
%\mbc{[Comment: referee 2 pointed out that the last paragraph of the abstract is too general and should be more aligned with what we write in the conclusions. So I try to adjust the above bold-face paragraph as follows:]} \mbc{
We find that a template-fitting technique performs best in the presence of high-quality and almost complete multi-band photometry, especially if radio sources that are also X-ray emitting are treated separately, using specific templates and priors. When we reduced the quality of photometry to match that available for the EMU all-sky radio survey, the quality of the template-fitting degraded and became comparable to some of the machine learning methods. Machine learning techniques currently perform better at low redshift than at high redshift, because of incompleteness of the currently available training data at high redshifts.
%\ray{ref 2 also complained about ``template-fitting techniques'' (plural) where only one was used, so I've changed this to singular above}

\end{abstract}


\section{Introduction}\label{Sec:intro}

%\ray{ Blue bits in the text are comments and FYI from me, plus notes to myself, and will be deleted in the final draft.}\\
%\rayq{ Red bits in the text are questions to co-authors from me.}

Next-generation radio surveys will generate catalogues of tens of millions of galaxies \citep{norris17b}. Much of the science generated by these surveys depends on the availability of identifications and redshifts for each radio source. However, in spite of recent advances, even multi-object spectroscopy can only  provide redshifts for a small fraction of these galaxies. Instead, only photometric redshifts (hereafter photoz's) 
%are therefore crucial to modern precision cosmology and galaxy evolution studies since they only 
can provide the necessary redshift information for the large samples of radio sources.
% of objects 
%which are required by statistical  approaches. 

%\mbc{[changed from referee 1 request] 
In this context we focus
%focused the attention 
\pointless{on an extensive radio survey, the Evolutionary Map of the Universe (EMU), which has been designed to use} 
the new ASKAP radio telescope to make a deep ($\sim 10\mu Jy$/beam target rms) radio survey covering the entire Southern Sky and extending  as far North as $30^{o}$ \citep{norris11}. EMU aims to detect about $70$ million sources, about half of which are expected to be star-forming galaxies and the rest Active Galactic Nuclei (AGN). EMU's key science goals include tracing the evolution of galaxies from the early Universe to the present day, and using the distribution of radio sources to explore the large-scale structure and cosmological parameters of the Universe. These  goals require redshift estimates for a significant fraction of the sources. 
%\mbc{[changed from referee 1 request] 
In particular, AGN represent an important challenge to photometric redshift techniques \citep[see][for a review]{salvato18}. 

%{\bf Note by Duncan: Is there a published paper to point to for these numbers? If not then this paper may need to justify them?}

Since the early pioneering work \citep{baum1962,butchins1981,loh1986}, many different methods to evaluate photoz's have been implemented and extensively tested on a wide range of data. A comprehensive review of these techniques has recently been published by \cite{salvato18}. 

All these methods use some a-priori knowledge gained either from spectroscopy or from physical assumptions, to deduce the function that maps the parameter space of photometric observables onto the spectroscopic redshift distribution. Methods can be roughly divided into two categories: (\textit{i}) SED (Spectral Energy Distribution) template fitting, and (\textit{ii})  machine learning (ML) techniques.

Template fitting techniques \citep[e.g.][]{ arnouts1999, benitez00, bolzonella11, brammer10, duncan18a} rely on a library of template galaxy spectra which are then shifted to different redshifts and fitted to the observed photometric data points. The various implementations differ in the way the template library is assembled (from real galaxy spectra or from synthetic spectra), in the possible inclusion of emission lines (crucial to model AGN and starburst spectra),  and on the ability to use priors and in the fitting procedure adopted. 
%\mbc{[changed from referee 1 request] The way to extract photometric bands, as well as the amount of collected bands} can also profoundly affect the photoz accuracy. 
%\ray{[further changes] 
\pointless{The techniques used to extract the photometric measurements, the number of  bands, and the simultaneity of the measurements} 
can all profoundly affect the photoz accuracy. 
For example, single band extractions give worse redshifts than combined extractions. Most template fitting techniques do not use wavelengths longer than the infrared, since the templates are not well defined at longer wavelength. Notable exceptions are \cite{rowanrobinson} who used early IRAC data up to 8$\mu$m from the SWIRE survey, \cite{aretxaga} who used radio and far-infrared data, and \cite{pearson13} who used Herschel and sub-mm data. 
Here we use the Le Phare \citep{arnouts1999, ilbert2006} code (see Section \ref{SEC:Le Phare}) with optical and near-infrared data.

Machine learning (ML) based techniques first introduced by \cite{tagliaferri2002,firth2002}, use the a-priori knowledge from a ``training set'' of objects (also known as a Knowledge Base or KB) for which accurate spectroscopic redshifts are available. 
%\mbc{[comment on referee 1 request: here KB is already well defined.]} 
Implementations  
%differing in the adopted interpolative method, 
include  random forest, neural networks, nearest neighbours, 
%\mbc{
\pointless{support vector machines, Gaussian process regression,} 
and self-organised maps. Recent developments include the ability to generate %\mbc
\pointless{probability density functions (PDFs) for the redshifts, rather than a single most likely point estimation \citep{amaro17,amaro18, duncan18b}}, 
and the ability to take account of spectral variability \citep{pasquet17}. \cite{masters2015} also discuss the effect of the biases introduced by the a priori information.

Several  studies have compared the relative merits of ML and template approaches
\citep{hildebrandt2008, dahlen2013, abdalla11}. 
%In all methods, the range of application is defined by the parameter space sampled either by the spectroscopic training set, or by the adopted library of  templates. 
%%\ray{the rest of the intro is a bit verbose  - I will shorten and re-order it}
For example, ML methods cannot effectively estimate photometric redshifts for objects fainter than the spectroscopic limit, or for rare objects that are not represented in the training sample. 
\pointless{Throughout this paper, and for all foreseeable ML applications, the training data and the target data must be limited to the same limiting magnitude.}  
Template fitting methods can estimate redshifts of rare or peculiar objects as long as a representative template is included in the library. Key advantages of the template fitting methods are: (i) they can be extrapolated beyond the spectroscopic limit and, \pointless{(ii) they can provide an estimate of the galaxy spectroscopic type and thus morphology, providing by default (iii) also a quality of the fit. Template fitting procedures  also provide a redshift probability distribution function, which has only recently become available from ML codes}.

However, if the KB  samples the observed parameter space well, and the properties of sources in the KB closely match those in the target  data, ML methods can be more accurate than template fitting. Furthermore, they  do not require an a priori hypothesis about the underlying physics, and they can use  non-photometric information such as morphology, radio polarization, photometric gradients \citep{gieseke2011,norris2013} or even the spectroscopic type provided by template fitting procedures \citep{cavuoti2017}.

\pointless{The complementary nature of the two techniques have also inspired approaches that use a hybrid of  SED fitting and machine learning in a single collaborative framework, which improve the accuracy of photoz estimates  \citep[e.g.][]{cavuoti2017b, duncan18b}. }

The quality of photoz's might be expected to improve with the depth of the data \pointless{and thus a smaller photometric error},
%fundamental for template fitting techniques
 the number of the bands available, and a decreasing spacing between the filters 
\cite[e.g.][]{budavari2008,benitez09}.
While this is usually true for template fitting methods, this is not necessarily true for 
ML methods where additional parameters, or features, may increase the noise and decrease performance. 
For example, increasing the number of bands may increase the incompleteness of the data, and reduce the density of training points in the parameter space, thereby increasing the noise in the  photoz's.  
Instead, the choice of the features to be used  should be optimized by a feature selection phase \citep{brescia2013, disanto18}. 
\pointless{\cite{polsterer2014} and \cite{cavuoti2014b} have shown that, assuming that the automatic feature selection can identify crucial parameters,
%beyond those traditionally adopted and identified by standard astronomical reasoning}.
%\cite{polsterer2014} and \cite{cavuoti2014b} have shown that  
feature selection should be data-driven rather than driven by astronomical reasoning. }

Most studies comparing the performance of photoz methods have been  based on optically-selected samples, \pointless{omitting outlier sources such as AGN and Starburst galaxies, which}
%which under-represent the AGN and starburst galaxies 
 dominate radio-selected samples. Furthermore, the median redshift of sources found in EMU is expected to be z $\sim$ 1.2, which is very different from the median redshift z $<$ 0.2 of most wide optical surveys \citep{norris11}. Most of the high-redshift radio sources are high-z radio-loud AGN, which are relatively rare in optical surveys because they are faint at optical wavelengths. 
For example, about half the sources in modern radio surveys host an AGN \citep{norris2013}, most of which are detected only in very deep X-ray surveys
 \citep{LaFranca2012nx, smolcic17}.
%most of which are not detected in X-ray \citep{LaFranca2012nx, smolcic17}.

\pointless{In template fitting techniques, the choice of the library of templates plays a crucial role. While a 
%more or less 
standard library can be used for computing photoz for normal galaxies at any redshift and depth, alternative templates must be considered when working on radio or X-ray selected sources. In particular, for X-ray sources, the library of templates change with the depth of the X-ray survey \citep[e.g.][]{Salvato:2011mz,hsu2014}}.

%Template fitting techniques typically use a standard library of templates that are applicable to most optically-selected samples. This is not true for X-ray and radio surveys, on the other hand, as they have a significant fraction of AGN whose properties change as a function of survey depth \citep[e.g.][]{Salvato:2011mz,hsu2014}.
 % and this in turn defines the libraries and priors which need to be used.  
%cant find citation arianna2017 that was given here
Furthermore, current templates generally do not use wavelengths beyond $\sim$ 5 $\mu$m whereas  ML methods can use radio, IR, and X-ray data to improve the photoz accuracy. 

\pointless{An additional challenge that is faced when computing photoz for large radio surveys such as EMU, is that the photometry will heterogeneous, using  different photometric catalogs of different depths. This is in contrast to deep optical pencil-beam surveys, where  images are registered to the same astrometric frame and the photometry is then measured with the same aperture, taking  into account the different point spread functions.
%The quality of photometry is also poorer in large surveys: %\mbc{
For example, Table~\ref{Tab:all sky} compares the photometric data that is avaialable for COSMOS and that is predicted will be available for EMU.}

\pointless{As a result, comparisons of photometric redshift techniques performed on optically-selected samples do not necessarily reflect their performance on radio-selected samples. This is addressed by \cite{duncan18a, duncan18b}, whose work is complementary to this paper. Duncan et al. use a different regression technique from any used here (Gaussian process regression, or Kliging) but, more importantly, divide their data into several classes (star-forming, IR AGN, etc) and separately train their algorithms on these classes, and combine the results from the different sets, together with the results from template fitting, using a Bayesian estimator. This combination of results is very successful, but here our focus is different. While we want to quantify the quality of the photoz that will be available for EMU and compare it with what is available for pencil beam surveys, we also want to understand whether flagging the sources that are known to be X-ray or variable AGN can improve the results. Finally, we want to see whether using radio and X-ray fluxes can improve the performances of ML algorithms. We do this in the COSMOS field, comparing the performance of a number of photoz techniques, on a sample of sources detected in radio, at the same depth that will be available for COSMOS.}

%including a SED fitting technique and several machine learning (ML) techniques, to radio and x-ray selected samples in the COSMOS field.  Once more, the COSMOS data set provides an ideal benchmark to study this effect. 
 
Previous analysis of this field based on $31$ filters ($13$ broad bands, $6$ narrow bands and $12$ intermediate bands) with a typical depth of $26$ mag (AB) allowed template fitting methods to predict photoz's with an accuracy better than $0.015$ for normal galaxies \citep{ilbert2009} and for a sample of X-ray selected galaxies \citep{Salvato:2009zw, Salvato:2011mz}. 

In this paper, \S \ref{SEC:thedata} presents the data and overall experimental approach; \S 3 describes the individual methods, \S 4 describes the methods and tests, and \S 5 presents the 16 cases that we tested. In \S 6 and \S 7 
%\mbc{[changed from referee 1 request] 
\pointless{we compare and discuss the results from the various methods.}

\section{The data}
\label{SEC:thedata}

The data from the COSMOS project are publicly available\footnote{\url{http://irsa.ipac.caltech.edu/data/COSMOS/}} and %\mbc{[changed from referee 1 request] 
\pointless{provide} 
an ideal benchmark for this project \citep{Scoville:2007rw}. COSMOS covers a $2$ sq. deg. field: an area large enough 
\pointless{to reduce cosmic variance and to sample the bright and }
%\mbc{[changed from referee 1 request] 
rare objects. It has a large spectroscopic follow-up pursued using optical and near-infrared spectrographs \pointless{mounted on the largest telescopes available, curated into a master spectroscopic catalog by Salvato}, thus increasing the sampling completeness. It has deep, multi-epoch multi-wavelength observations from X-ray (XMM, Chandra) to radio (VLA), including  UV (Galex, \cite{zamojski07}, optical (broad band photometry from SDSS, CFHT, intermediate \pointless{and narrow} band photometry from SUBARU \citep{taniguchi07}), NIR and MIR
%\mbc{
\pointless{\citep[Ultravista and Spitzer-IRAC bands)]{mccracken10, sanders07}}.
The data are highly homogeneous, since images have been registered to a common grid and convolved to a common PSF and fluxes computed in a common aperture. 
\pointless{When this work started, the only multiwavelength COSMOS catalog available \citep{ilbert2009} reported magnitudes instead of fluxes like the latest version \citep{laigle16}. This added a complication to the computation of the photoz, for both SED fitting and ML algorithms.} All data have been corrected for extinction, and magnitudes (AB) are computed in a $3$ arcsec aperture.
%The full list of measured parameters made available for the experiments (not all of them were actually used) is given in the Appendix (Table \ref{TAB:parameters}).

The COSMOS survey is complemented by VLA \citep{schinnerer07, schinnerer10} XMM-Newton \citep{hasinger07,Brusa:2010lr} and {\it Chandra} \citep{Civano:2012ys} observations. 
VLA-COSMOS  provides radio coverage of the field with roughly the same depth and resolution expected for  EMU.  The COSMOS field is currently being surveyed to an even greater depth using the VLA\citep{smolcic17} but those data are not used in this paper. \cite{herreraruiz17} have observed all these sources with VLBI, and confirm that a significant fraction of them are high-redshift radio-loud AGN.
\cite{sargent2010} argue that 99.9\% of the VLA-COSMOS data have a secure optical counterpart, although, since some of the radio sources are likely to be very high redshift radio-loud AGN, which are extremely faint at optical/IR wavelengths, a small fraction of these are likely to be mis-identifications.

%The multiwavelength coverage of COSMOS is deeper and includes more optical and IR photometric bands than the expected 
%multiwavelength coverage of the entire Southern sky  by EMU. However, at X-ray wavelengths, eROSITA \citep{predehl} will provide observations to a depth similar to those of XMM in the COSMOS field. 
%(optical, NIR, MIR with a depth of XX, YY, ZZ, respectively (see Table \ref{Tab:all sky}). 
%XMM-COSMOS can instead be cut at the depth expected to be reached by eROSITA  {\bf (MARA: XXX)} in order to see 
%the accuracy  that can be reached by the various  methods for those sources that we know are AGN but for 
%which the X-ray detection will not be available. 
The VLA-COSMOS $1.4$ GHz sample consists of $2242$ sources with optical counterparts \citep{sargent2010}.  In this paper, we use as our primary set a subsample of $757$ sources that have reliable spectroscopic redshifts.  
%{\bf anybody know why we didnt use the full 797 that are available?}. 
We call this our ``spectroscopic KB''. Of the $757$ sources in the KB, $91$ have X-ray fluxes measured by XMM \citep{Brusa:2010lr} and an additional (i.e. excluding those also detected by XMM) $158$ have X-ray fluxes measured by {\it Chandra} \citep{Civano:2012ys,marchesi16}.

From this vast array of data, a total of $45$ different photometric measurements, or ``features'', were used in the experiments here and are listed in Table~\ref{TAB:parameters}. 
%%\ray{actually there are 45 listed in this table - need to sort this out}
\pointless{Except where stated otherwise, all data were used in all experiments.}

\section{Experimental approach}
\label{approach}
%\ray{I have written this section by going through the original emails and data, trying to reverse engineer what was actually done. In some cases I've had to make educated guesses, so please confirm or correct this as appropriate. }

We ran the experiment as a ``blind data challenge'', in which different groups (all of whom are co-authors of this paper) were invited to test their algorithms on the KB, with the challenge being run by a control group. 
Machine learning techniques require the KB to be split into two subsets called the 'training' and 'test' set respectively. These two subsets need to be disjoint and to map the parameter space 
%\mbc{[changed according to referee 2 request] 
\pointless{representatively}. 
For the experiments described here, the control group provided the testers with the training set, which includes redshifts, and the 
%\mbc
\pointless{}{blind}
test set, which does not include redshifts. The testers then used the training set to train their algorithms.  The trained algorithms were then applied to the test set to yield a set of estimated redshifts. These were then passed back to the control group who compared them with the true spectroscopic redshifts, to evaluate the algorithms in terms of standard statistical indicators.

%In the implementations of some ML methods, the KB is split in three disjoint subsets, comprising the training and test set plus an additional validation set, used to avoid overfitting of the data. %\rayq{However, to the best of my knowledge, none of the methods in this paper did this - does anybody disagree? If not I will delete this line.}.

As well as evaluating the test on the best data available, we are interested in the sensitivity of the tests to the number of bands available, the sensitivity of the photometry, whether radio flux density is used, and whether X-ray detected AGN are included in the training sets. These tests were therefore varied in four ways:

%\ray{Full details are in ``summary of which data used for which test.xlsx'' available separately on request}

\begin{itemize}
\item {\bf Variation 1: Bias:} 
%The properties of the KB affect the final accuracy of the derived photometric redshift (see \cite{masters2015} for an analysis of the COSMOS data set, and \cite{hildebrandt2008} for a more general analysis). 
In general, spectroscopy tends to be available on sources that are brighter than the population of sources for which we wish to obtain photometric redshifts. Here we test the ability of the techniques to train on a brighter sample of sources and extrapolate that training to a larger sample containing fainter sources.  We therefore performed two variations: 
In the ``Bright'' variation, the training set was chosen to be the optically brightest $50$\% of sources, selected in the $i$ band. In the  ``Random'' variation, the training  set was randomly chosen from the parent sample of $757$ objects. In both cases, because of an implementation detail, the fraction was not exactly half. In the ``Bright'' variation, $391$ sources were in the training set, and $366$ were in the test set. In the ``Random'' variation, $343$ sources were in the training set, and $414$ were in the test set. 
%\mbc{Connected to this Variation 1, there is a comment of referee 1 (see Section 3 comment in the referee report), which warn us that the training sample is not representative of all radio sources. But I don't understand the reason of such comment, since we perfectly know (and write also) such problem and that's why we exactly want to analyze the extrapolation of methods to fainter objects. I think we should better clarify such aspect in our answer to this point.} 
%\ray{I've inserted some text in the response}

%%\ray{these numbers are slightly different from table 1 - which is right?}

%In the first case we sorted in the  magnitude the 324 sources with spectra  and used the brightest  162   as training sample and the remaining 162 fainter as blind sample. In the second case we randomized the sources and distribute them in training and blind sample (\pointless{percentages?}.)
%{\bf Ray points out inconsistency with data in section (number of sources and number of X ray sources.}

\item {\bf Variation 2: Depth} 
As discussed in Sec.~\ref{Sec:intro}, all-sky photometry surveys will be much shallower than the deep data available in pencil-beam fields such as COSMOS. 
%\mbc{A commented note by Duncan here asks to be more specific on such mentioned bands. I agree with him also in the optic to accomplish referee 2, who recommend to improve the impact on incoming large surveys. Ray, may we be more detailed here?}
% \ray{
\pointless{For example, over most of the area of EMU, the only optical/infrared surveys (listed in Table \ref{sensitivity} along with their limiting magnitudes), available in the next year will be the 
%shallow  Galex all sky\footnote{available at \url{http://galex.stsci.edu/GR6/}},
%Galex is not optical/IR and does not detect a significant fraction of radio sources, so is not really relevant
SkyMapper\citep{wolf18}, ALLWISE \citep{wright10}, and VHS \citep{mcmahon13}, while smaller but still significant areas will be covered by DES \citep{des16} and other surveys.  
This will introduce incompleteness in these large surveys, and a significant fraction of objects will only have upper limits on their photometry in some bands. At X-ray wavelengths, eROSITA \citep{Merloni12} will provide an all-sky catalog that is expected to be about 30 times deeper than ROSAT \citep{Boller16}, which is still much shallower than XMM-COSMOS, for example.}
%{\bf Duncan's note: We need to be more specific about which bands, especially wrt upcoming large surveys with eg LSST, Skymapper, WFIRST? where are the most likely gaps?}. 
We simulated this effect by performing two variations. In the ``Deep'' variation, the training set used the deepest data available in the KB. For example, these data include the Spitzer-IRAC measurements \citep{sanders07}, which are available only in small regions of the sky.
In the ``Shallow'' variation, the training set used shallower data to simulate the data available to the EMU survey. Specifically, (\textit{i}) Spitzer-IRAC data was limited to bands $1$ and $2$ and to the depth of the ALLWISE data \citep{wright10}, and (\textit{ii}) optical photometry data for sources fainter than $i=22$ was removed, \pointless{i.e. keeping the sources that are about 1 magnitude fainter than what is expected for the final depth of SkyMapper.}
\pointless{For the same reason, we also included only broad band photometry in the ``Shallow'' data, removing narrow and intermediate band photometry.}
None of these reductions in the quality of the data affected the size of the training ($391$, $343$) and test ($366$, $414$) sets. \pointless{However, these changes typically removed optical photometry for about one third of sources, and infrared photometry for about two thirds of sources}.

%%\ray{In this test, experimenters applied their methods, after training on the ``shallow'' data,  to two data sets: ``allbands'' or ``deep'' (i.e. same as used for the other tests) and a ``allsky''  or ``shallow'' (which corresponds to the shallow data used for the training set). I suggest we only use the ``shallow'' tests, because
%\begin{itemize}
%\item as may be expected, all tests perform poorly when applied to the ``deep'' data after being trained on the ``shallow'' data. The results on ``shallow'' are significantly better.
%\item the effect of applying the methods to a data set deeper than the training set has already been tested in the ``bias'' variations, 
%\item training on EMU-like data and then applying the methods to COSMOS-like data doesn't have any application in the real world, and makes it harder to understand what was done
%\end{itemize}}

\item {\bf Variation 3: Radio}: template fitting techniques typically do not use wavelengths longer than the mir-infrared, since the templates are not well defined at long wavelengths. However, ML methods can in principle use radio data too. To our knowledge, no systematic study of whether the radio fluxes may be used to help constrain photoz's has been performed using ML methods. 
To see if the inclusion of radio data made a difference, testers were asked to incorporate radio data in their tests in the ``radio=Y'' variation, and to ignore them in the ``radio=N'' variation. This did not affect the size of the training and test sets.
 
\item {\bf Variation 4: X-ray AGN}:  In previous template fitting in the COSMOS \citep[e.g.][]{Salvato:2009zw}, 
%Lockman Hole, (E)Chandra Deep Field South, and EGS fields  
X-ray detected sources were treated differently, by using different libraries and priors. 
Here we test how ML techniques are affected by the presence of these sources and whether knowing in advance that they are AGN can help to improve the results.
%The larger the size of the field, the larger is the fraction of outliers (bright AGN)  and the lower is the reached accuracy (MARA NEEDS TO RE-WRITE THIS PART). 
%{\bf Duncan's: note  Surely its more that if theres a bright AGN then the bands where the AGN is bright should be excluded? in other words, a SED prior based bands choice?}
To see if the presence of AGN affected the results,  X-ray AGN detected by XMM were excluded from the training sets  in the ``X-ray=N'' variation, and included in the ``X-ray=Y'' variation. However, it should be noted that since most radio-loud AGN are not detected in X-rays, this test does not remove all AGN. Furthermore, Chandra data in this field were not available at the time of these tests and so were not used in this process.
As a result of this variation, the size of the training set used in the ``X-ray=N'' tests was reduced to $302$ and $278$ objects respectively in the ``bright'' and ``random'' samples, and the size of the test set was correspondingly increased to $455$ and $479$.
\end{itemize}

These four variations are not independent and so each of them must be combined along with the other variations.
The combination of these variations resulted in a  set of $16$ 
%tests
%\ray{
\pointless{experiments}, 
listed in Table \ref{TAB:summary_exp}, which shows the size of the training and test sets and the number of features.
 In all cases (except for Le Phare), the tests were trained on the training set provided, and then applied to the entire test set, which includes members of the training set. The control group then removed the sources in the training set from the test set before evaluating the results. \pointless{Le Phare was tested only in the variation 1,2,4, with X-ray detected AGN either treated as galaxies (X-ray=N), or separately (X-ray=Y).}

\begin{figure*}\centering
\includegraphics[,scale=0.4]{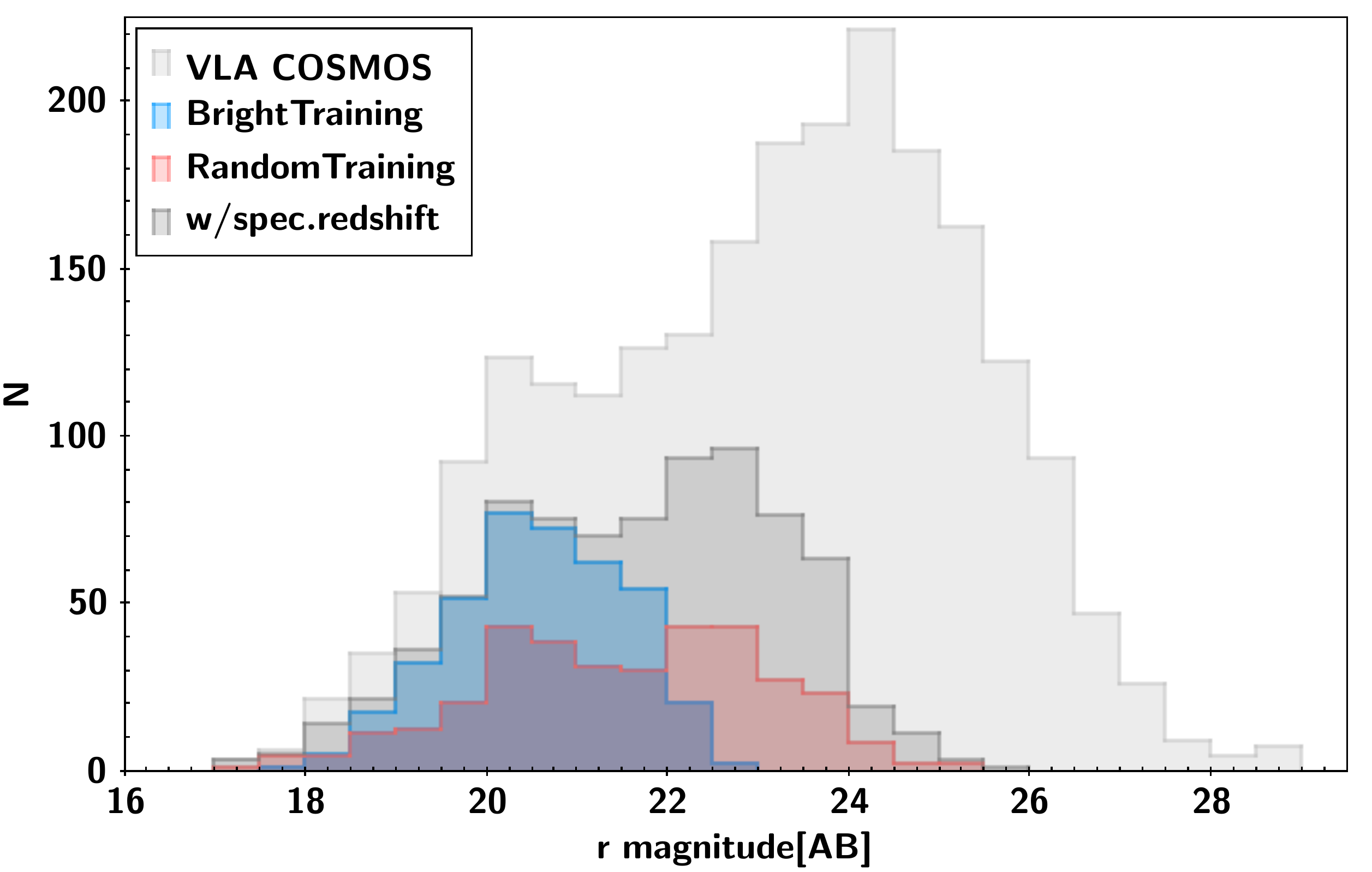} a)
\includegraphics[,scale=0.4]{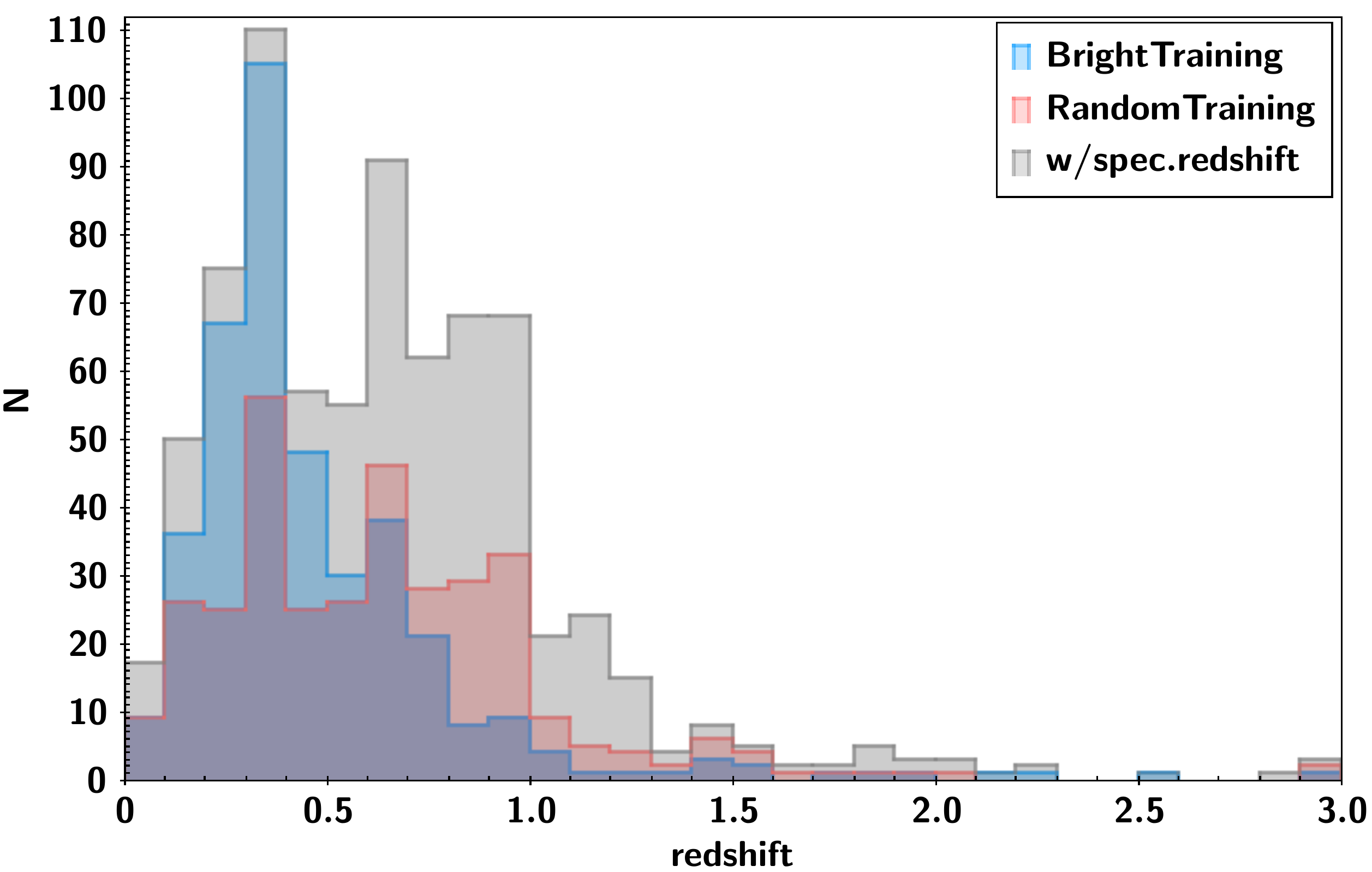} b)\\
\caption{Histograms of the brightness and redshift distributions of the whole KB and the bright and random samples, with respect to the distribution in magnitude and redshift for the entire sample.  
Panel a) the R magnitude distribution of the entire VLA COSMOS data set, the subset of those with redshifts (which constitutes the whole KB in this paper), and the bright and random training samples. Panel b):  the redshift distribution of the entire KB in this paper, and the bright and random training samples. 
}
\label{FIG:histo}
\end{figure*}

\pointless{In Figure \ref{FIG:histo} we show the resulting magnitude and redshift distribution for the bright and random training and test sets. We also report the magnitude distribution of the entire sample.}

%\mbc{[Comment: referee 2 explicitly ask to improve basic information about the different training and test sets. In particular to provide a better overview on used data that could infer knowledge about the critical aspects referred to the difficulties of supervised learning methods. The referee suggests to put here redshift histograms of various training and test sets and colour-redshift plots to help in the analysis of the presented variations. I think it is a very good suggestion...]}

 For consistency in this paper, we adopt the following terminology.  An ``experiment'' is one of the sixteen listed in Table \ref{TAB:summary_exp}. A ``test'' occurs when one of these experiments is undertaken by one of the six ``methods'' listed in Sec.~\ref{SEC:methods}.
 
\begin{table}\centering

\resizebox{\textwidth}{!}{
\begin{tabular}{llllccccc}
\hline
Id.    & CODE &  KB bias   & depth & Radio  & X-ray & \# training & \# test & \# training\\
 & & & & & & sources & sources & features\\
\hline
A1   & BDNY & Bright       &  Deep             & N                 & Y  & 391 & 366 & 44\\
B1   & BDYY & Bright       &  Deep             & Y                 & Y  & 391 & 366 & 45\\
C1   & BDNN & Bright       &  Deep             & N                 & N &  302 & 455 & 44\\
D1   & BDYN  & Bright       & Deep              & Y                 & N  &  302 & 455 & 45\\
E1   & BSNY  & Bright       & Shallow      & N                 &Y  & 391  & 366 & 44 \\
F1   & BSYY  & Bright       & Shallow      & Y                 & Y  & 391   & 366 & 45 \\
G1   & BSNN & Bright        & Shallow     & N                 & N &  302 & 455 & 44\\
H1   & BSYN  & Bright        & Shallow     & Y                  & N&  302 & 455 & 45\\
A2   & RDNY  & Random       & Deep            & N                 & Y & 343 & 414 & 44\\
B2   & RDYY  & Random    & Deep             &  Y             & Y& 343 & 414 & 45\\
C2   & RDNN  & Random      & Deep            & N                &  N & 278 & 479 & 44 \\
D2   & RDYN  & Random     &  Deep             & Y               &  N & 278 & 479 & 45\\
E2   &  RSNY  & Random    &  Shallow    & N               & Y    & 343 & 414 & 44\\
F2   & RSYY  & Random    & Shallow     & Y                & Y& 343  & 414 & 45\\
G2   &  RSNN & Random    & Shallow     & N               &  N & 278 & 479 & 44\\
H2   & RSYN & Random    & Shallow     & Y               & N & 278 & 479 & 45\\
\hline
\end{tabular}
}
\caption{Summary of the $16$ experiments. 
Column $1$: identification code; 
column $2$: mnemonic code;  
column $3$: Bright or Random training set; 
column $4$: shallowness of optical/IR data; 
column $5$: radio fluxes used (Y) or not used (N) in training; 
column $6$: bright X ray detected AGN included (Y) or not included (N) in the training set;
column $7$: number of sources in the training set;
column $8$: number of sources in the test set;
column $9$: number of features in the training  set. The test set has the same number of features minus one, corresponding to the 
%\mbc{target (
\pointless{spectroscopic redshift}. 
%%\ray{some of these numbers need to be changed by 2 as the KB has 757 not 759. But first check the data tables to see where the discrepancy has come from.}
%{\bf General consideration: while working on the paper I found very useful to label the experiments not with the present coding but with a four digits code such as BDYY (BRIGHT, DEEP, YES RADIO YES X-RAY). I added it to the table. I would recommend to remake the plots using this coding and following the order adopted to describe the results from MLPQNA, which makes much more sense.}
}
\label{TAB:summary_exp}
\end{table}

%Of these 2242 sources, only  324 (14.4\%) are detected either in XMM- or Chandra-COSMOS \citep[see][for the X-ray to optical/NIR association]{Brusa:2010lr,Civano:2012ys}. Of these 324 sources, only XXX will be detected by eROSITA. 
% 231 of which are X-ray detected by XMM/Chandra and YYY detected by eROSITA.
%{\bf Important note from Duncan: \\
%It would be useful here to show two representative samples from COSMOS, one selected in the radio and another selected in the X-ray, or optical, or infrared. Then, show that the radio selection does indeed give  a clearly different source type/redshift distribution. This would help justify this paper as a standalone paper to investigate the issues in finding photometric redshifts specific to radio surveys. Otherwise, a referee could well just say ?this has been done before in theoretical and we don?t need to do it again in the radio?. \\

%ADDITIONAL COMMENT by Giuseppe:\\
%Along the same line, it could be useful at least for a couple of esperiments (possibly already present in literature) to compare the results
%obtained here in this paper with those obtained for an optically selected sample.
%}
%
%{\bf GENERALREMARK: different methods end up using different datasets for training in the same experiments.
%Each team should be more specific about the pruning criteria adopted. This is also needed to justify the different figures encountered in the
%following tables.}

\section{The methods}\label{SEC:methods}

We used the Le Phare template fitting method as a baseline. and compared  four  different machine learning based methods with it: the k-nearest neighbours (kNN), 
%the Self Organizing Maps (SOM), 
two different implementations of Random Forest (RF), and the Multi Layer Perceptron with Quasi Newton Algorithm (MLPQNA). Here we discuss the implementation of each of them.

For each ML test, the method used the training sets whose size is shown in Table \ref{TAB:summary_exp}, and the test sets consisted of all members of the KB that were not included in the training set. The size of the test set for each experiment is shown at the top of Table \ref{TAB:results}. 
\pointless{The number of redshifts estimated is generally lower than this number, for two reasons. First, each algorithm has different ways of handling missing values. The Dameware platform, used by RF-NA and MLPQNA, removes features or objects for which data are missing, resulting in the reduced size of the data sets shown in Table 2. kNN also omitted objects with missing values, but where possible used subspaces of features that \emph{were} detected to measure an alternative kNN distance. The R platform, used by RF-JHU, uses the \textit{rfimpute} function which replaces missing values by a weighted average of those with similar redshift using another random forest.} 
Furthermore, before analysing the results, we removed from the test set any invalid results in which the method had failed, as indicated by a \pointless{$NaN$ (Not-a-Number)} in the estimated redshift. The actual number of valid results in the test set for each test is indicated by N in each entry of Table \ref{TAB:results}.

%\noindent {\bf Duncan's note: New note  This is a nice description, and could in some ways replace the description thats in the abstract. On a related note, it sounds like pure SED fitting is a subset of this approach - could we describe it that way instead of as a distinct approach?
%
%\noindent Answer: to Duncan's comment. I believe that SED fitting is a completely different approach and, in particular, that the roles played by biases 
%in the parameter space are completed different in the two families of methods. Therefore I would rather leave things as they are keeping the two families disjoint.  
%}.

\subsection{Le Phare}\label{SEC:Le Phare}
The Le Phare code is public\footnote{\url{http://www.cfht.hawaii.edu/~arnouts/Le Phare/Le Phare.html}} \citep{arnouts1999,ilbert2006} and has been extensively tested \citep[e.g.,]{hildebrandt2008,dahlen2013,cavuoti2017, santini15} for photoz and stellar mass computations. It is also the code used by the COSMOS collaboration. In this way we can easily compare the results  between radio and optically selected galaxies. 
%Its reliability and accuracy have been subject to careful analysis \citep{hildebrandt2008,dahlen2013,cavuoti2017} 
\pointless{Le Phare can account for the intergalactic medium, and for the contribution of the emission lines to the templates. 
The code also allows the use of various types of priors. Here we used the absolute magnitude, which enables better constraints on the photometric redshifts of AGN \citep[e.g.,]{Salvato:2009zw, Salvato:2011mz}.}
The photometric redshift can be computed either by minimising the $\chi^2$ (z$_{\rm BEST}$), or by marginalising over the best solution for each of the templates in use ($z_{\rm ML}$).  In both cases, in addition to the  (z$_{\rm BEST}$) and ($z_{\rm ML}$), $1 \sigma$ and $3 \sigma$ upper and lower errors are computed, 
\pointless{together with} 
the probability distribution function P(z). 
%\mbc{[Comment:] referee 2 asked to improve a critical discussion about Le Phare method, since it is the only one used here as SED fitting method. In particular some words about sensitivity to template set and mag zeropoint corrections. In the following I tried to put something:] 
\pointless{Le Phare shares the properties of most common SED template fitting methods \citep[e.g.,][]{hildebrandt2010, dahlen2013}. For example, in most such methods, a residual bias  may be caused by an imperfect fit of the template to the data, inaccurate photometry, or insufficient data.} %Moreover, as is common in  SED fitting methods,  increasing the number of photometric bands may offer a limited improvement, because of residual uncertainty caused by the recalibration of zero points. 
However, Le Phare has frequently shown a better robustness than some other algorithms to such problems \citep{hildebrandt2010}.

\pointless{In this work, we used the library  of templates and extinction laws from \cite{ilbert2009} for all the sources in the  experiments where the information on the X-ray nature of the source was ignored. In the experiments where the X-ray information was used, the X-ray selected sources were treated separately, using the same libraries and extinction laws as in \cite{Salvato:2011mz}, who also used X-ray and morphological information before setting the priors and choosing the correct set of templates. Then, depending on whether the source was point-like or extended in optical images, a prior on absolute magnitude was imposed when running Le Phare.}

\pointless{For the extended sources, M$_{\rm g}$ could vary between -8 and -24 (typical range for normal galaxies and low luminosity AGN, dominated by the host). For the point-like sources, we allowed a range in M$_{\rm g}$ between -20 and -30, which is typical for bright AGN and QSO, also at high redshift.}
%we used the library of templates and extinction laws as  \cite{ilbert2009}. 
%For the X-ray detected sources we used the same libraries and extinction laws as in \cite{Salvato:2011mz}, who  also used X-ray and morphological information
%s where taken into account  
%before setting the priors and choosing the correct set of templates.
\pointless{For all the experiments we allowed a redshift solutions up to $z=7$, in steps of $0.01$.}
%RPN: I dont agree with the following statement, since radio flux and an IR upper limit can be sufficient to constrain a high-z quasar (see Orenstein+2018): Note that for the computation of real sources in EMU, a lower redshift limit should be considered, depending on the depth of the data available. Higher redshift will be allowed for sources in the DES area, while for the regions of sky where only shallow data are existing, the redshift should not be allowed to be higher than 2-3.}
%It is important to note that  template fitting methods can use mid-IR and radio data for computing stellar masses or SFR, once the redshift is known, but these bands cannot be used to compute photometric redshifts since current templates degenerate at these wavelengths. 

Since template fitting does not use a training set, and the overall results do not depend on the size of the test set, it would be meaningless to run Le Phare for those different experiments that differ only by varying the training set. We therefore ran Le Phare on the entire sample of $757$ objects, in four different experiments as shown in Table 3.

%\noindent {\bf  GENERAL COMMENT: different groups have used different pruning criteria to produce their data sets for training 
%and test. While I am sure we must give the details in the paper, I am undecided on wher to put it, i.e. in this section or rather 
%in the individual experiments. PLEASE COMMENT and PROVIDE THE NEEDED info. 
%
%\noindent MARA: add something about the actual implementation of the method used for the experiments and on the criteria used to 
%prune the data set.}.

\subsection{kNN}

kNN is a standard regression method in machine learning, and takes into account the similarities of objects in the $n$-dimensional feature space defined by the input parameters. kNN is widely used for computing photometric redshift \citep{gieseke2011,polsterer2013, cavuoti2017}, and has also  been used to estimate redshift uncertainties for SDSS sources \citep{oyaizu08}.

The principle of kNN is that the observed parameter space (OPS) defined by the input features is populated using the objects in the spectroscopic KB. Typically, the features are photometry measured at a number of optical and infrared wavelengths, but as radio photometry becomes deeper in next-generation radio surveys, radio features are also likely to be important.  In addition to photometric measurements, the feature space can also include other source characteristics such as polarization or morphological parameters.

The photometric redshift for an object not in the KB is then evaluated by  looking at the $k$ nearest neighbours of the object in the OPS, where
 ``nearest neighbour'' of a source is defined by the  Chebyshev distance $d$ to other objects in the parameter space, defined as: 
\begin{equation}
d^a=\sum_i | x_i - x_0 |^a
\end{equation}
where  $x_i$ and $x_0$ are, respectively  the coordinates in the $n$-dimensional feature space (i.e. the OPS) of the $i^{-th}$ point in the KB, and of the point for which we need to derive the estimate; $a$ is an exponent which can be adapted to the problem. 
In the case $a=2$, the Chebyshev distance reduces to the traditional Euclidean distance, while  $a=1$ leads to the so called  Manhattan distance. 

``$k$'' is an integer number  typically about $10$, but the optimal value can be inferred by  running the routine in a loop with $k$ varying within a certain range and then comparing the results based on a specific evaluation metric. 
The optimal value for $a$ can also be identified empirically by looping through different values of $a$. 

Once the $k$ nearest neighbours with a spectroscopic redshift have been identified, the redshift of the queried source can  be obtained as, e.g. the mean or the median of the neighbour redshifts.

The computational time needed to execute kNN can be dramatically shortened by implementing specific spatial data structures, e.g. kd-trees \citep{gieseke2011,polsterer2013}, enabling its application  to massive data sets \citep{zinn2012b, luken18}.

%\noindent {\bf Add something about the actual implementation of the method used for the experiments}.

The kNN algorithm works best if the reference data set are shaped such that the feature space is populated homogeneously, i.e. avoiding strong concentrations in a certain region, or sparsely-populated regions. In the tests described here,  we made no correction for any excess sources in any of the given training samples. 

Instead, to distribute the  photometric data over the parameter space  for each band, 
\pointless{we replaced the magnitudes $mag_i$ and $mag_{i+1}$ by a colour $c_i = mag_i - mag_{i+1}$. This operation reduces the effective number of dimensions $n$ of the feature space  by 1. We then  normalized the colours, to distribute the data uniformly in the feature space, by replacing each colour $c_j$ by $(c_j/c_{max})$, where $c_{max}$ is the maximum value of $c_j$. } 
%\mbc{Instead, to distribute the photometric data over the parameter space for each band, we  first derived standard colours from the given magnitudes, which causes the reduction of the effective amount of dimensions of the feature space from $n$ to $n-1$. [Comment: the rest of the paragraph is really strange. As written, it seems that we replace $c_{max}$ by $1$ and then all other colours by $(c_j/c_{max})$, which results to divide by $1$ all other colours, without any effect...???].
 
The  redshift of each object in the test set was then estimated by taking the mean redshift of the $k$ nearest neighbours of the object. For this work, the  optimal value of $k$ was inferred by choosing the value of k that gave the best fit in the training set, and was allowed to vary from test to test, but was typically around $10$.

\pointless{In this implementation of kNN, missing data were handled by constructing  different independent sub-feature spaces, populated by the remaining data.  As a result, when the data are sparse (as in the shallow experiments) the number of training objects in any one feature space drops even more, resulting in poor performance when data are sparse and poorly-sampled. 
}
\begin{table}
\centering
\caption{Numbers of objects and features used in the MLPQNA and RF-NA training sets, after deleting invalid data. }
%%\ray{some of these numbers are larger than the numbers in Table 1 and are therefore erroneous.}
\begin{tabular}{|c|c|c|c|c|c|c}
%\tableline
\hline
Experiment & objects & features\\
\hline
A1 & 53 & 28 \\
B1 & 53 & 29 \\
C1 & 16 & 28 \\
D1 & 16 & 29 \\
E1 & 391 & 7 \\
F1 & 391 & 8 \\
G1 & 302 & 8 \\
H1 & 302 & 9 \\
A2 & 316 & 29 \\
B2 & 316 & 30 \\
C2 & 269 & 29 \\
D2 & 269 & 30 \\
E2 & 270 & 5 \\
F2 & 270 & 6 \\
G2 & 228 & 5 \\
H2 & 228 & 6\\
\hline
%\tableline
\end{tabular}
%\tablecomments{}
%\end{center}
\label{TAB:MLPQNA}.
\end{table}

\subsection{Random forest}

Random Forest is a  popular non-parametric regression technique which learns by generating a forest of random decision trees, by following the variations in the parameter space of the training sample objects. Random Forest combines two successful techniques (CART \citep{breiman1984} and  Bagging \citep{breiman1996}) with a novel approach to dimensionality reduction \citep[random subspace sampling;][]{ho1998} to produce an ensemble classifier \citep{breiman2001}. This method has been successfully applied to photometric redshift estimation \citep[e.g.,][and references there]{carliles2010,carrasco15, cavuoti2017, fotopoulou18, salvato18}. %%\ray{add more references}.

The problem is posed as the estimation of the conditional mean of the response values (redshift in this case) conditioned on feature values (photometry). The only assumption about the population is that its distribution is continuous. The algorithm first generates a set of bootstrap samples 
%selected with replacement and uniformly at random 
from the training set.  Each bootstrap sample is then used to train a separate randomized regression tree.  
Each tree recursively partitions the training set. 
%\ray{
%\pointless{To maximize the likelihood of containing the correct redshift}
%\rayq{is this an acceptable plain-english substitute for ``according to the optimal reduction in empirical risk'' as requested by ref 2?}
%\pointless{according to the optimal reduction in empirical risk} 
%\mbc{the bold text is asked to be simplified and clarified, according to a request from referee 2
%chosen over all possible split points in a randomly selected subset of feature dimensions at each node. 
%\textcolor{red}{Mara:the sentence above does not make sense.}
Partitioning stops \pointless{when a node} contains a minimum specified number of training set objects, typically between five and ten. The mean response over the points in the node is that individual tree's estimate of the response for any new test points that are classified into that node. The ensemble estimate for a given point is then the mean of the individual tree estimates for that point.

Some advantages of this technique are: (\textit{i}) the training time scales relatively well with feature dimensionality, naturally tending to condition on those features which are most informative, while the regression time is essentially independent of dimensionality; (\textit{ii}) it is computationally relatively fast after the training phase even with low-dimensional input spaces; (\textit{iii}) it behaves somewhat like kNN, but with the benefit of being invariant to scale differences between dimensions; (\textit{iv}) it assumes very little about the underlying population distribution; and, finally, (\textit{v}) it has very few user-defined parameters, and thus has a gentle learning curve, typically performing well even with default parameter values.

The primary caveat with this technique is that, as a non-parametric technique, it relies heavily on data, so that data sparsity may become an issue as the dimensionality of the feature space increases, as is the case in the present setting. We used two different implementations of RF. 

The RF-NA implementation used the DAMEWARE platform\footnote{http://dame.dsf.unina.it/dameware.html} \citep{brescia2014}, with $1000$ trees without limiting the depth of the tree (i.e. the nodes were expanded until all leaves are pure). \pointless{It used the raw photometry, rather than deriving colours from the photometry.} Unfortunately the input data for RF-NA had to be heavily censored to remove objects or features that contained missing values, resulting in the numbers of objects and features shown in Table~\ref{TAB:MLPQNA}.

The RF-JHU implementation used the R package \citep{R13}, with $500$ trees, \pointless{working on the raw photometry, rather than deriving colours from the photometry.}
%\textcolor{red}{Max, is this also true for RF-NA?} 
\pointless{Missing data were replaced where possible by interpolated data using the \emph{rfimpute} function,
 which replaces missing values by a weighted average of those with similar redshift using another random forest, resulting in more available data for some tests, compared to other techniques that used heavily censored data.} 
\subsection{MLPQNA}

The MLPQNA algorithm is a type of neural net consisting of a multilayer perceptron (MLP) in which the learning rule uses the Quasi Newton Algorithm (QNA) to find the stationary point of a function. MLPQNA makes use of the  Limited memory - Broyden Fletcher Goldfarb Shanno (L-BFGS) algorithm \citep{byrd1994}, originally designed for problems with a large number of features. 

%%%%%%%%%%%%%% NEW
%\mbc{
\pointless{The implemented MLPQNA model uses Tikhonov regularization \citep{tikhonov1995}, based on the weight decay as regularization factor. When such factor is accurately chosen, then generalization error of the trained neural network can be improved, and training can be accelerated. In real problems the best decay regularization value is unknown. It must be heuristically experimented within the range of $0.001$ (weak regularization) up to $100$ (very strong regularization) and has a strong impact on the computing time (the lower its value, the greater the computing time). In order to achieve the weight decay rule, the method minimizes the generic merit function $f = E + \lambda S/2$, where $E$ is the training error function, $S$ is the sum of squares of network weights and the decay coefficient $\lambda$ controls the amount of smoothing applied to the cyclic training process. The optimization is then performed until the training error is below the constant value imposed by the user (generally $0.001$). In particular, to estimate the photo-z regression result at each training iteration, MLPQNA uses the following composite merit function, based on the Least Square Error + Tikhonov regularization,}
$$f = \sum\limits_{i=1}^{N_{sources}} \frac{(y_i - t_i)^2}{2} + \frac{\lambda \left \| W \right \|^2}{2}$$

\pointless{where $y$ and $t$ are, respectively, output (estimated photo-z) and target ($z_{spec}$) for each input source, while $\lambda = 0.1$ and $W$ is the MLP weight matrix.}

\pointless{The learning rule used to update the MLP weight matrix at each training iteration makes use of the Quasi Newton rule, a Newton's method based on the calculation of the Hessian of the training error, more effective in avoiding the local minima of the error function and more accurate in the error function trend follow-up, thus achieving a powerful capability to find the absolute minimum error of the optimization problem \citep{brescia2013}.}
%%%%%%%%%%%%%%

As for the RF-NA method, \pointless{it used the raw photometry, rather than deriving colours from the photometry, and} the input data had to be heavily censored to remove objects or features that contained missing values, resulting in the numbers of objects and features shown in 
Table~\ref{TAB:MLPQNA}.
\pointless{The analytical details of the MLPQNA method, as well as its performances on different data sets, have been extensively discussed elsewhere \citep{brescia2013,cavuoti2015,cavuoti2012,cavuoti2014b}.
For these tests we used the MLPQNA implementation that is available as a public service under the DAMEWARE platform \citep{brescia2014}.}

%\subsection{SOM}
%
%Self Organizing Maps (or SOM) are among the most popular tools for unsupervised clustering. 
%A SOM can be considered as a collection of nodes arranged in a grid of arbitrary dimension, although for visualization purposes two dimensions are most common. Each node is attached to a vector of weights w with the same dimension as an input training vector, t. The implementation used here is described by \cite{geach2012}.
%
%\noindent {\bf Jim Geach: please add a short description of the actual implementation. Leave mathematical details to references.}

%{\bf DUNCAN's Note: 
%
% - Its not obvious how the various specific methods map back to the general description given in section 3.0.  
%In other words, which ?parts? of the complete possible chain of photos method each one uses.
%
%\noindent ANSWER: PLEASE BE MORE XPLICIT
%
% - At this point in the paper I?m curious whether we are testing different methods with the same training and test sets, one method with different training and test sets, or all methods with ?their best? training and test sets. Each possibility would have different ramifications. 
%
%\noindent ANSWER: I would say that while training and test set are originally the same, the different pruning operated by individual teams (by rejecting NAN or bad data) lead to slightly differnet test sets. In this sense I believe that the best approximation is that all methods have
%tailored the same training and test set to match their specific needs.
%}

\clearpage
%\onecolumn
%\begin{landscape}
\begin{table}
\resizebox{\textwidth}{!}{\centering
\begin{tabular}{llllllllllllllllll}
\hline

Experiment         &          &A1        &B1        &C1        &D1        &E1        &F1        &G1        &H1        &A2        &B2        &C2        &D2        &E2        &F2        &G2        &H2        \\
\hline
Code               &          &BDNY      &BDYY      &BDNN      &BDYN      &BSNY      &BSYY      &BSNN      &BSYN      &RDNY      &RDYY      &RDNN      &RDYN      &RSNY      &RSYY      &RSNN      &RSYN      \\
Training set size  &          &       391&       391&       302&       302&       391&       391&       302&       302&       343&       343&       278&       278&       343&       343&       278&       278\\
Max test set size  &          &       366&       366&       457&       457&       366&       366&       457&       457&       416&       416&       481&       481&       416&       416&       481&       481\\
\hline
kNN                &N=        &       366&       366&       293&       293&       366&       366&       293&       438&       414&       414&       322&       322&       414&       414&       322&       322\\
                   &NMAD=     &      0.15&      0.15&      0.13&      0.14&       0.1&      0.48&       0.1&err       &      0.05&      0.05&      0.05&      0.04&      0.23&      0.24&      0.22&      0.22\\
                   &$\eta$=   &        56&        58&        58&        59&        31&        95&        28&        95&        18&        18&        11&        11&        49&        52&        49&        52\\
                   &$\beta$=  &        44&        42&        27&        26&        69&         5&        46&         5&        82&        82&        60&        60&        51&        48&        34&        32\\
\hline
RF-JHU             &N=        &       366&       366&       438&       438&       366&       366&          &       438&       414&       414&       467&       467&       414&       414&       467&       467\\
                   &NMAD=     &      0.11&      0.12&      0.12&      0.12&        43&      0.45&          &err       &      0.07&      0.07&      0.07&      0.07&      0.09&      0.09&       0.1&       0.1\\
                   &$\eta$=   &        28&        27&        28&        30&        95&        95&          &        95&        15&        15&        16&        16&        20&        19&        21&        19\\
                   &$\beta$=  &        72&        73&        69&        67&         5&         5&          &         5&        85&        85&        82&        82&        80&        81&        77&        79\\
\hline
RF-NA              &N=        &       366&       366&       293&       293&       366&       366&       293&       293&       414&       414&       322&       322&       414&       414&       322&       322\\
                   &NMAD=     &      0.13&      0.12&      0.16&      0.17&      0.11&      0.09&      0.12&      0.12&      0.07&      0.07&      0.06&      0.06&      0.13&      0.13&      0.11&       0.1\\
                   &$\eta$=   &        33&        25&        86&        83&        28&        22&        35&        33&        14&        15&         8&         7&        36&        36&        28&        25\\
                   &$\beta$=  &        67&        75&         9&        11&        72&        78&        42&        43&        86&        85&        62&        62&        64&        64&        48&        50\\
\hline
MLPQNA             &N=        &       366&       366&       293&       293&       366&       366&       293&       293&       414&       414&       322&       322&       414&       414&       322&       322\\
                   &NMAD=     &       0.2&      0.25&      0.15&      0.14&      0.13&      0.12&      0.08&      0.09&      0.06&      0.06&      0.05&      0.05&      0.12&      0.14&      0.11&      0.12\\
                   &$\eta$=   &        80&        88&        36&        31&        40&        40&        22&        27&        17&        19&        14&        13&        36&        38&        27&        32\\
                   &$\beta$=  &        20&        12&        41&        44&        60&        60&        50&        47&        83&        81&        58&        58&        64&        62&        49&        46\\
\hline
Le Phare           &N=        &       757&          &       571&          &       509&          &       549&          &       757&          &       571&          &       509&          &       549&          \\
                   &NMAD=     &      0.02&          &      0.01&          &      0.08&          &      0.08&          &      0.02&          &      0.01&          &      0.08&          &      0.08&          \\
                   &$\eta$=   &         5&          &         3&          &        22&          &        23&          &         5&          &         3&          &        22&          &        23&          \\
                   &$\beta$=  &        95&          &        73&          &        52&          &        56&          &        95&          &        73&          &        52&          &        56&          \\

\hline
\end{tabular}
}
\caption{Results of the $16$ experiments. Line $2$ of the header gives the code as described in \S \ref{approach}: Bias (Bright/Random), IR Depth (Deep/Shallow), Radio (Y/N), X-ray (Y/N). Column $1$: method name; column $2$: metric: N=number of redshifts  estimated, $\sigma$=standard deviation of estimated-true, $\eta$=percentage of outliers, $\beta$= overall success rate, expressed as a percentage, as defined in the text.
%\textcolor{red}{MARA: please report the values of Lephare also for the Bright experiments. even if they are the same. So that it is easier to compare the results.}
%%\ray{Need to find the other numbers for G1}
}
\label{TAB:results}
\end{table}

%\twocolumn
%\end{landscape}
\newpage
\clearpage

\section{Results of the experiments}
\begin{figure*}\centering
\includegraphics[,scale=0.4]{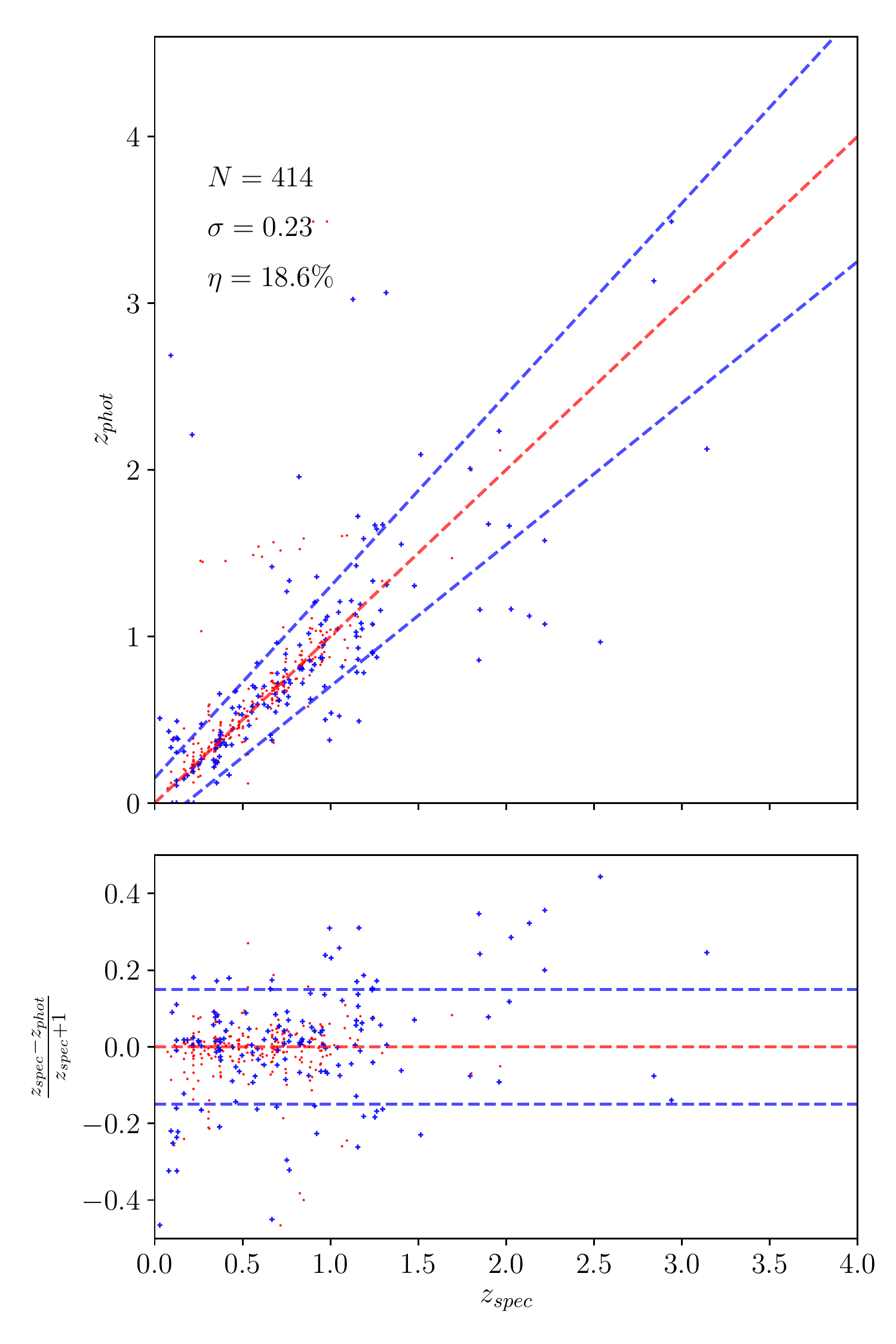} a)
\includegraphics[,scale=0.4]{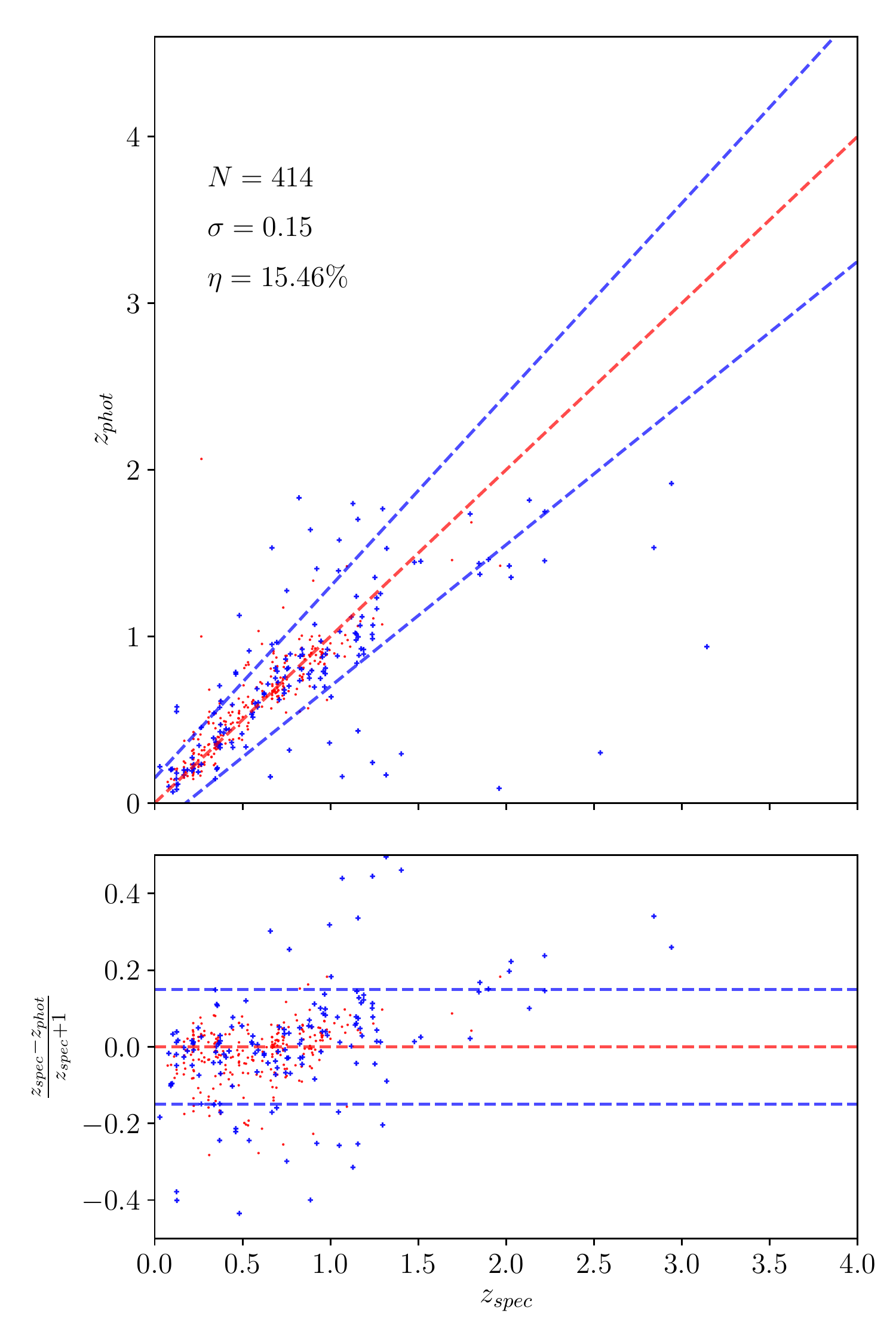} b)\\
\includegraphics[,scale=0.4]{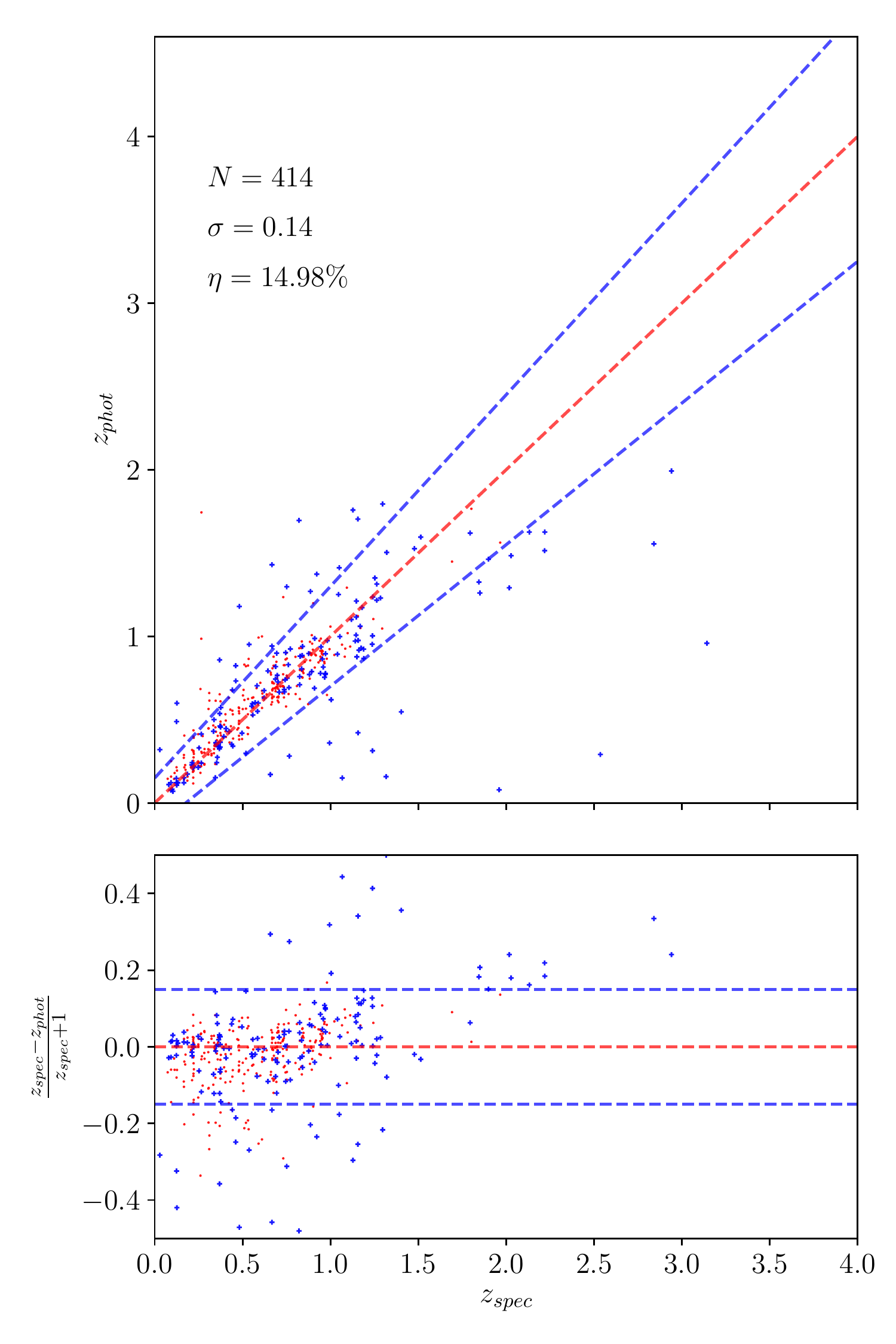} c)
\includegraphics[,scale=0.4]{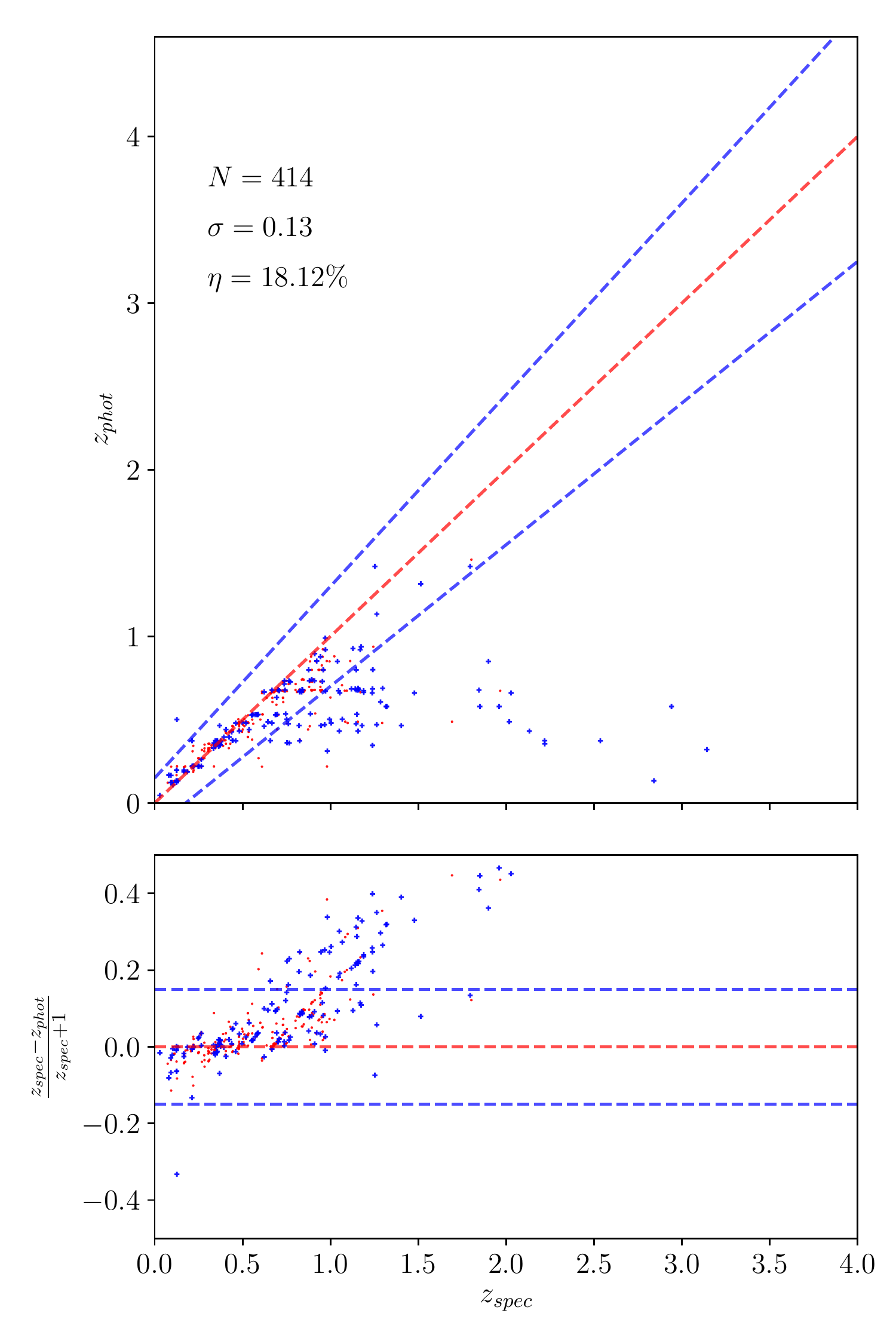} d)
\caption{Summary of the results obtained in the experiment B2/RDYY with the various methods. 
Panel a): MLPQNA.
Panel b): RF-NA. 
Panel c): RF-JHU.
%Panel d): SOM; 
%Panel e) Salvato; 
Panel d): kNN.
}
\label{FIG:blind_RDYY}
\end{figure*}

We summarise the results of the tests in Table~\ref{TAB:results} and present a plot for each test in Figures \ref{FIG:LePhare} to \ref{FIG:blind_RSYN_shallow}. We define the error on each measurement $\Delta z = z_{spec} - z_{phot}$, and we define an ``outlier'' as being a measurement for which $|\Delta z| \geq 0.15 * (1+z_{spec})$. 
The upper panel of each plot  shows the distribution of $z_{spec}$ vs. $z_{phot}$, and the lower panel shows the normalized residuals
$\frac{\Delta z}{z_{spec}+1}$ vs $z_{spec}$. The dashed blue lines mark the position of the outlier region defined by $|\Delta z|  \geq 0.15 * (1+z_{spec})$. The dashed red line marks the locus of $z_{spec}=z_{phot}$. 

In the Table we list: the number $N$ of points used in the test set;   $\sigma$ = standard deviation $(\Delta z )$ ; the normalised median absolute deviation  =  NMAD = $ 1.4826 * (median( |\Delta z |)$ ; and the fraction of outliers $\eta$ for which $ \Delta z > 0.15 * (1* z_{spec} )$. \pointless{Because the fraction of outliers does not take into account the reduction in the sample size due to missing or bad data, we also define an overall success rate $\beta$ as the number of correct (i.e. non-outlier) redshifts divided by the total sample size, expressed as a percentage.}
%We also list the kurtosis 
%\begin{equation}
% g_2 = \frac{\mu_4}{\mu_2^2} - 3 = \frac{\frac{1}{n} \Sigma ^n_{i=1} (x_i - \bar{x})^4}{(\frac{1}{n} \Sigma ^n_{i=1} (x_i - \bar{x})^2)^2} - 3
% \end{equation}
 %and the skew.
% \begin{equation}
% \gamma_1 = \frac{\mu_4}{\sigma^3} 
% \end{equation}
% Both the Kurtosis and Skew use the centered moments of x, which are calculated as $ \mu_k = E[(x - \mu)^k] $. 

\pointless{In each plot we identify the sources that were classified as AGN  on the basis of mid-infrared colours \citep{chang17, donley12}, or an X-ray detection by Chandra or XMM.}

\subsection{Results from the Le Phare experiments}
\begin{figure*}\centering
\includegraphics[,scale=0.4]{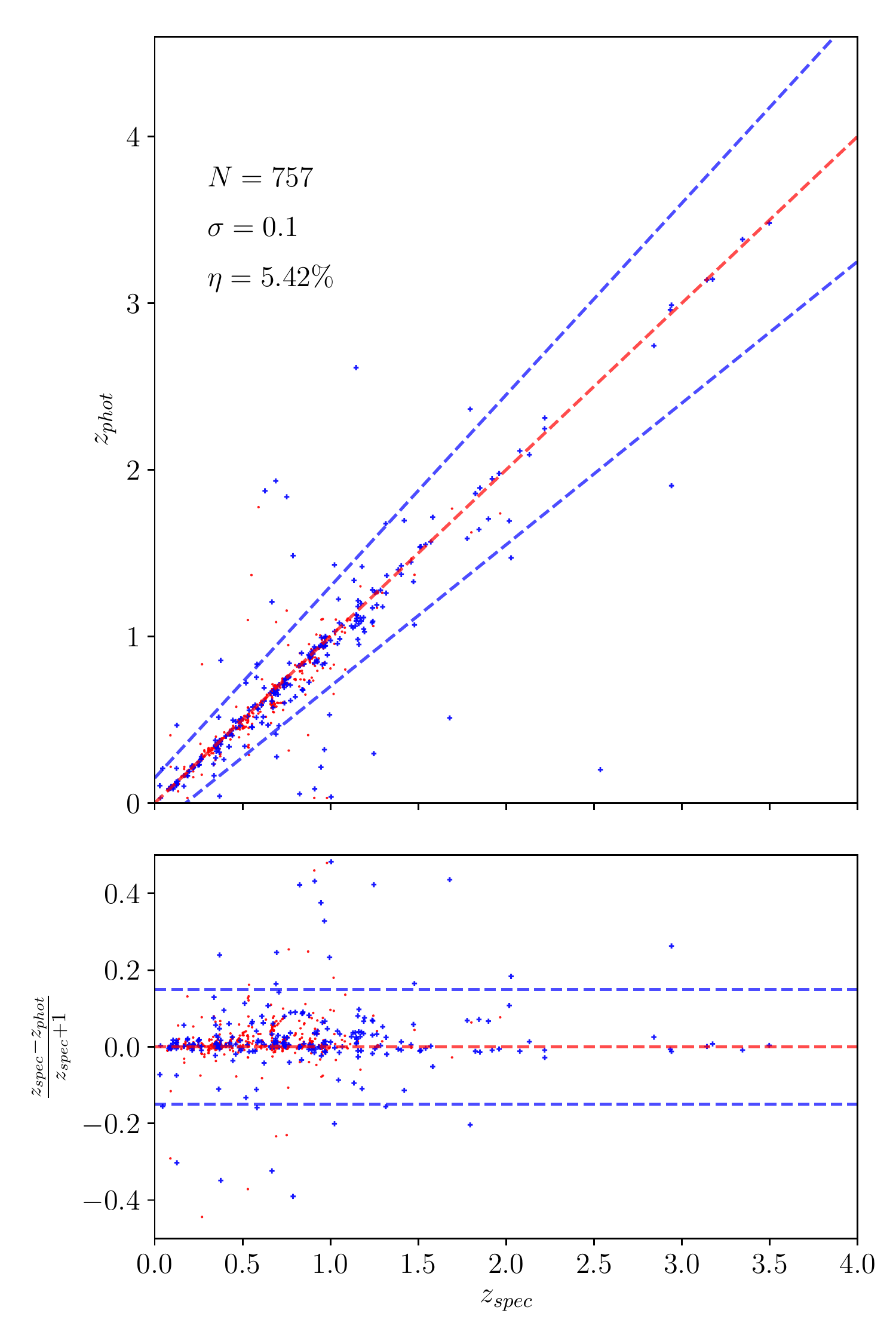} a)
\includegraphics[,scale=0.4]{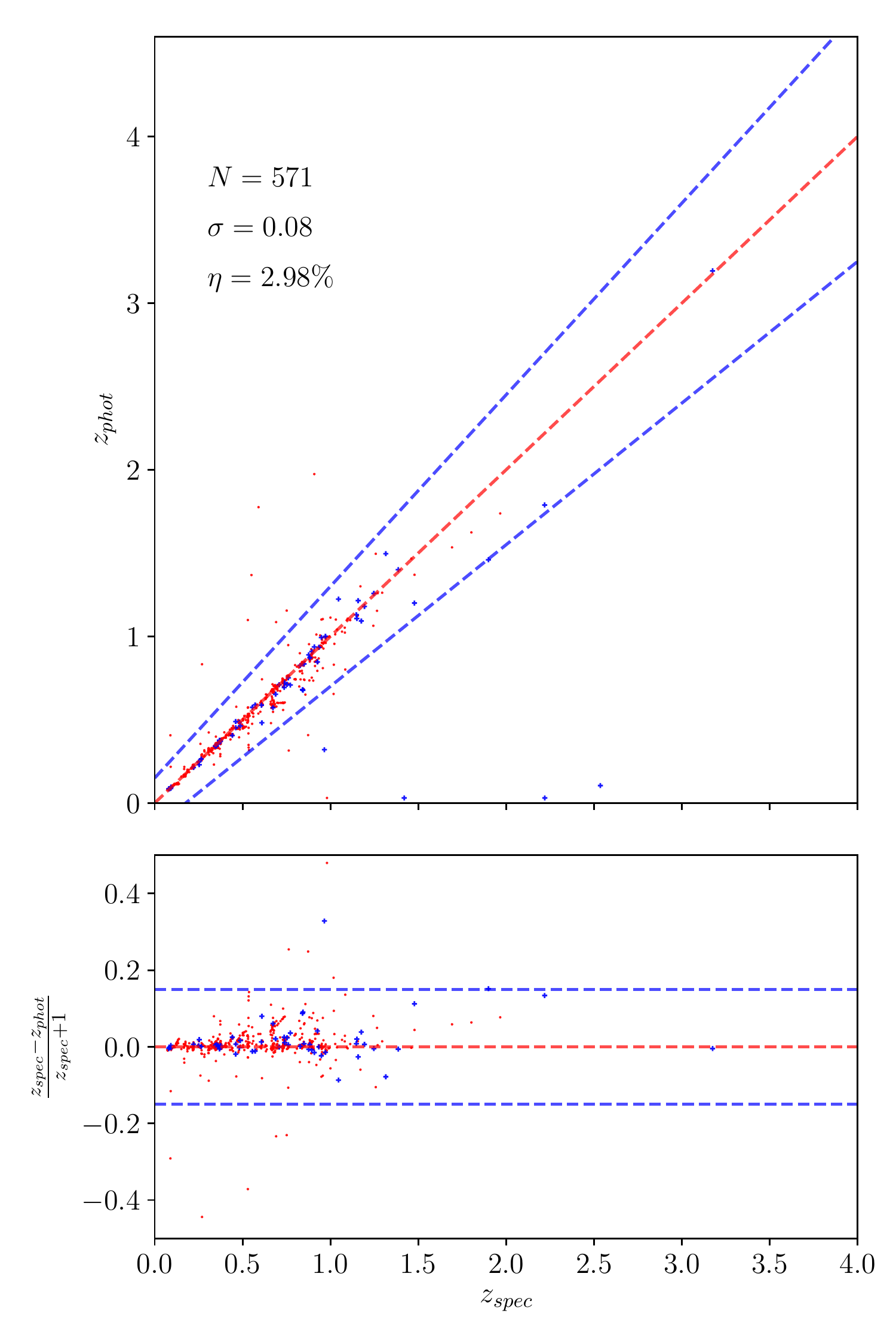} b)\\
\includegraphics[,scale=0.4]{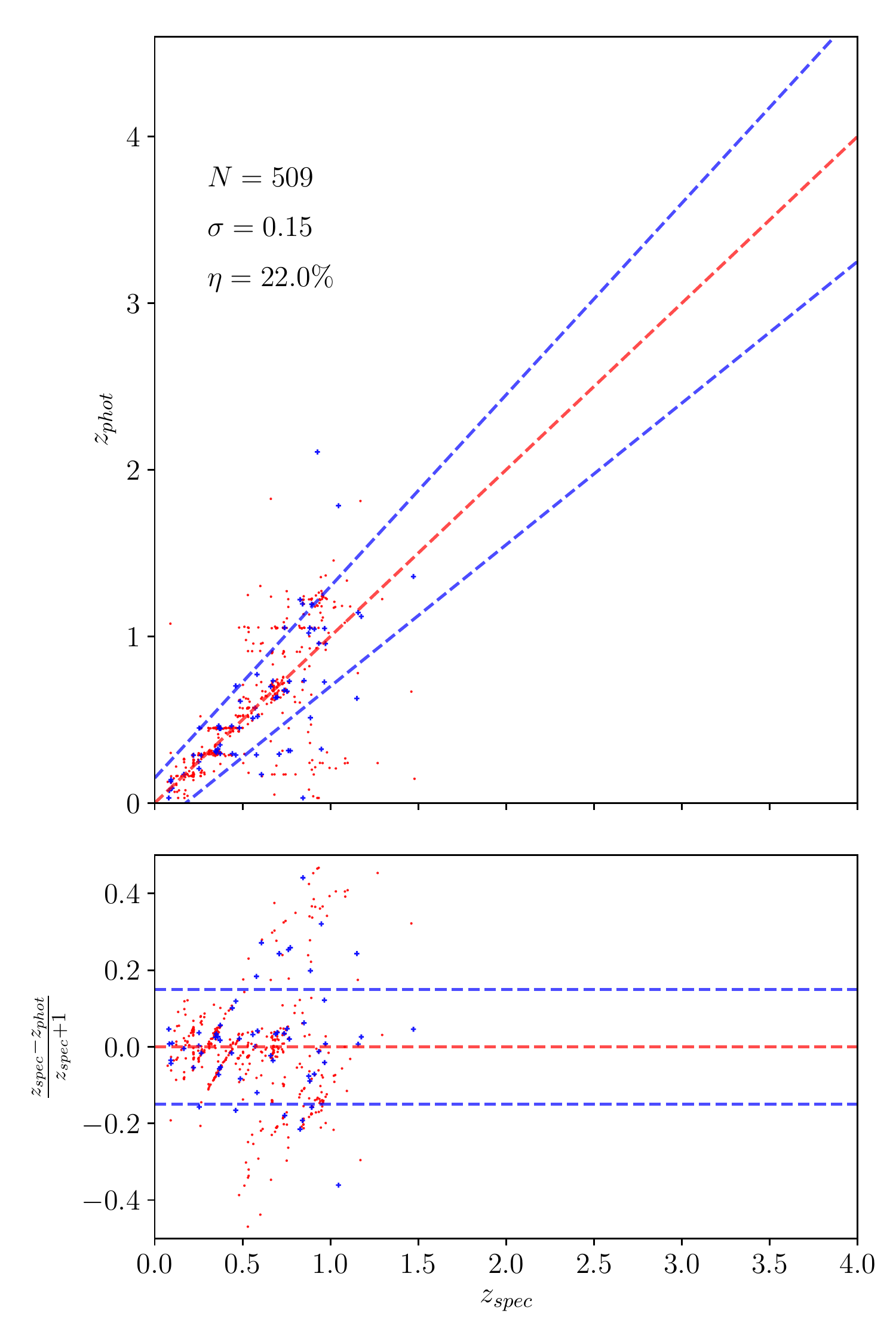} c)
\includegraphics[,scale=0.4]{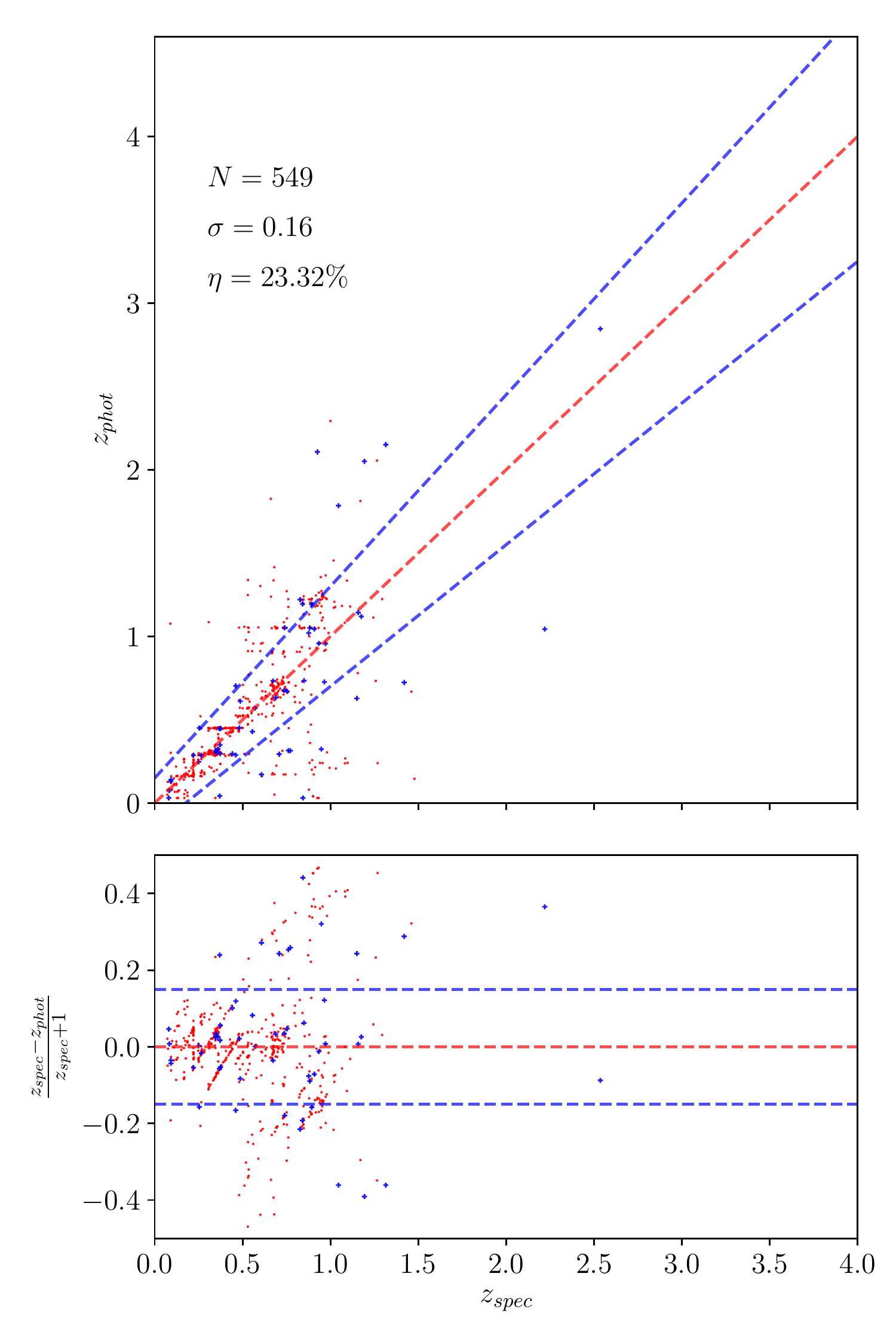} d)
\caption{Summary of the results obtained from Le Phare in four experiments.  \pointless{In this and all subsequent figures, blue crosses indicate AGN and red dots indicate non-AGN, as described in the text. The dashed blue lines mark the position of the outlier region defined by $|\Delta z|  \geq 0.15 * (1+z_{spec})$. The dashed red line marks the locus of $z_{spec}=z_{phot}$. } 
Panel a) Experiment A2/RDNY, using all available data. Panel b) Experiment C2/RDNN, using all data but excluding X-ray sources. Panel c): Experiment E2/RSNY, as (a) but using optical/IR data with reduced sensitivity. Paned d) Experiment G2/RSNN as (b) but using optical/IR data with reduced sensitivity \pointless{and eliminating the narrow and intermediated band photometry, as it will not be available for the entire EMU survey. As expected, the limited availability in photometry, produce systematic errors, when key features like the  4000\AA break are falling between the bands.}.
%\textcolor{red}{Mara:the plots are nice but I would write inside each plot also the name of the algorithm used and the experiment, so that one does not need to look continuously to the caption.and I would increase BY A LOT the size of the symbols, fonts of the axis and legends. Blue and black are also difficult to distinguish. }
}
\label{FIG:LePhare}
\end{figure*}

Le Phare does not use a training set, and instead fits the data to astrophysically-derived templates.  We therefore ran Le Phare on the entire sample of $757$ objects. The templates do not include radio photometry, and thus for none of the experiments with radio=Y we have a result from Le Phare. The results are shown in Figure~\ref{FIG:LePhare}.

Predictably, Le Phare performs better than any of the other techniques when using all available data, but progressively degrades as the quality of the available data is reduced. \pointless{In particular, when only shallow, broad band photometry is available, systematics appear --photoz constant over a large range of redshift ranges-- because the key features such as the 4000\AA\  fall within bands.}
%\textcolor{red}{Mara: we should comment on the differences between X-ray=Y and X-ray=N, but we need first to see the plot.}
%
%\noindent{\bf NOTE: A simple inspection of the plots shows that the usual statistical indicators fail to provide a good estimate of how the various methods work. Look for instance to Figure 1, where most plots basically do not provide any useful results but some statistical indicators still show appreciable results. Should we use other indicators?}
%
%
%{\bf IN THE CASE WE DECIDE TO PUT HERE THE DETAILS ABOUT THE PRUNING, PLEASE PROVIDE INFORMATION for all EXPERIMENTS. }
%

\subsection{Results from the BRIGHT biased experiments}

\begin{figure*}\centering
\includegraphics[,scale=0.4]{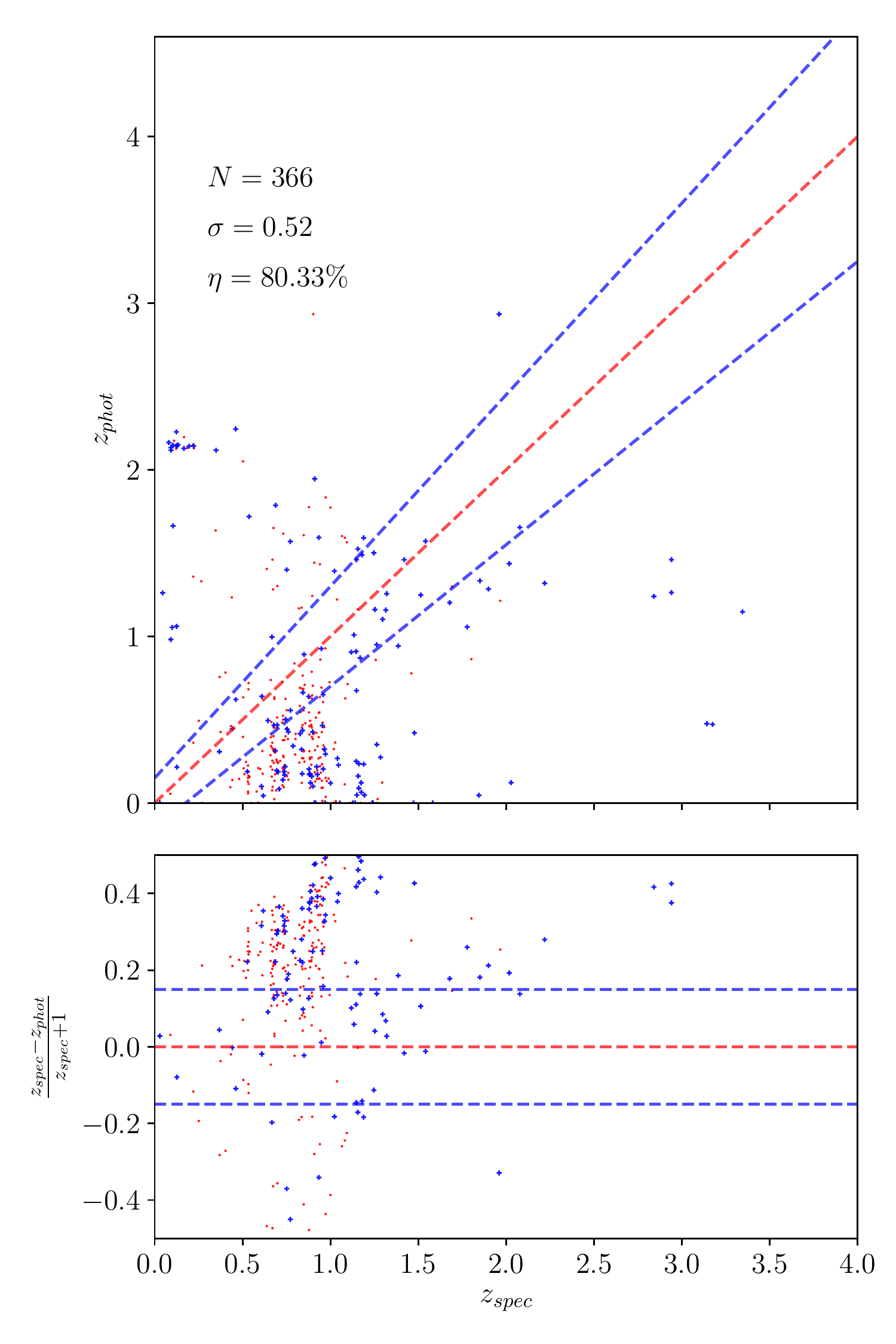} a)
\includegraphics[,scale=0.4]{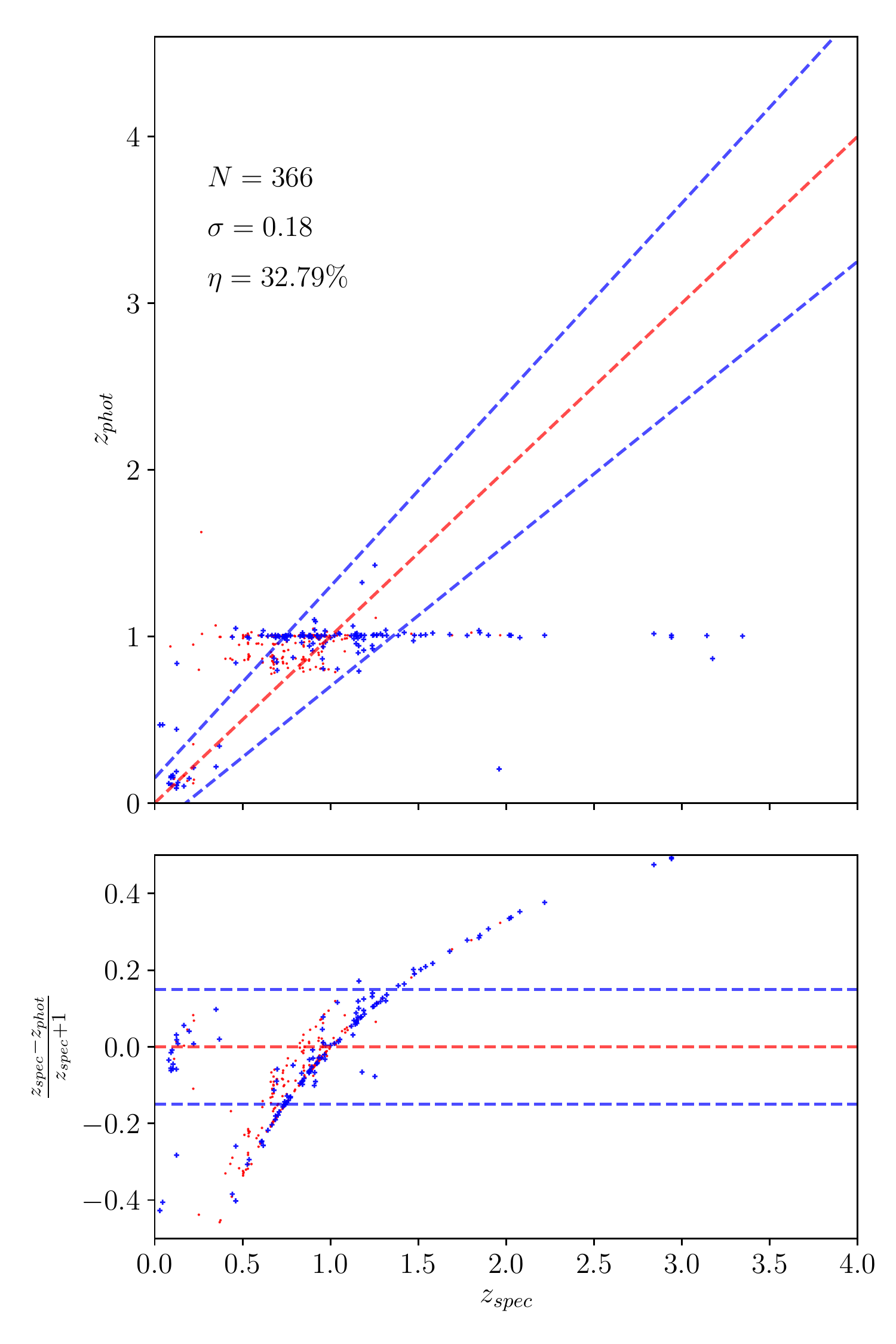} b)\\
\includegraphics[,scale=0.4]{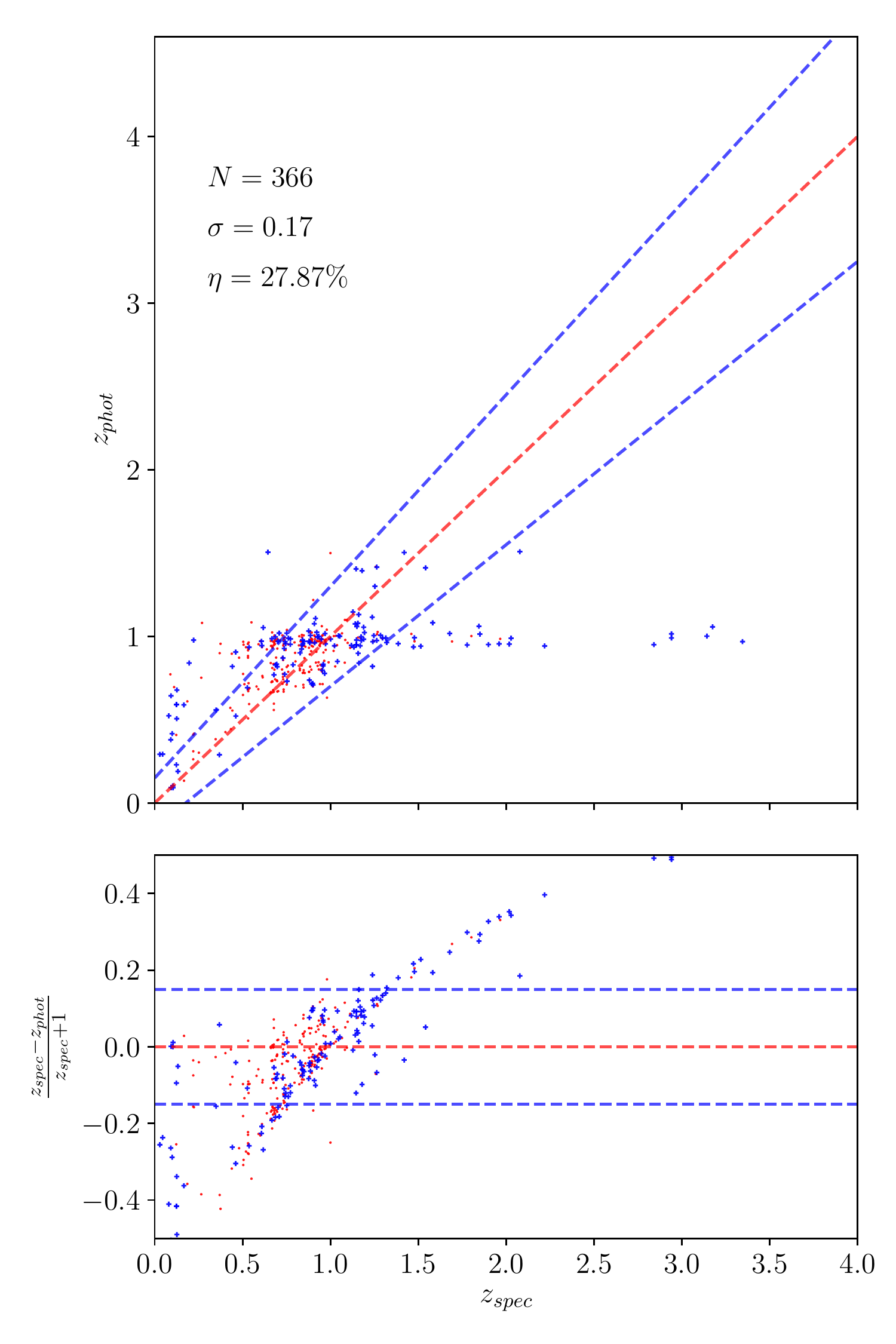} c)
\includegraphics[,scale=0.4]{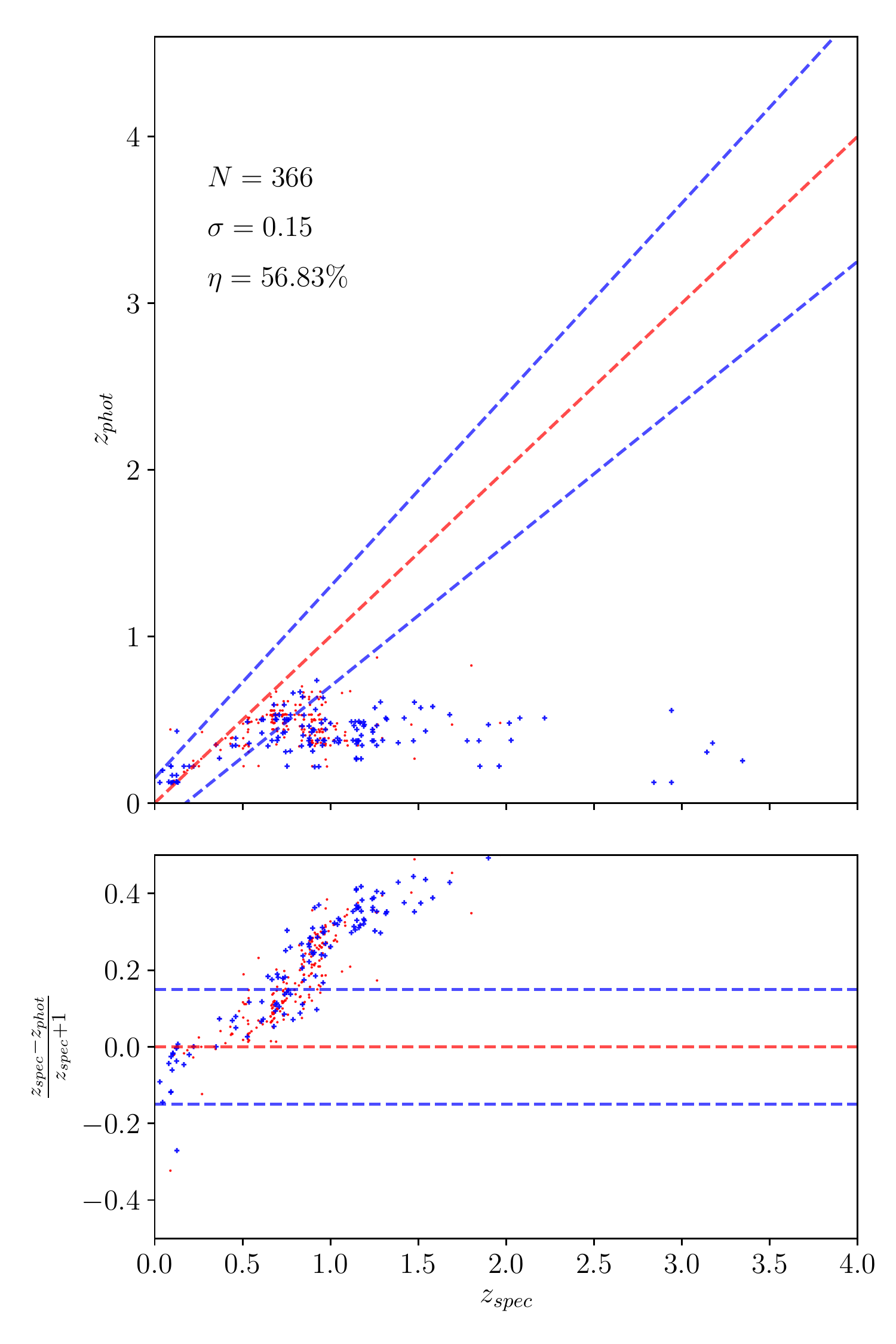} d)
\caption{A representative sample of results obtained when the training set is selected from a brighter distribution of galaxies than the test set, in experiment A1/BDNY.
Panel a): MLPQNA. 
Panel b):  RF-NA. 
Panel c): RF-JHU. 
Panel d): kNN.}
\label{FIG:biased}
\end{figure*}

Experiments A1 to H1 train each algorithm on a brighter data set, then apply it to the full data set. If this were an effective strategy, it would be invaluable to future surveys, since spectroscopy is often not available on fainter objects. However, no 
%\mbc
\pointless{supervised} machine learning techniques perform well in circumstances in which the training set differs significantly from the test set, and this is reflected in the results shown in Table~\ref{TAB:results}. This experiment demonstrates that, training on a brighter part of the galaxy distribution does not provide enough information for the algorithm to be able to extrapolate to fainter objects. \pointless{While this result may appear trivial, it is a warning that larger and deeper spectroscopic surveys are essential to train algorithms to measure redshifts for surveys such as EMU} . Given the poor quality of the results, we do not show the plots for each experiment, but instead show a representative sample in Figure~\ref{FIG:biased}.

\clearpage

\subsection{Results on the RANDOM DEEP experiments}

As may be expected, compared to their performance on the BRIGHT experiments, the machine learning methods perform better on a randomly selected training set, which matches the test set.
Here we summarize the results of these experiments on the RANDOM DEEP experiments, in which the full-depth data in the KB are used.

%{\bf the SOM results are obviously wrong. I'm writing this section assuming we manage to fix this}

\subsubsection{Experiment A2: RDNY}
\begin{figure*}\centering
\includegraphics[,scale=0.4]{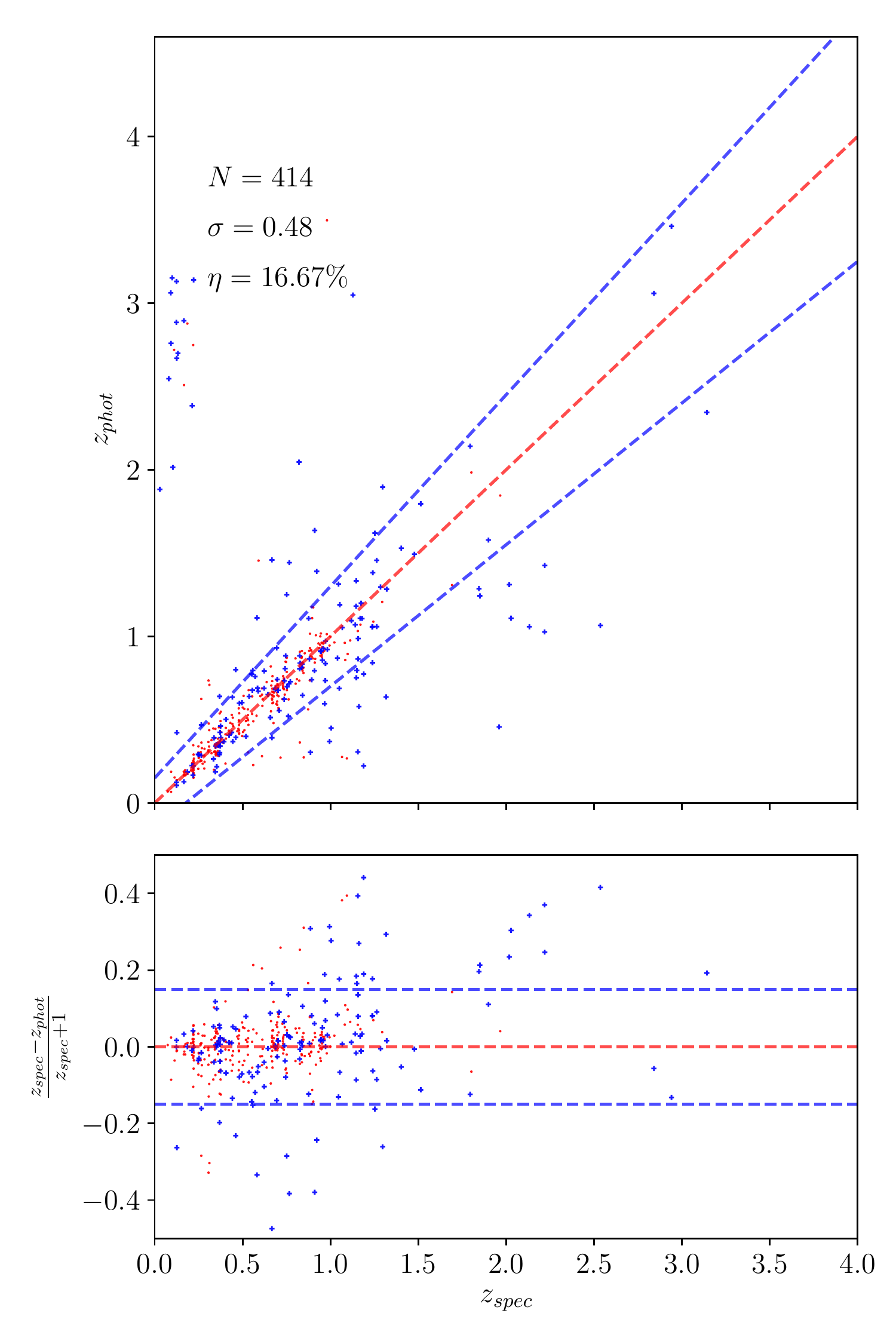} a)
\includegraphics[,scale=0.4]{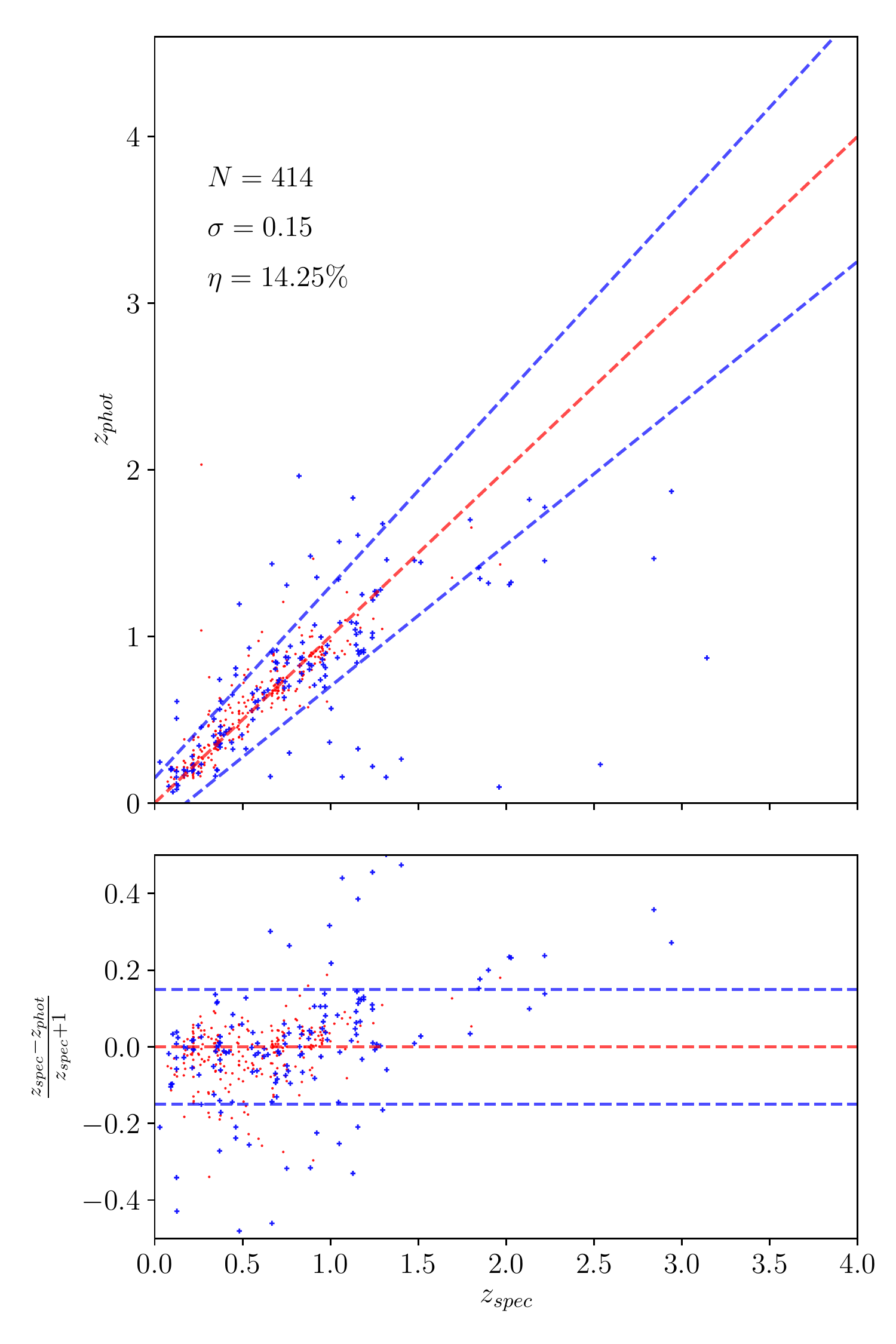} b)\\
\includegraphics[,scale=0.4]{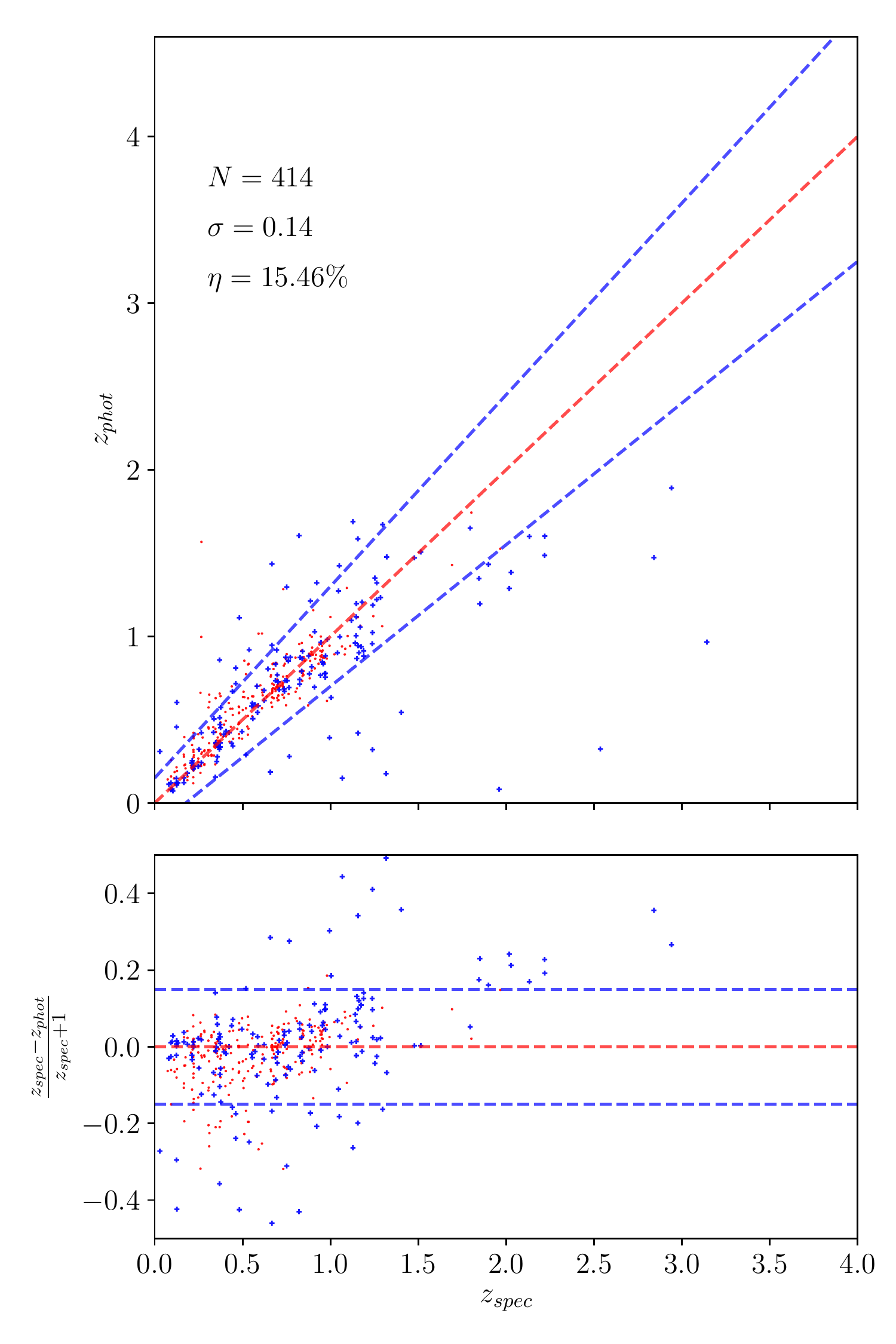} c)
\includegraphics[,scale=0.4]{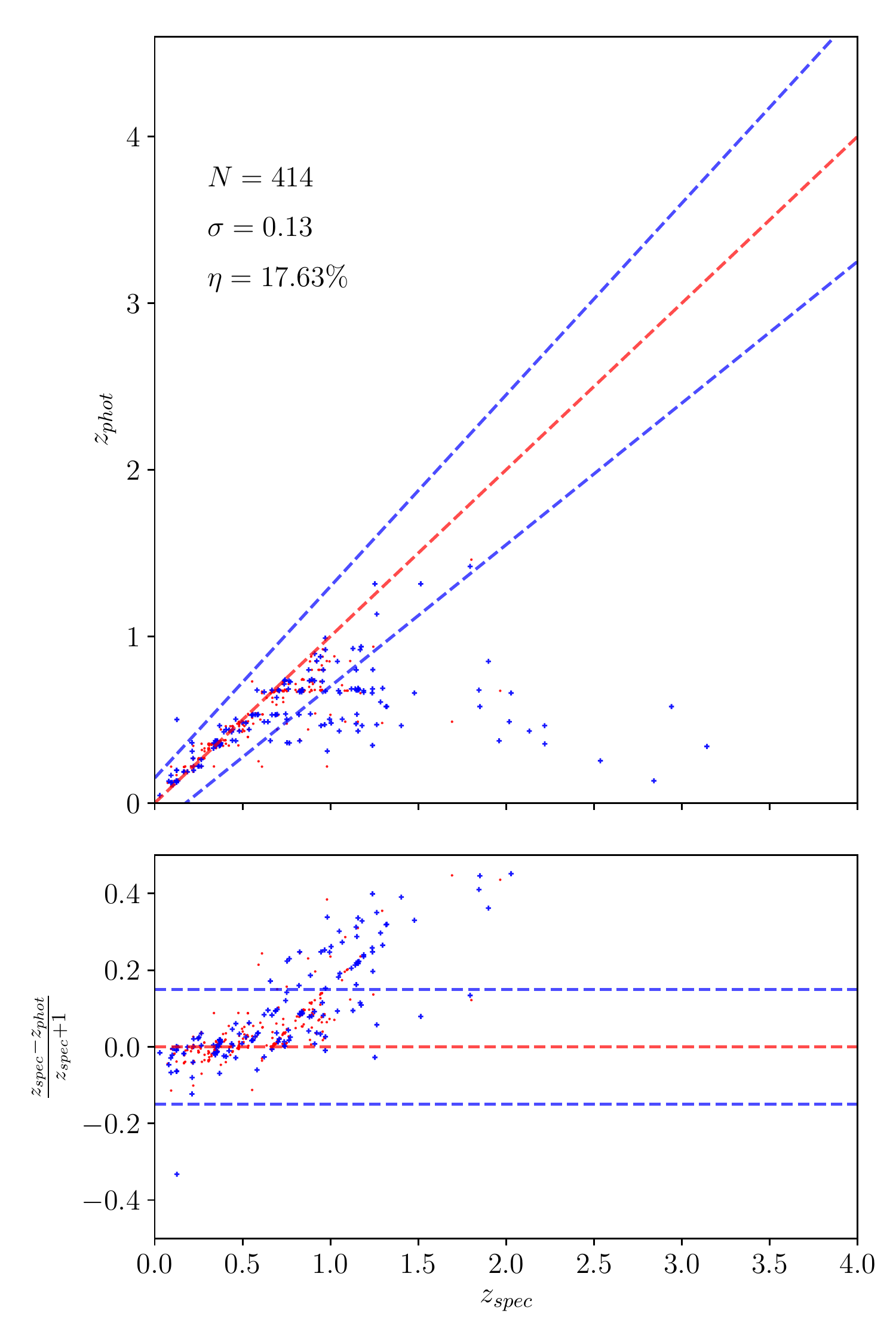} d)
\caption{Summary of the results obtained in the experiment A2/RDNY with the various methods. 
Panel a): MLPQNA.
Panel b): RF-NA.
Panel c): RF-JHU.
%Panel d): SOM;
Panel d): kNN.}
\label{FIG:blind_RDNY}
\end{figure*}

%\noindent \underline{Experiment A2: RDNY}
In this experiment, the full-depth COSMOS data, including X-ray sources, are used, and radio data are not used to train the algorithms.
%\textcolor{red}{Mara: Just to make sure that we understand each other. X-ray=Y means that the X-ray sources have been either treated a part, or removed from the training sample. Do you agree? Ray: No it's the opposite of that. Section 3 says that X-ray=Y means that the X-ray sources were included in the training set.}
As shown in Fig.~\ref{FIG:blind_RDNY}, all ML methods perform well, \pointless{with an overall succss rate $\beta$ in the range 82--86\%}. The best performing overall are kNN and MLPQNA, followed by RF-JHU, which has a smaller fraction of outliers but a slightly larger $\sigma$. kNN performs overall well, but seems to be affected by systematics, with the largest around $z \sim 1$,
%performs well at  $z < 1$, but shows a systematic offset  above that, 
presumably because of the paucity of neighbours at high redshift.
%MLPQNA performs best at high redshift.
\pointless{MLPQNA tends to seriously overestimate the redshift of the very nearby sources.}

%As may be expected due to their absence in the training set, Mir-AGN are poorly reproduced and tend to be outliers.  

\subsubsection{Experiment B2: RDYY}
This experiment (B2) differs from the previous one (A2) only in that the radio data are  included in the training set. In most cases the results of B2 are almost indistinguishable from those of A2. For MLPQNA the standard deviation is halved by the inclusion of radio data, \pointless{bringing back to the correct redshift the sources that in the previous experiment had an overestimated redshift.}
%and its scatter at low redshift is noticeable improved, although there are more extreme outliers. 
\pointless{In most cases, for all the methods, the majority of the outliers are the sources that {\it a posteriori} were found to be AGN}. We conclude that adding a single radio photometry measurement to an existing excellent optical/IR photometric data set \pointless{can limit the fraction of high-redshift outliers. At low redshift, UV data can be used for this purpose, but for EMU, the GALEX data available is too shallow to be useful in most cases.} 
%RPN I dont agree with the following statement: Thus, it will be extremely important for radio galaxy evolution studies via luminosity function, to verify with further analysis, whether Radio data can be used as surrogate of UV data.} 
%makes very little difference in most cases, but the improvement in MLPQNA should be explored further.

\subsubsection{Experiment C2: RDNN}

This experiment (C2) differs from A2 in that XMM X-ray sources are excluded from the training and test sets. 
%\pointless{ideally, we wanted also to train on AGN alone, but we found out that the sample was too small for this analysis}. 
\pointless{All methods, with the exception of RF-JHU, had a lower outlier rate on this variation than on A2, but also a lower sample size, resulting in a lower overall success rate. RF-JHU performed similarly in the two experiments, and ended up with a significantly higher success rate than the other methods. }

\begin{figure*}\centering
\includegraphics[,scale=0.4]{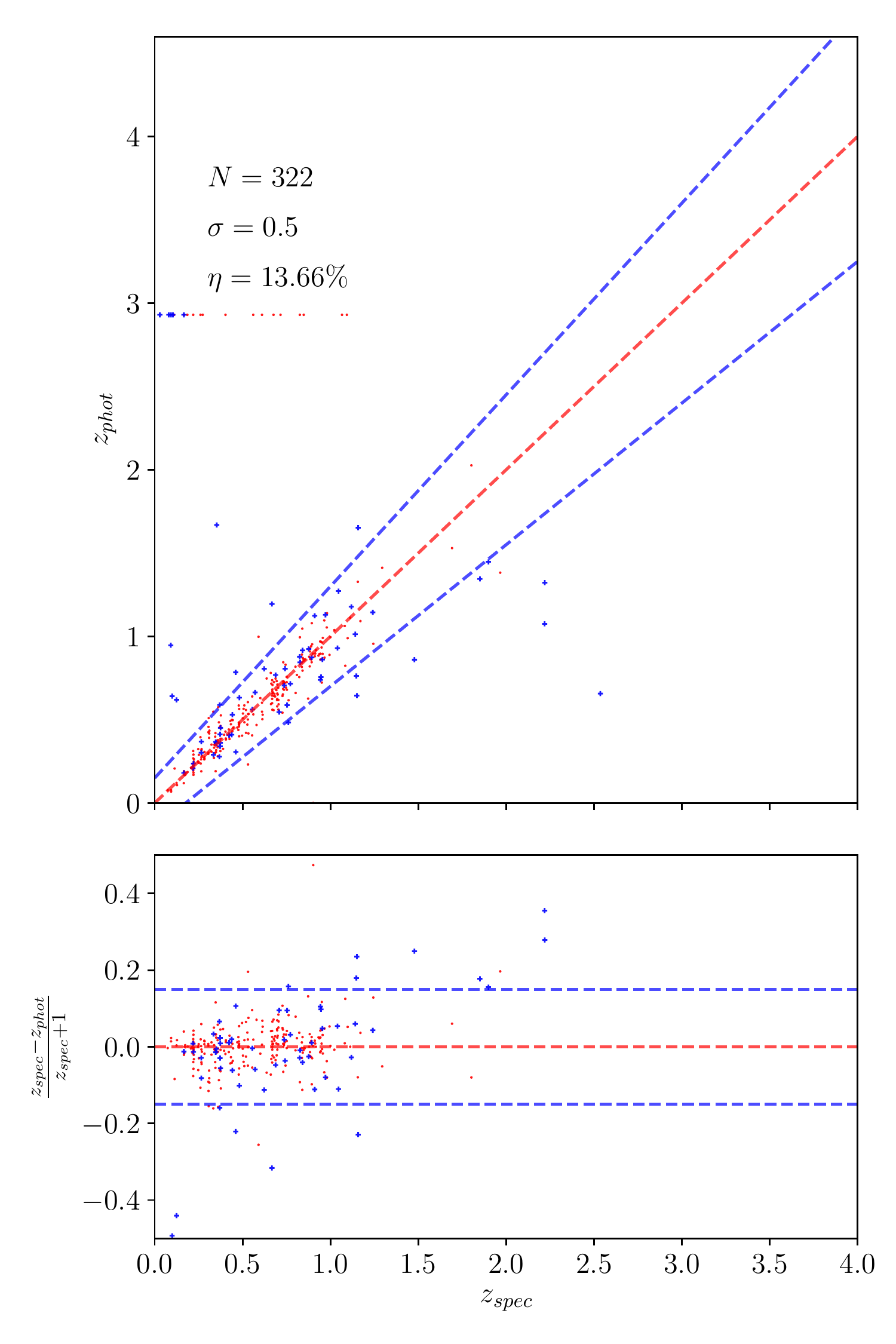} a)
\includegraphics[,scale=0.4]{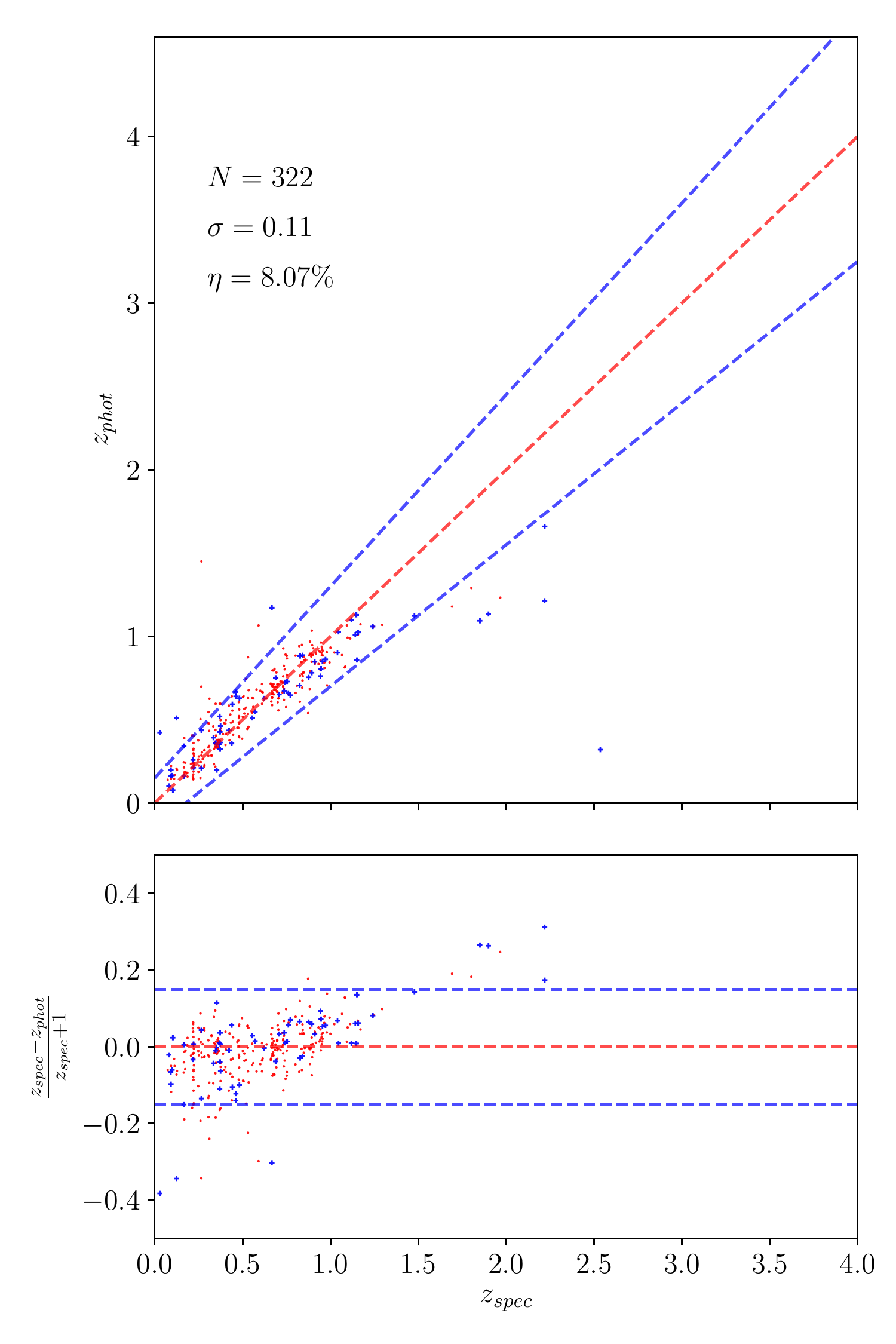} b)\\
\includegraphics[,scale=0.4]{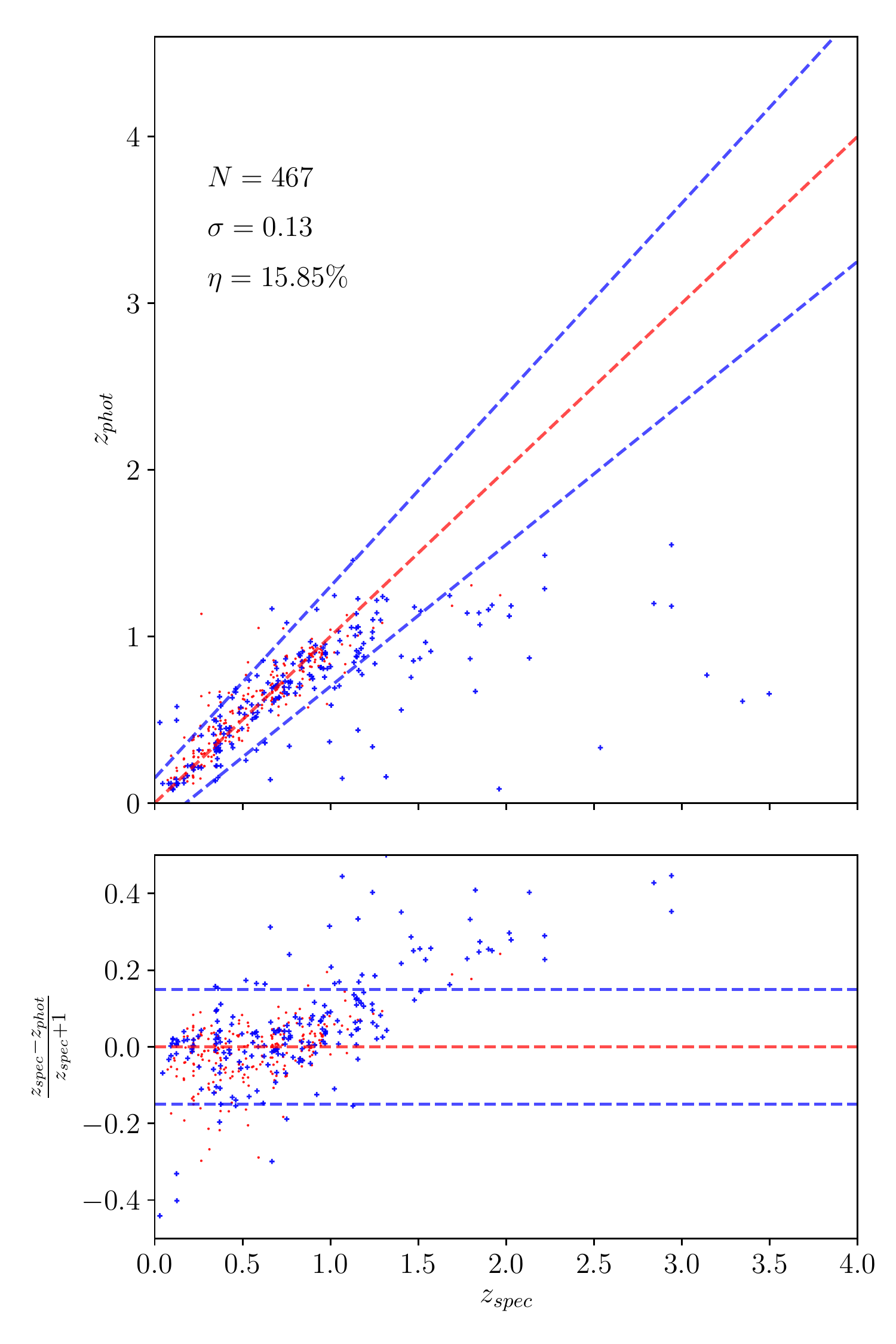} c)
\includegraphics[,scale=0.4]{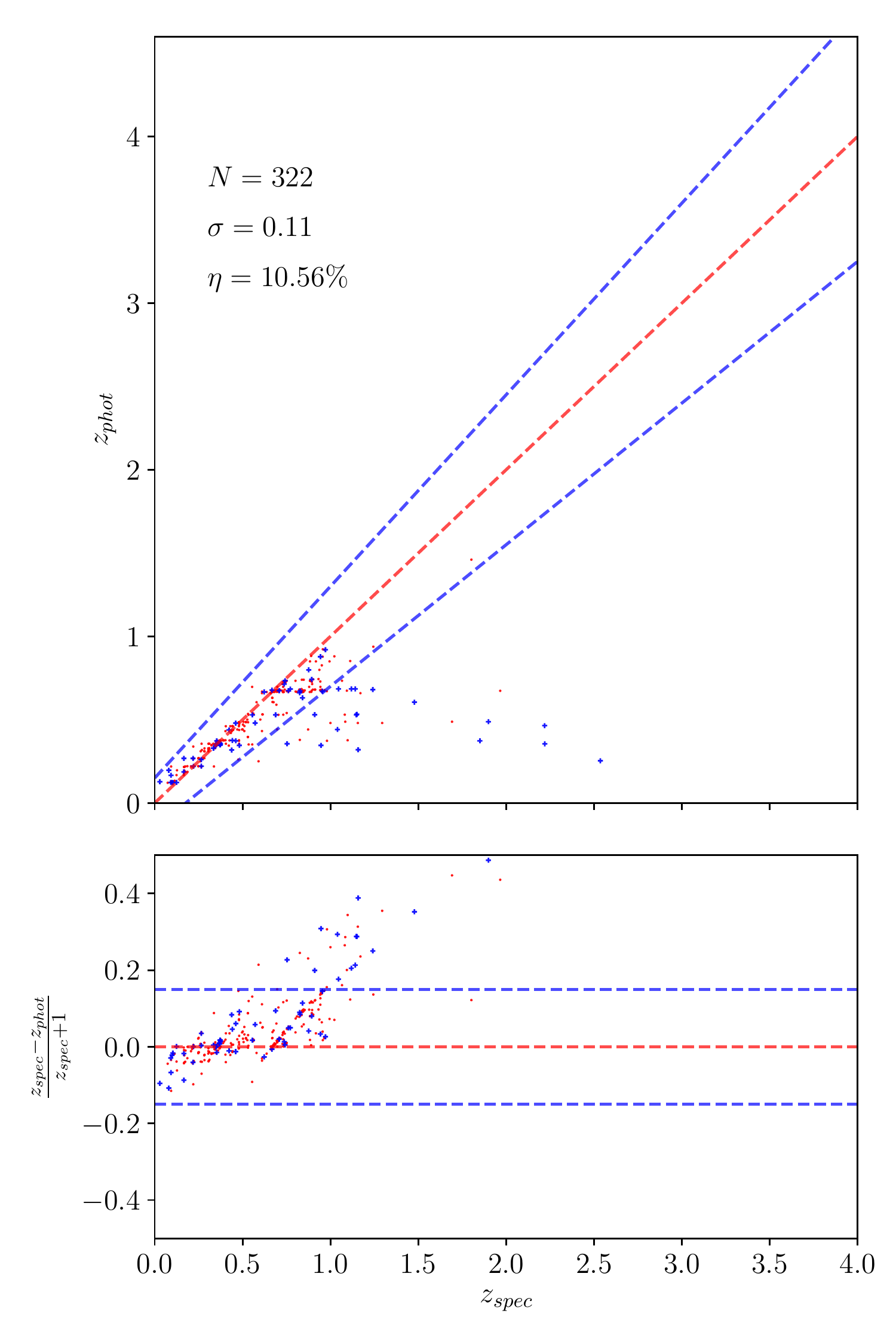} d)
\caption{Summary of the results obtained in the experiment C2/RDNN with the various methods. 
Panel a): MLPQNA.
Panel b):  RF-NA.
Panel c): RF-JHU.
%Panel d): SOM;
Panel d): kNN.
}
\label{FIG:blind_RDNN}
\end{figure*}

\pointless{As shown in Fig.~\ref{FIG:blind_RDNN} the lowest outlier rate $\sigma_{NMAD}$, is obtained with MLPQNA, RF-NA and kNN. RF-NA has almost half the fraction of outliers than MLPQNA  which shows a systematic pattern at $z_{phot}\sim 2.9$. kNN tends systematically to put the sources at $z_{phot}\sim 0.6$ for all the sources above $z_{spec}\sim 0.6$, probably for the same reason discussed in the first experiment.}

%\noindent {\bf Number of points in Geach experiment is very low. I rejected the plot.}

\subsubsection{Experiment D2: RDYN}
\begin{figure*}\centering
\includegraphics[,scale=0.4]{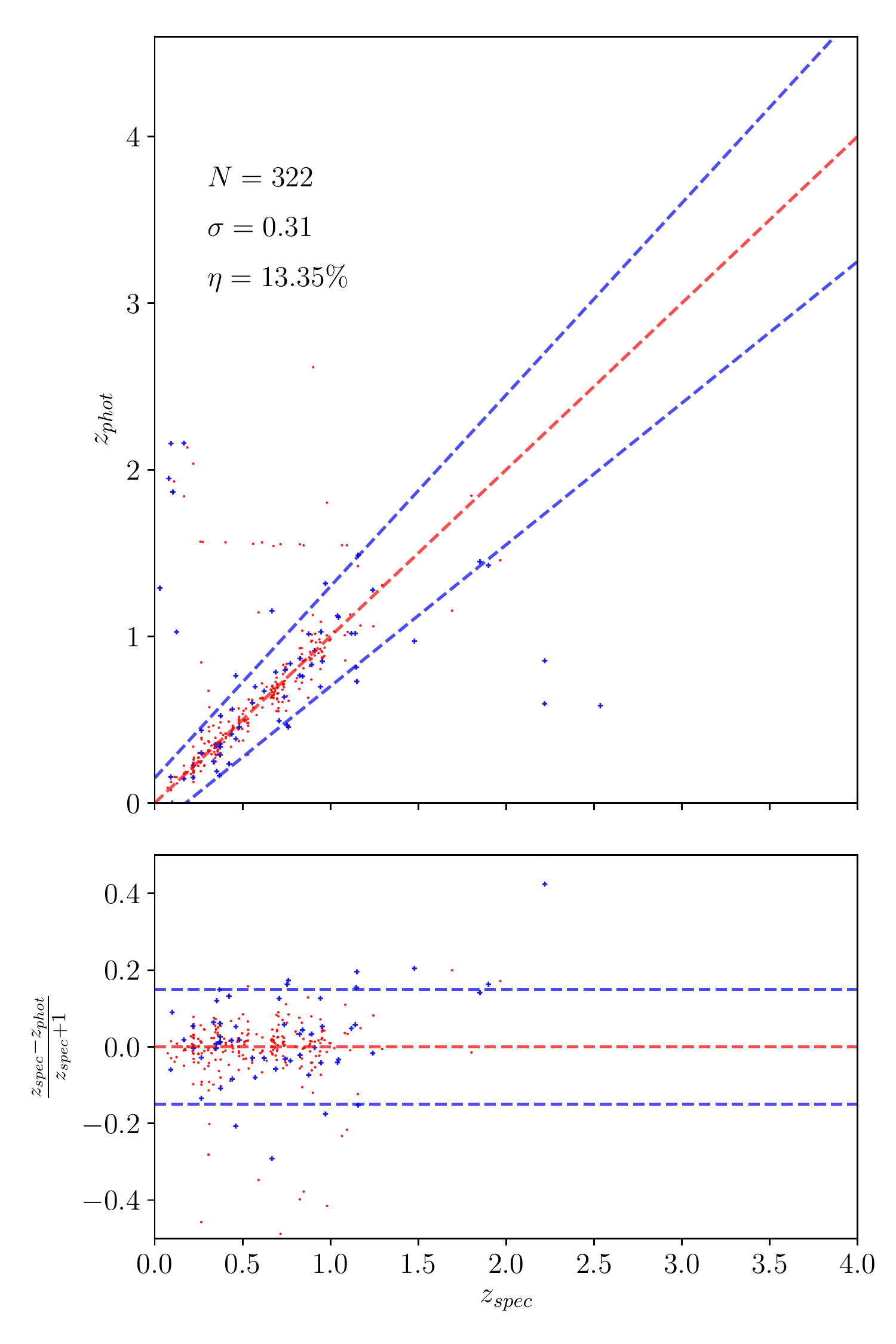} a)
\includegraphics[,scale=0.4]{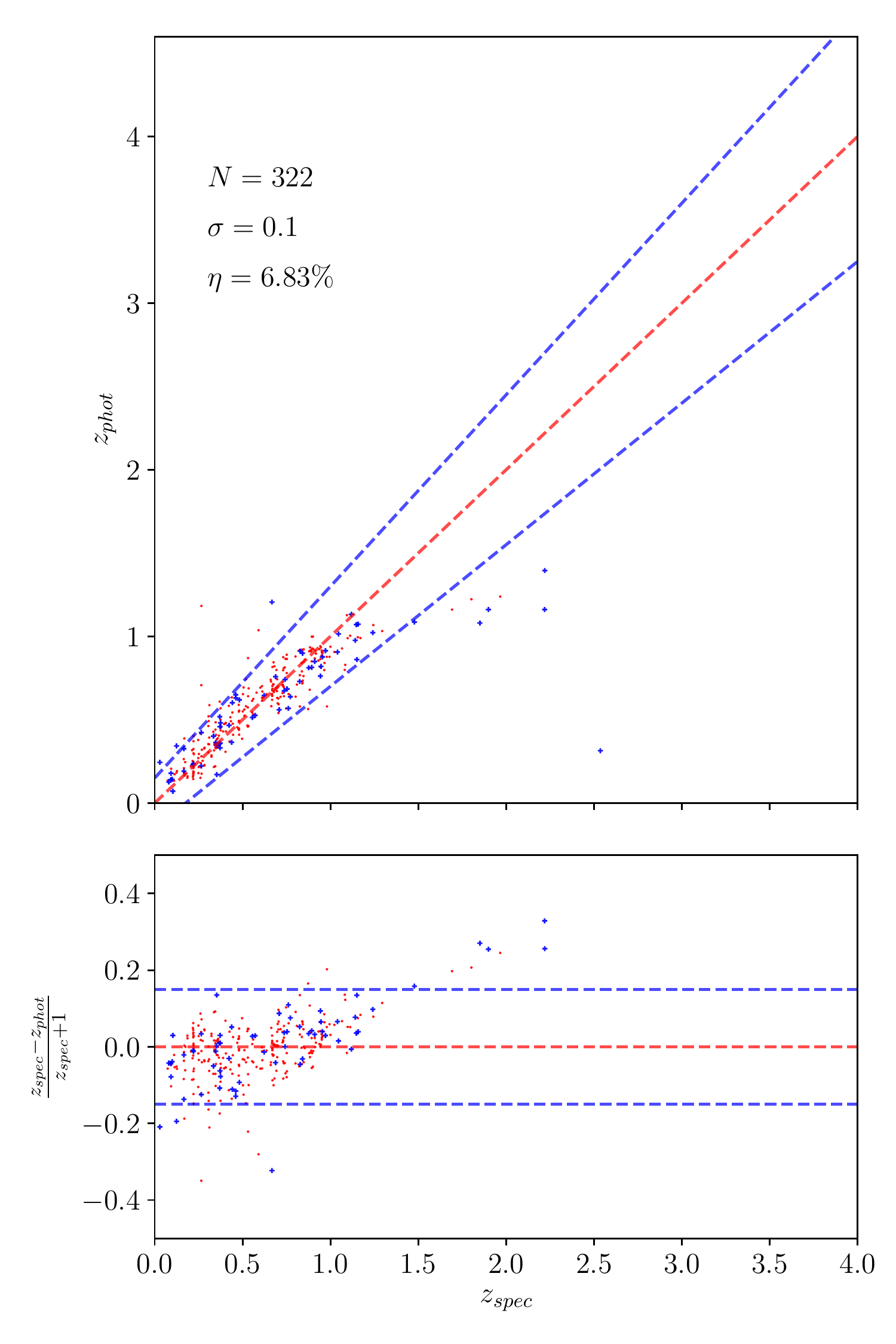} b)\\
\includegraphics[, scale=0.4]{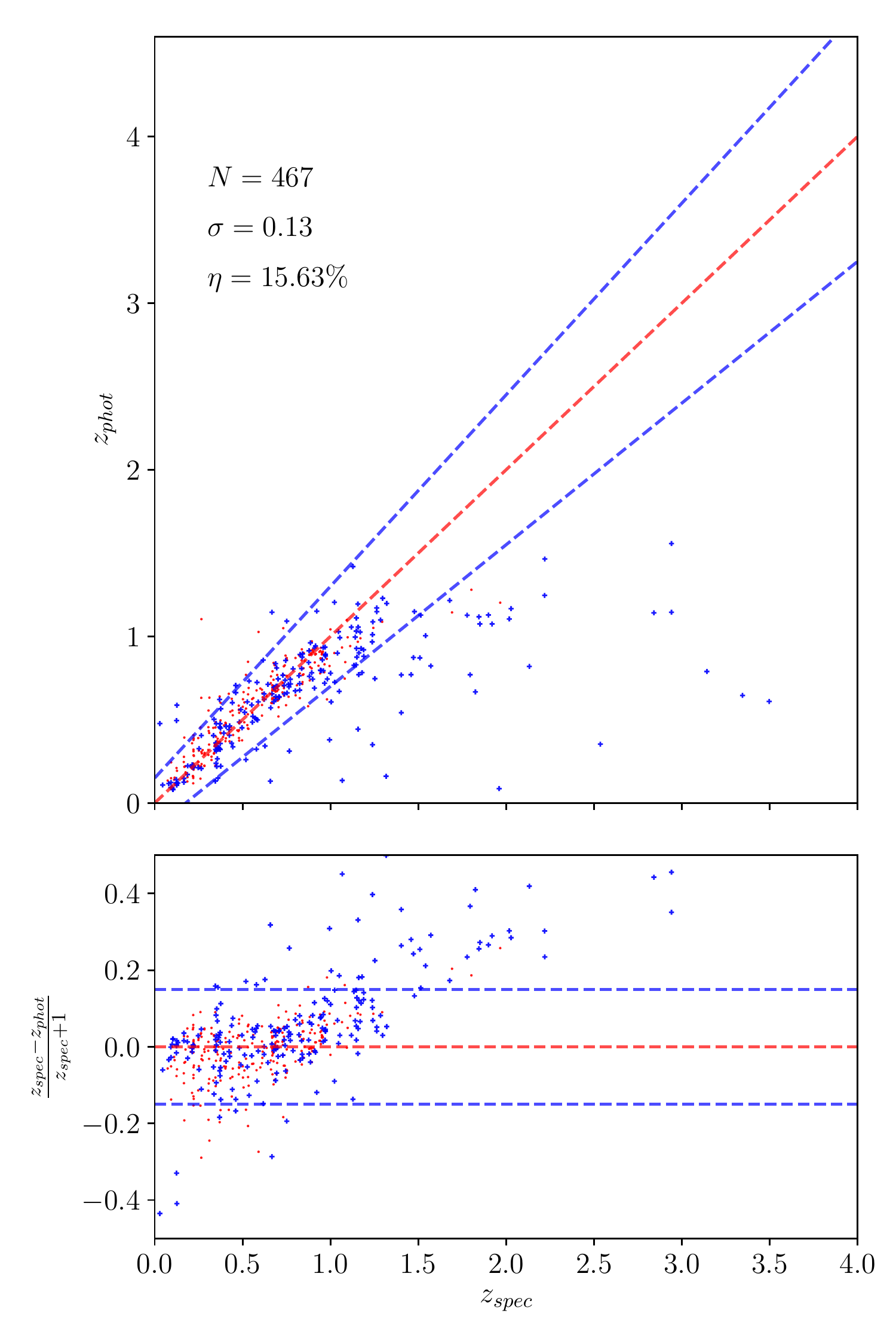} c)
\includegraphics[, scale=0.4]{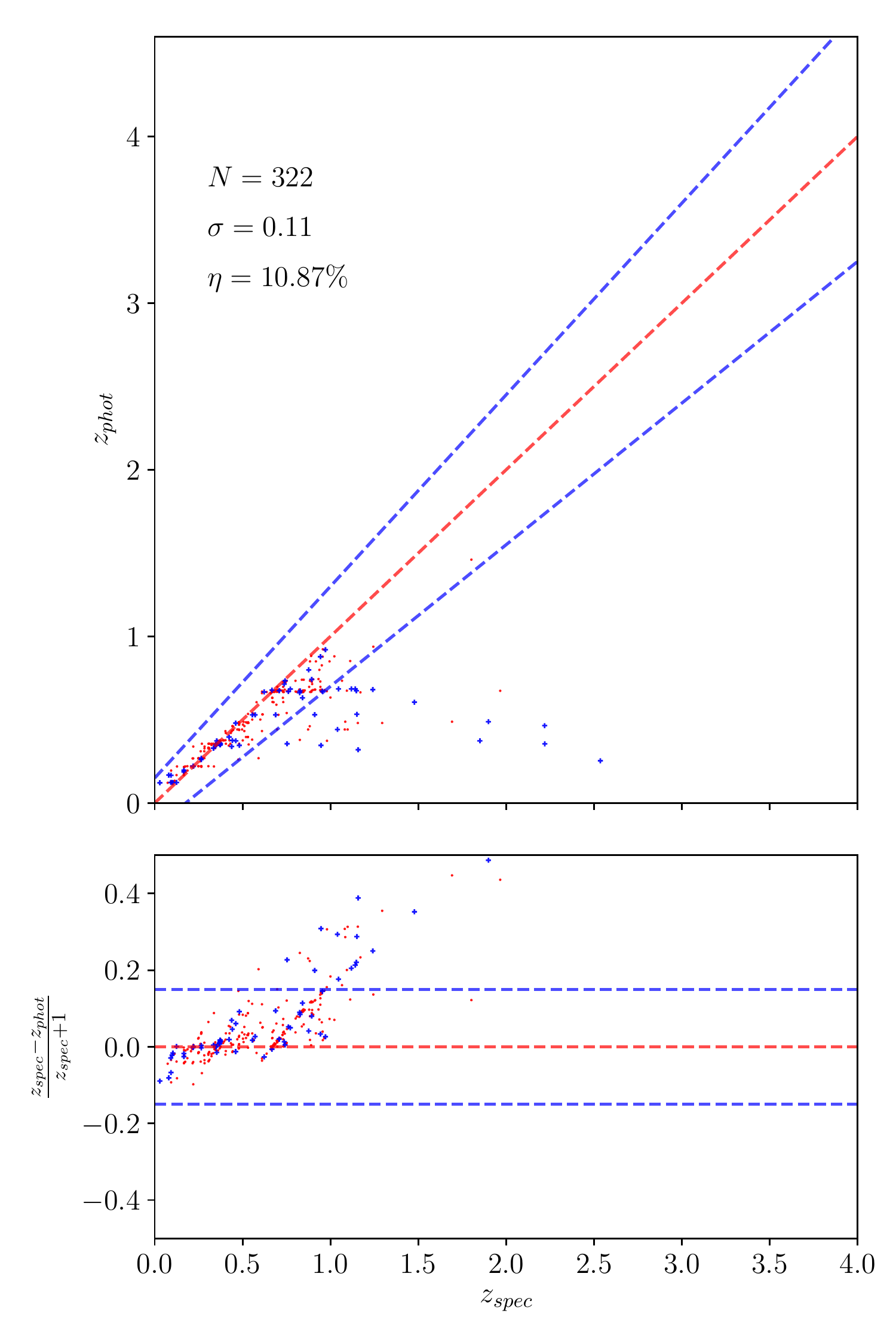} d)
\caption{
Summary of the results obtained in the experiment D2/RDYN with the various methods. 
Panel a): MLPQNA.
Panel b):  RF-NA.
Panel c): RF-JHU.
%Panel d): SOM;
Panel d): kNN.
}
\label{FIG:blind_RDYN}
\end{figure*}

D2 differs from C2 in that radio fluxes are used in the training process. As shown in Fig.~\ref{FIG:blind_RDYN}, all tests performed very similarly or slightly worse than C2. The lowest  outlier rates are achieved by MLPQNA, RF-NA and kNN, whilst RF-JHU assigns the greatest absolute number of non-outlier redshifts because of its lower failure rate. 
%In the case of MLPQNA a sizeable fraction of all outliers are either Mir AGN or Chandra AGN. 
All methods perform poorly at redshifts higher than $\sim1.0$, where training points become scarce.
%\textcolor{red}{I completely disagree on the reading of this plot. RF-JHU perform best, as it manage to assign a redshift to 155 sources more than the other methods. We should include the number of failures when computing the fraction of outliers.}

\subsection{Results on the RANDOM SHALLOW experiments}

In the RANDOM SHALLOW experiments, we deliberately degrade the quality of the optical/IR photometric data to simulate the photometric data that will be available for EMU. % 
%Following sentence moved to section 3.\pointless{For the same reason, we also considered only broad band photometry, dropping narrow and intermediate band photometry.}
%In the RF-JHU, SOM, and kNN methods, both the training and the test set were drawn from this degraded KB. In the MLPQNA and RF-NA methods, the training set was drawn from this degraded KB but the test set was drawn from the full-depth KB. Because of this mismatch between the test and training sets, it is likely that MLPQNA and RF-NA methods are not performing to the best of their ability in these tests. 

%{\bf the SOM results are obviously wrong. I'm writing this section assuming we manage to fix this}

\subsubsection{Experiment E2: RSNY}
\begin{figure*}\centering
\includegraphics[, scale=0.4]{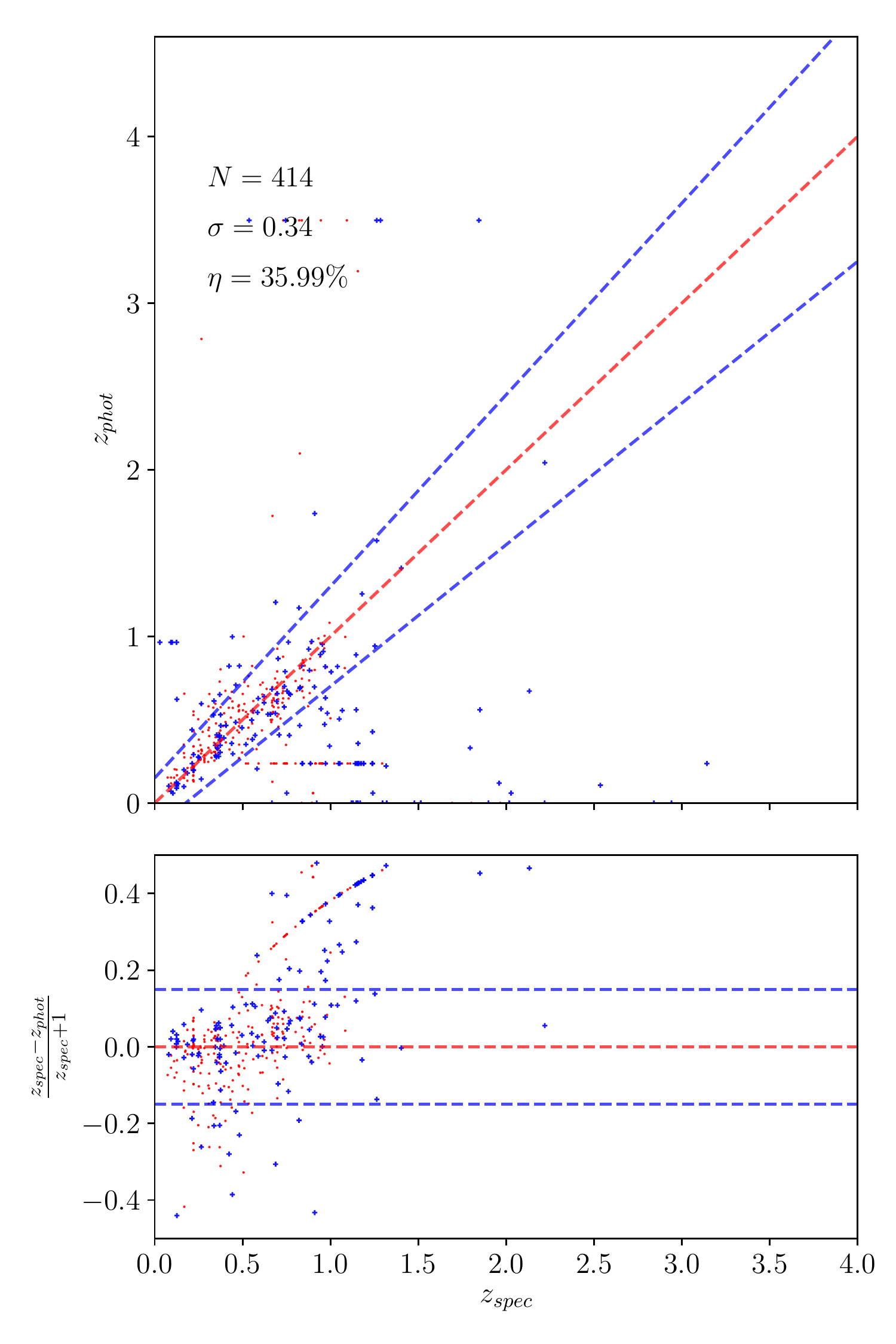} a)
\includegraphics[, scale=0.4]{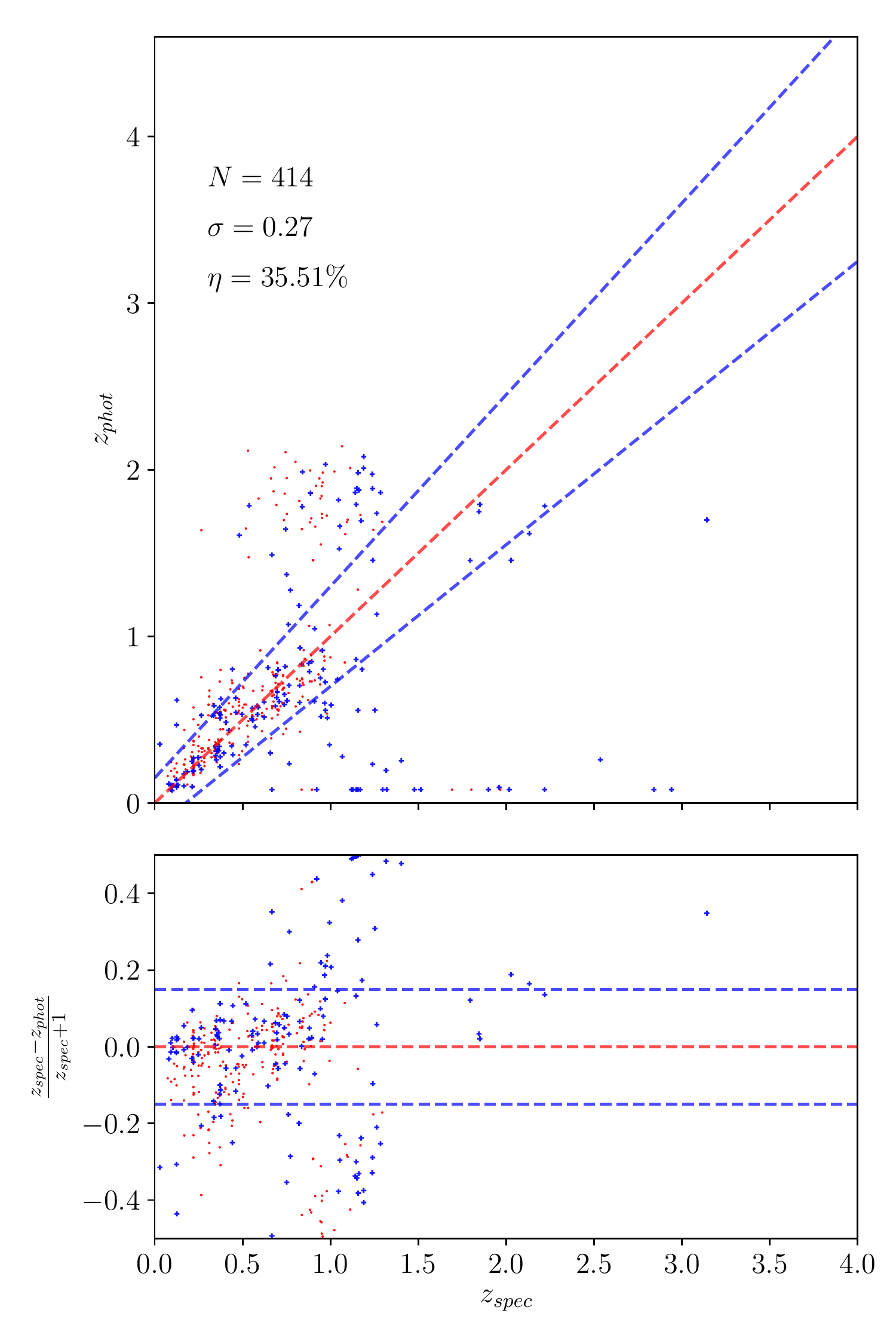} b)\\
\includegraphics[, scale=0.4]{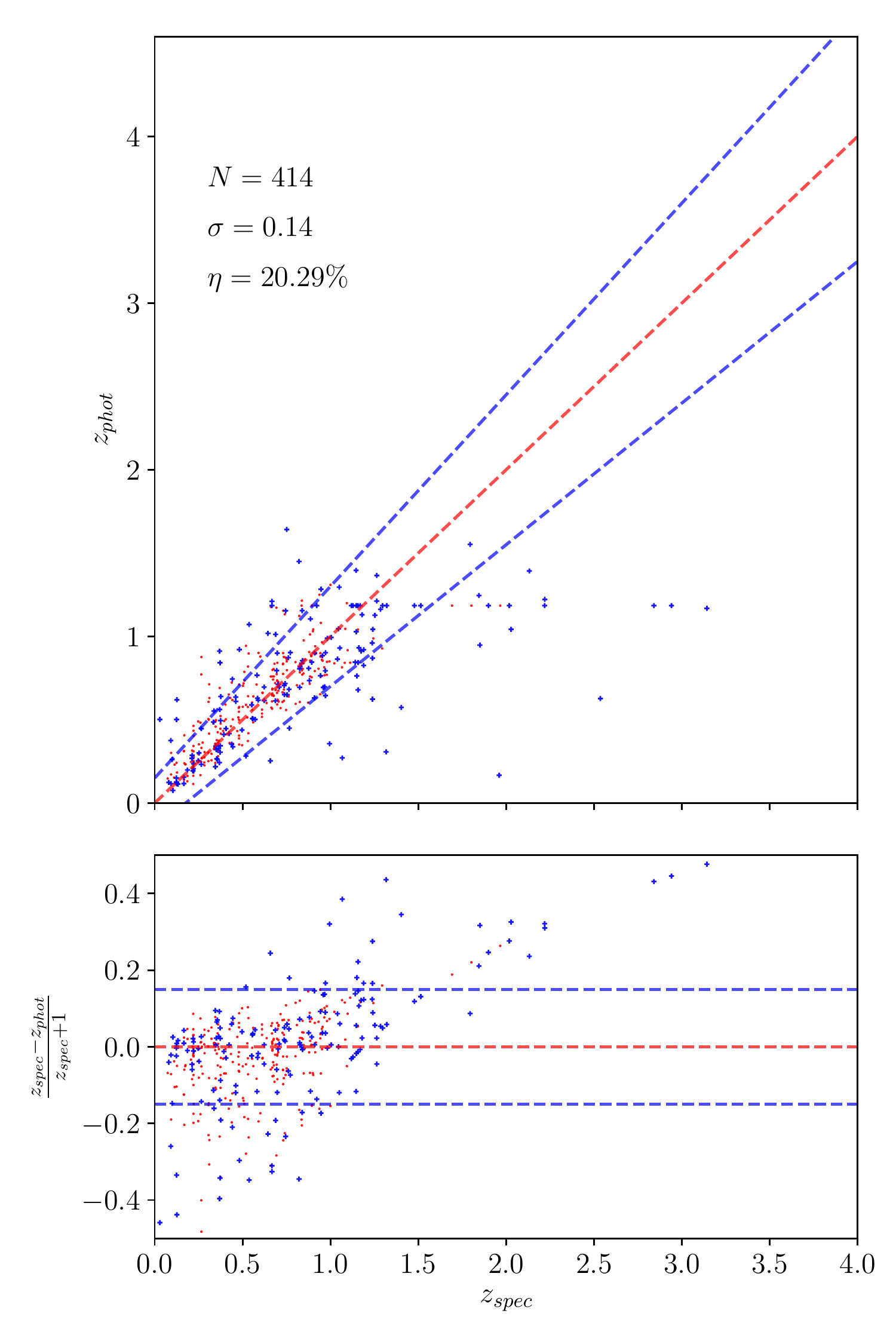} c)
\includegraphics[, scale=0.4]{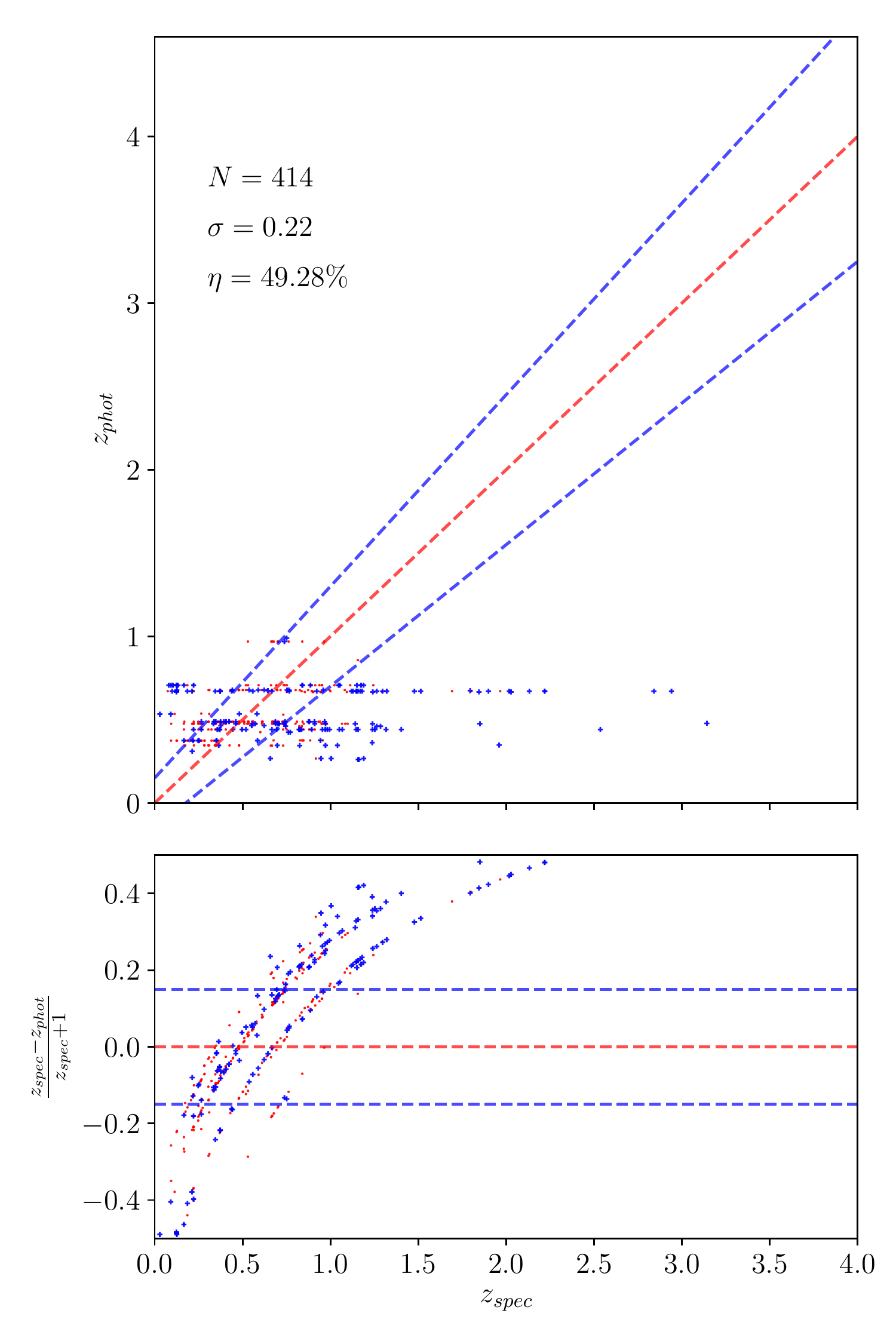} d)
\caption{
Summary of the results obtained in the experiment E2/RSNY with the various methods. 
Panel a): MLPQNA.
Panel b):  RF-NA.
Panel c): RF-JHU.
%Panel d): SOM;
Panel d): kNN.
}
\label{FIG:blind_RSNY_shallow}
\end{figure*}

In this experiment, the training sample included the XMM detected sources and we used degraded photometry data, without using the Radio data.
Naturally the results are not as good as the full-depth data, but the results of E2, shown in Figure~\ref{FIG:blind_RSNY_shallow}, still represent a valuable source of redshifts for large radio surveys. The RF-JHU algorithm performs the best, with $20$\% outliers and $\sigma = 0.1$, $\beta = 82\%$, but it fails above $z = 1$.

\subsubsection{Experiment F2: RSYY}
\begin{figure*}\centering
\includegraphics[, scale=0.4]{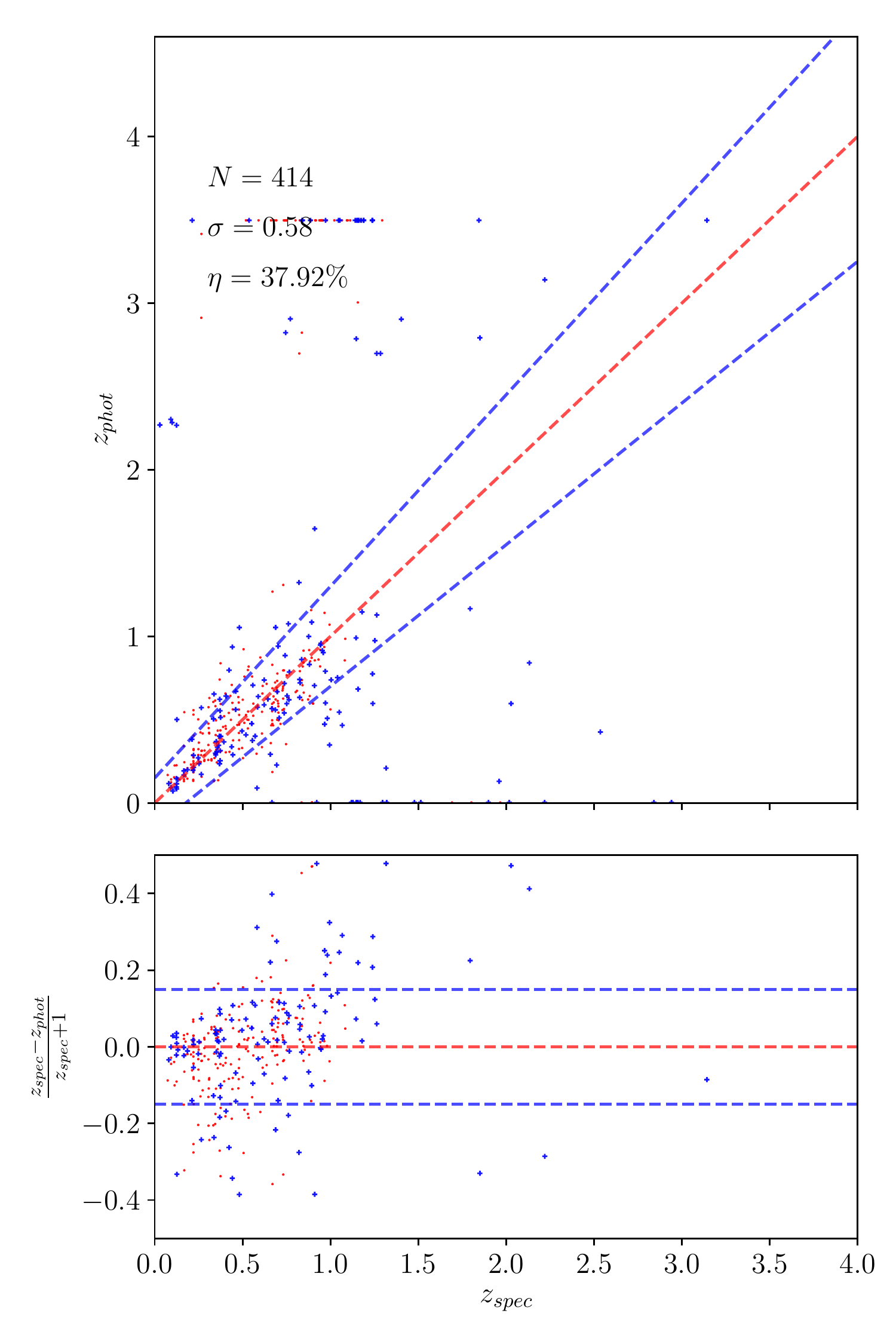} a)
\includegraphics[, scale=0.4]{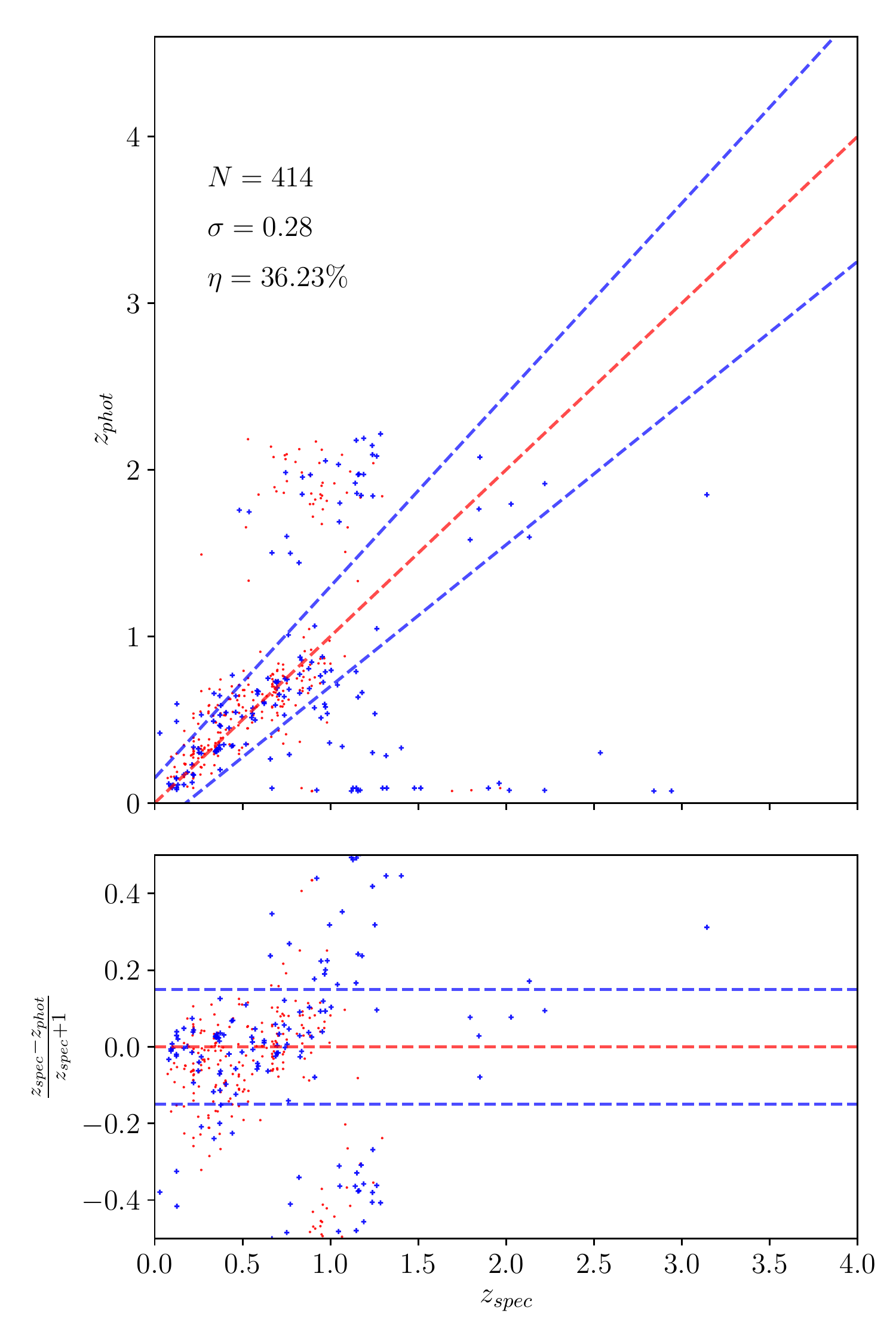} b)\\
\includegraphics[, scale=0.4]{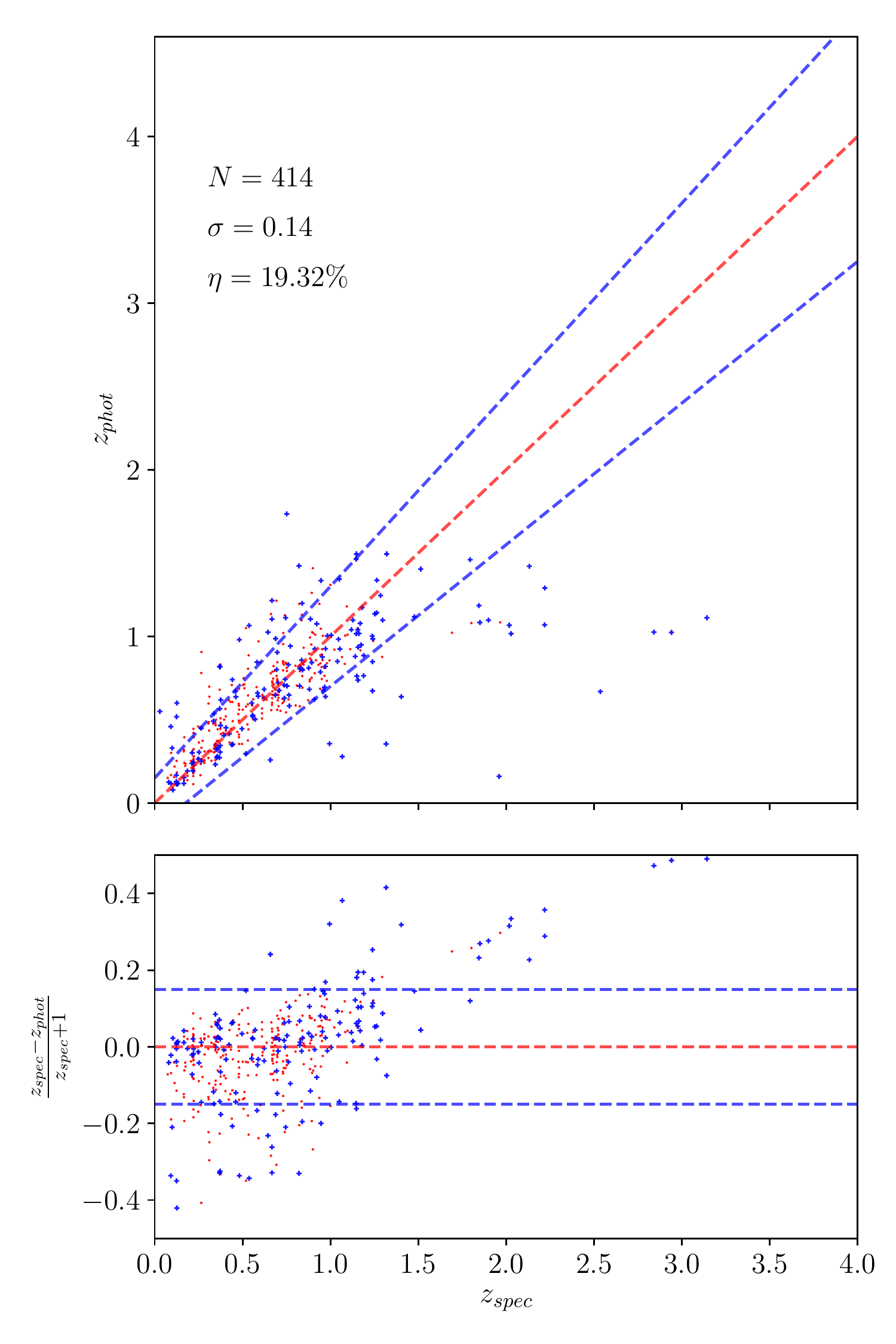} c)
\includegraphics[, scale=0.4]{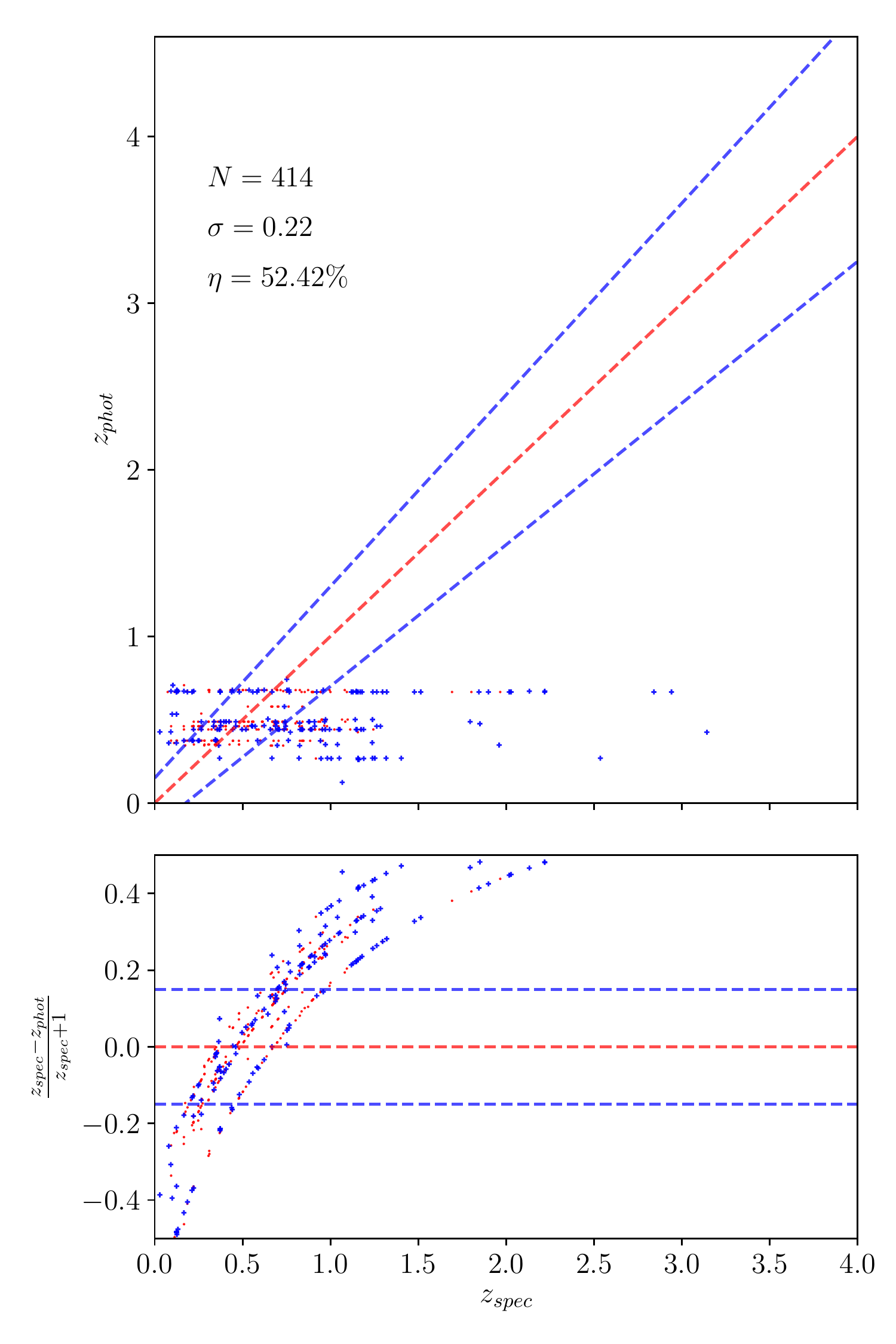} d)
\caption{Summary of the results obtained in the experiment F2/RSYY with the various methods. 
Panel a): MLPQNA.
Panel b): RF-NA.
Panel c): RF-JHU.
%Panel d): SOM;
Panel d): kNN.
}
\label{FIG:blind_RSYY_shallow}
\end{figure*}

F2 differs from E2 by including radio data. The results, shown in Figure~\ref{FIG:blind_RSYY_shallow}, show that all methods have a success rate that is not significantly different from E2, 
% I don't think the following statement is justified by the data. Its an enormous extrapolation from the evidence available here! \pointless{This suggests that the inclusion of  radio data can't compensate for the lack of information provided by a large number of deep photometric points.}
confirming that the addition of a single radio photometric point does not provide useful information.

\subsubsection{Experiment G2: RSNN}
\begin{figure*}\centering
\includegraphics[, scale=0.4]{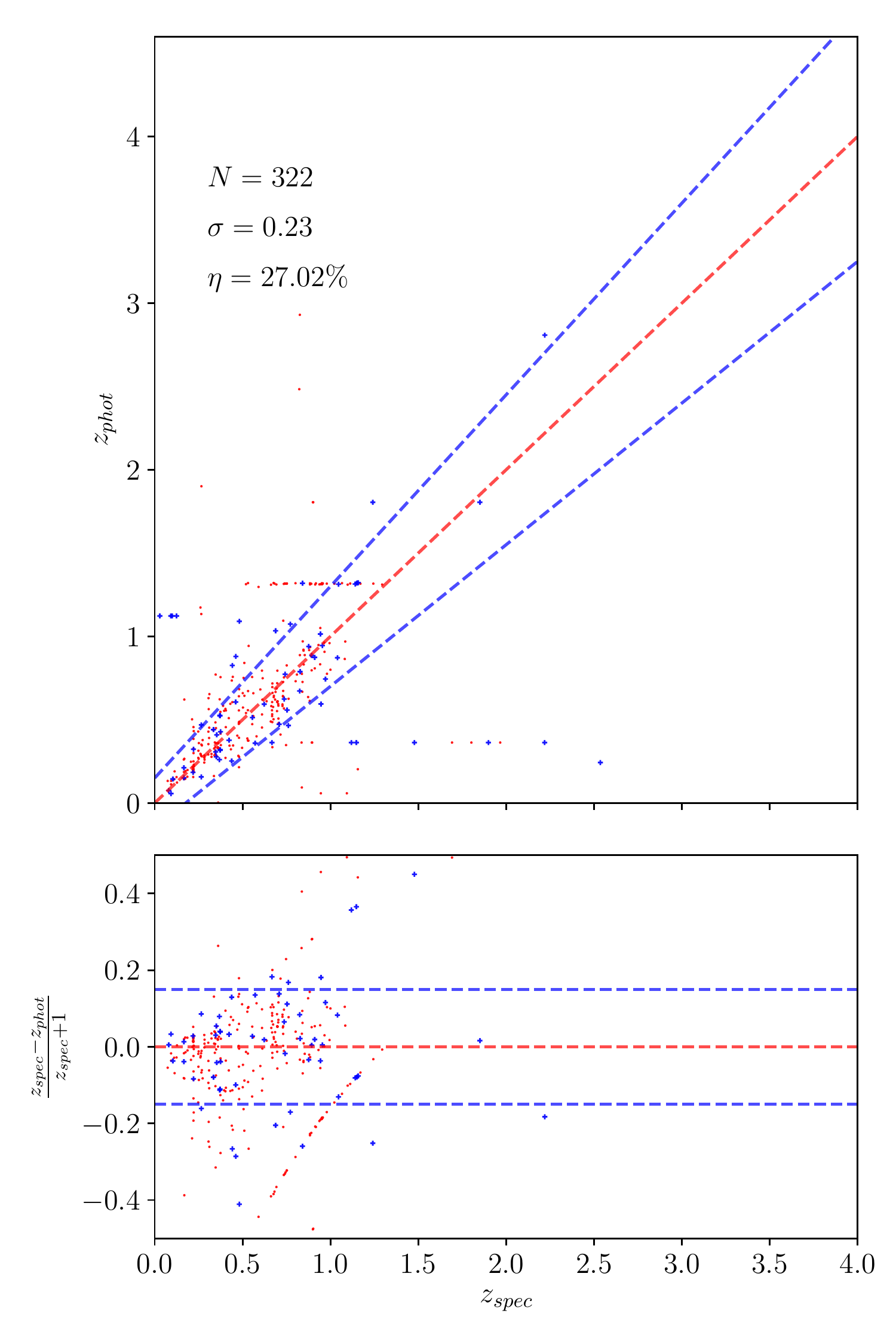} a)
\includegraphics[, scale=0.4]{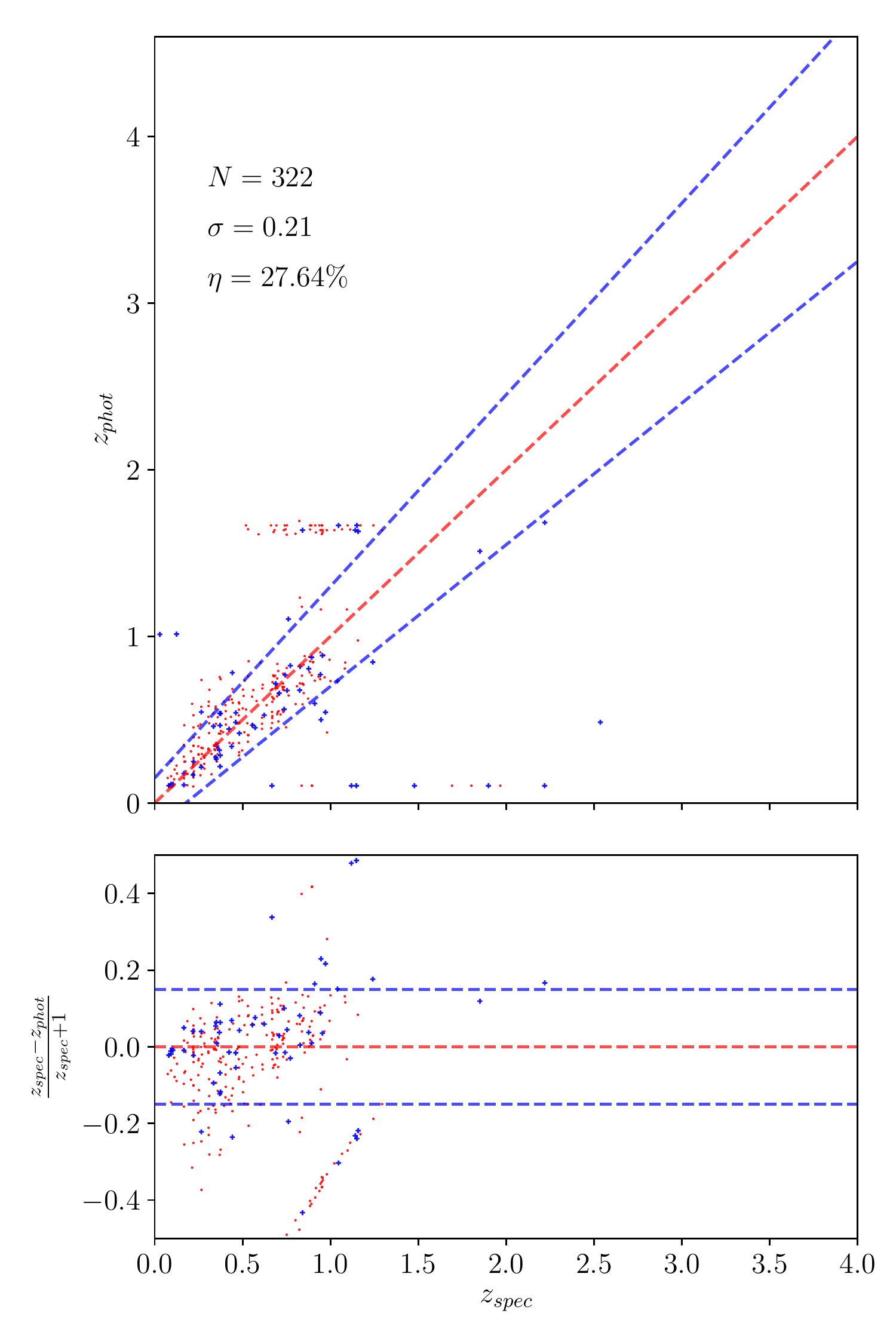} b)\\
\includegraphics[, scale=0.4]{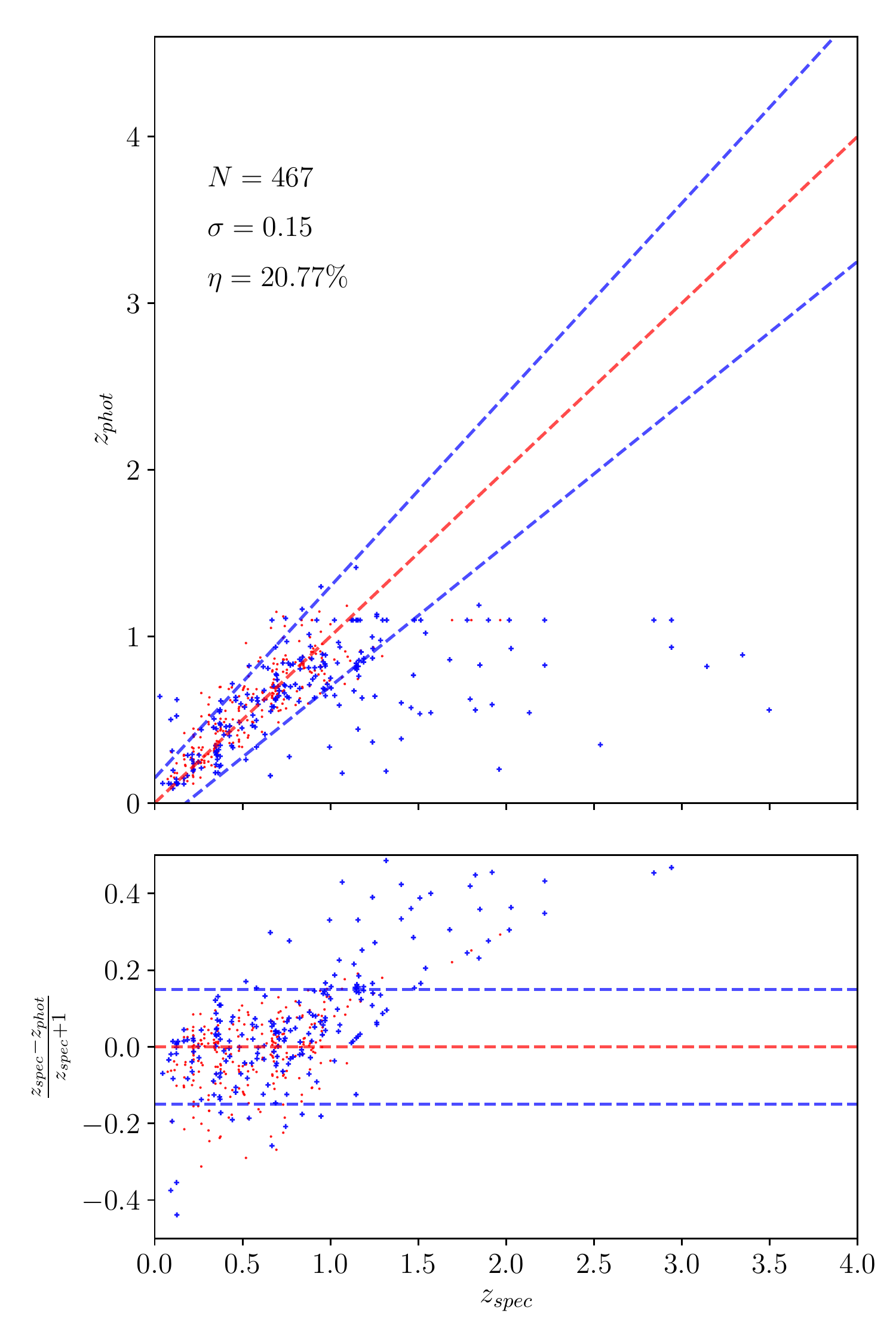} c)
\includegraphics[, scale=0.4]{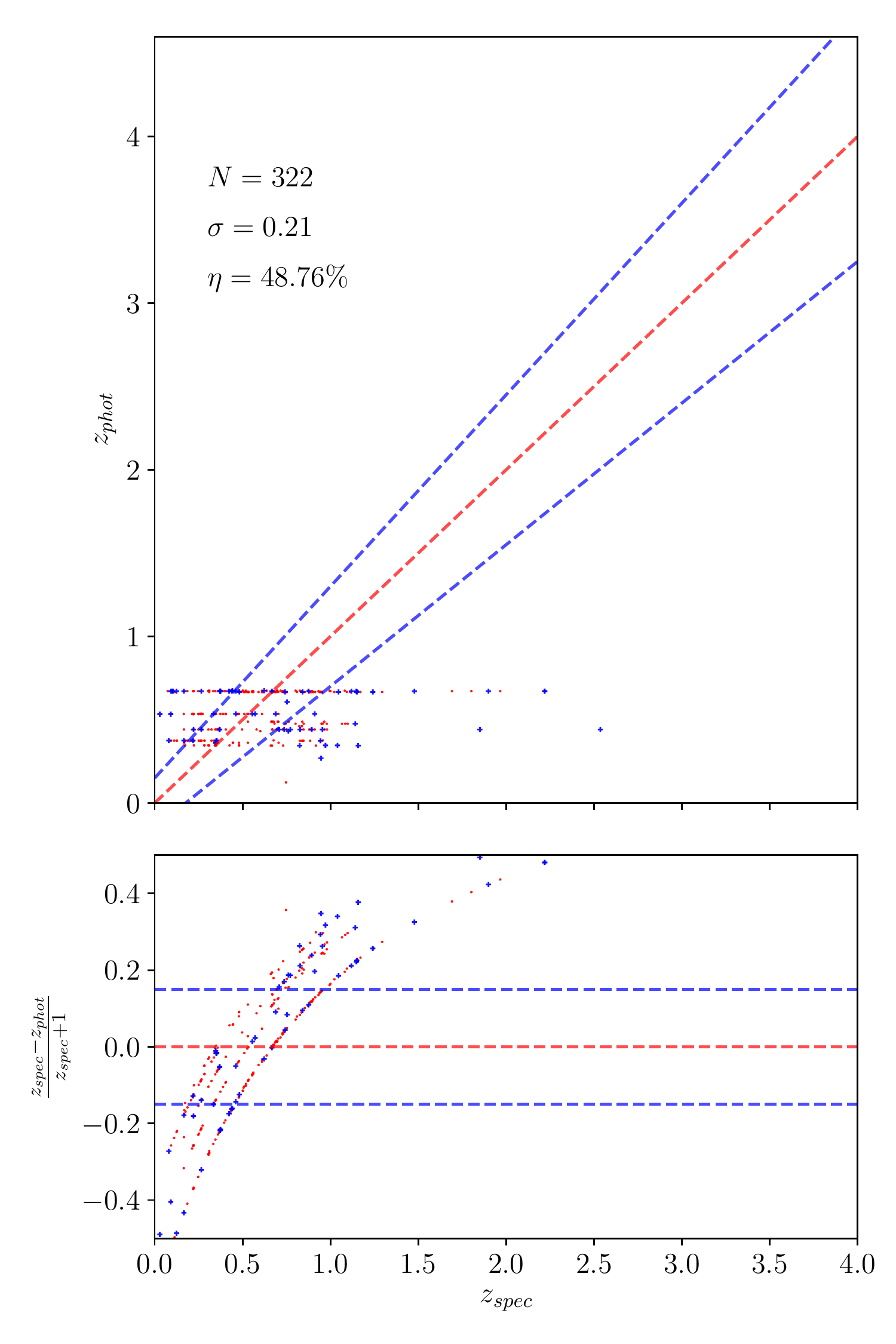} d)
\caption{
Summary of the results obtained in the experiment G2/RSNN with the various methods. 
Panel a): MLPQNA.
Panel b):  RF-NA.
Panel c): RF-JHU.
%Panel d): SOM;
Panel d): kNN.
}
\label{FIG:blind_RSNN_shallow}
\end{figure*}

G2 differs from F2 by excluding radio and X-ray data. The results, shown in  Figure~\ref{FIG:blind_RSNN_shallow}, show that MLPQNA and RF-JHU outlier rates are improved by omitting these data, although their success rates are lower. \pointless{Thus, when the photometric data set is poorly sampled, the results will be improved by  flagging AGN. For example, flagging all the EMU sources that are detected by eROSITA, will improve the quality of the photometric redshifts for the remaining sources.}
%whilst the other results are roughly the same. 

%\begin{figure*}\label{FIG:blind_RSNN}
%\includegraphics[, scale=0.20]{g2_brescia_cavuoti_longo_mlpgna_photoz_RSNN_blind_scatterPlot.pdf} a)
%\includegraphics[, scale=0.20]{g2_brescia_cavuoti_longo_rf_photoz_RSNN_blind_scatterPlot.pdf} b)
%\includegraphics[, scale=0.20]{g2_deep_carliles_photoz_RSNN_blind_scatterPlot.pdf} c)
%\includegraphics[, scale=0.20]{g2_deep_geach_photoz_RSNN_blind_scatterPlot.pdf} d)
%%\includegraphics[, scale=0.20]{g2_salvato_zphot_RSNN_blind_scatterPlot.pdf} e)
%\includegraphics[, scale=0.20]{g2_deep_zinn_photoz_RSNN_blind_scatterPlot.pdf} f)
%\caption{
%Summary of the results obtained in the  experiment G2/RSNN with the various methods. 
%Panel a) MLPQNA;
%Panel b)  RF-NA;
%Panel c): RF-JHU;
%Panel d): SOM;
%Panel e) kNN.
%}
%\end{figure*}

\subsubsection{Experiment H2: RSYN}
\begin{figure*}\centering
\includegraphics[, scale=0.4]{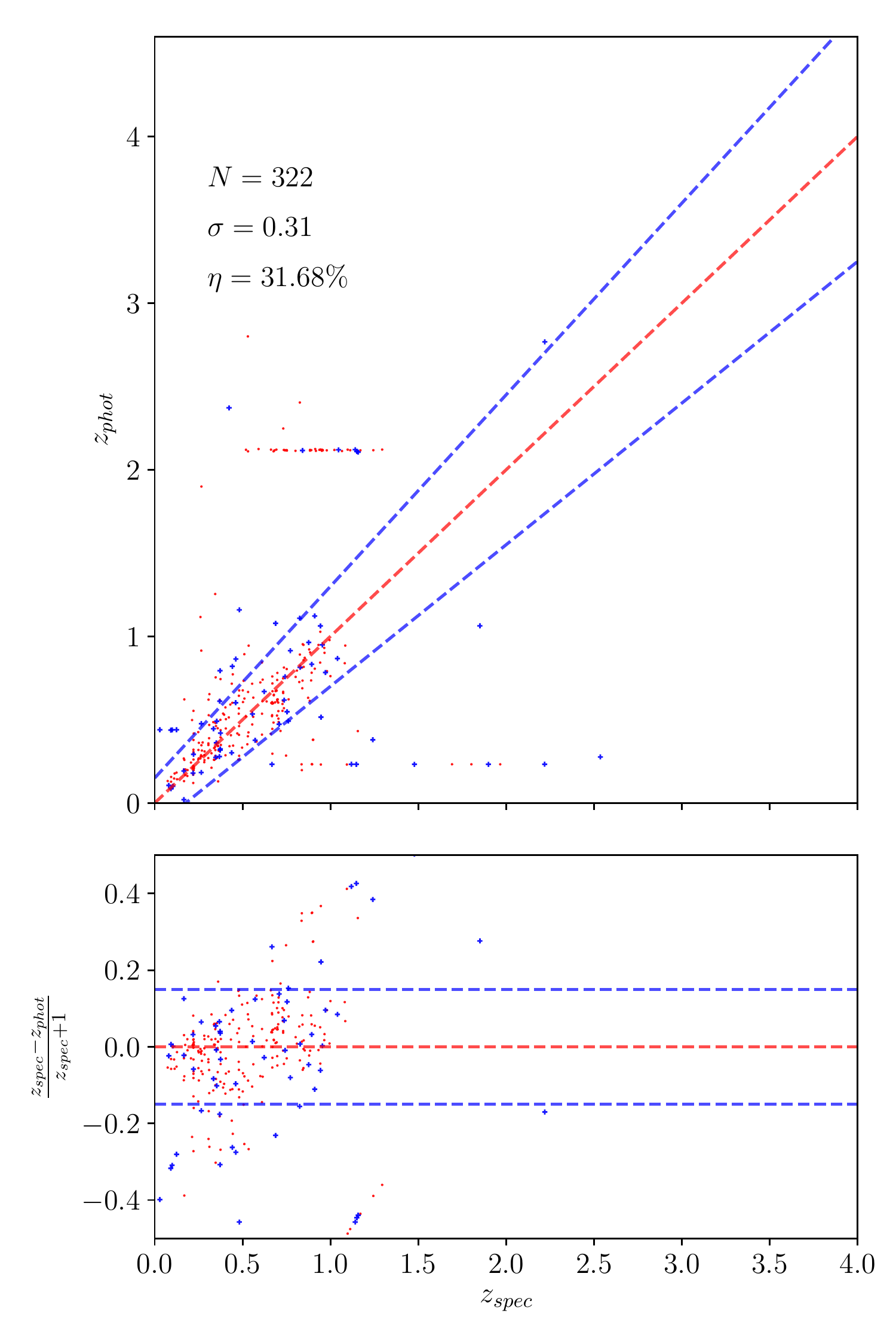} a)
\includegraphics[, scale=0.4]{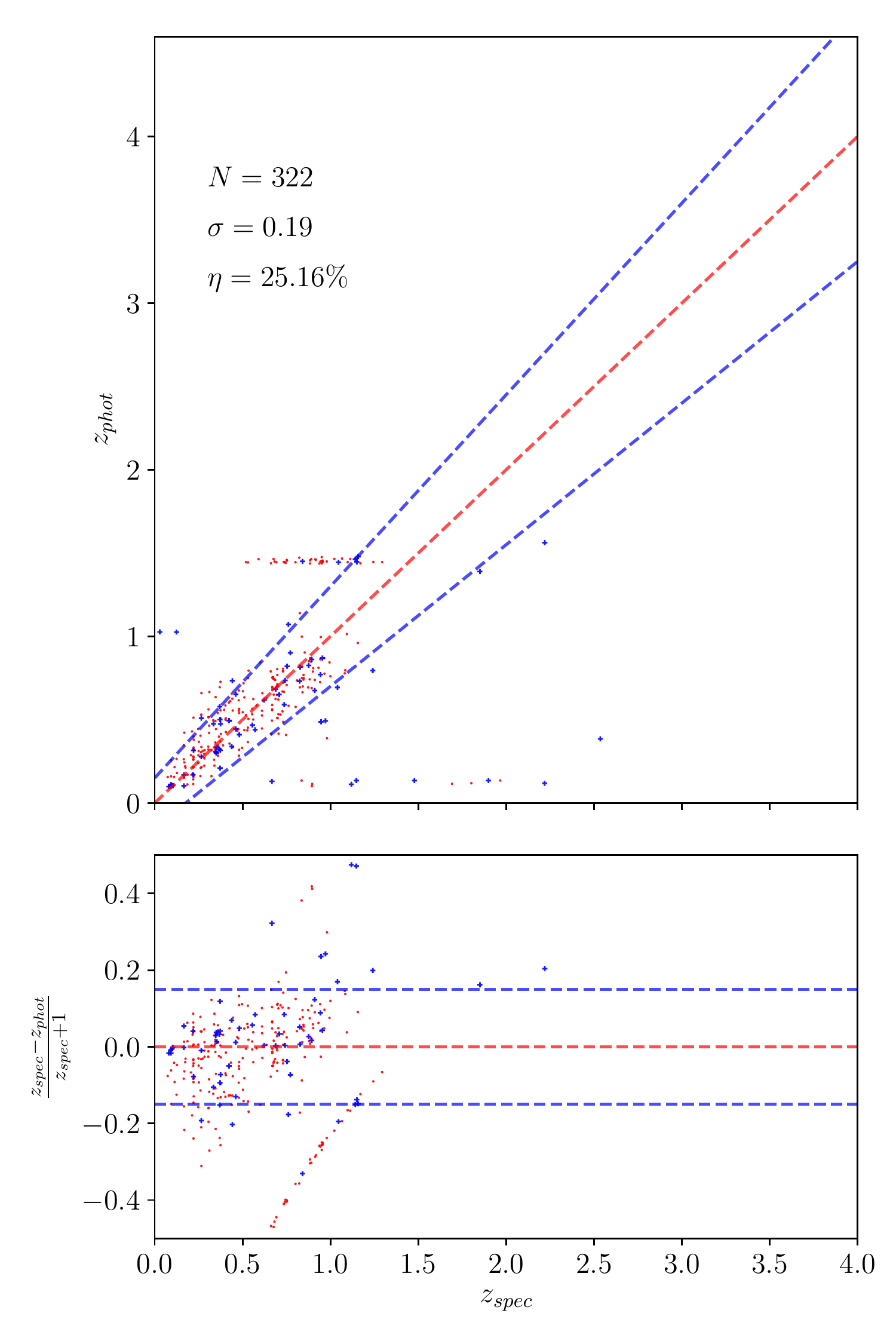} b)\\
\includegraphics[, scale=0.4]{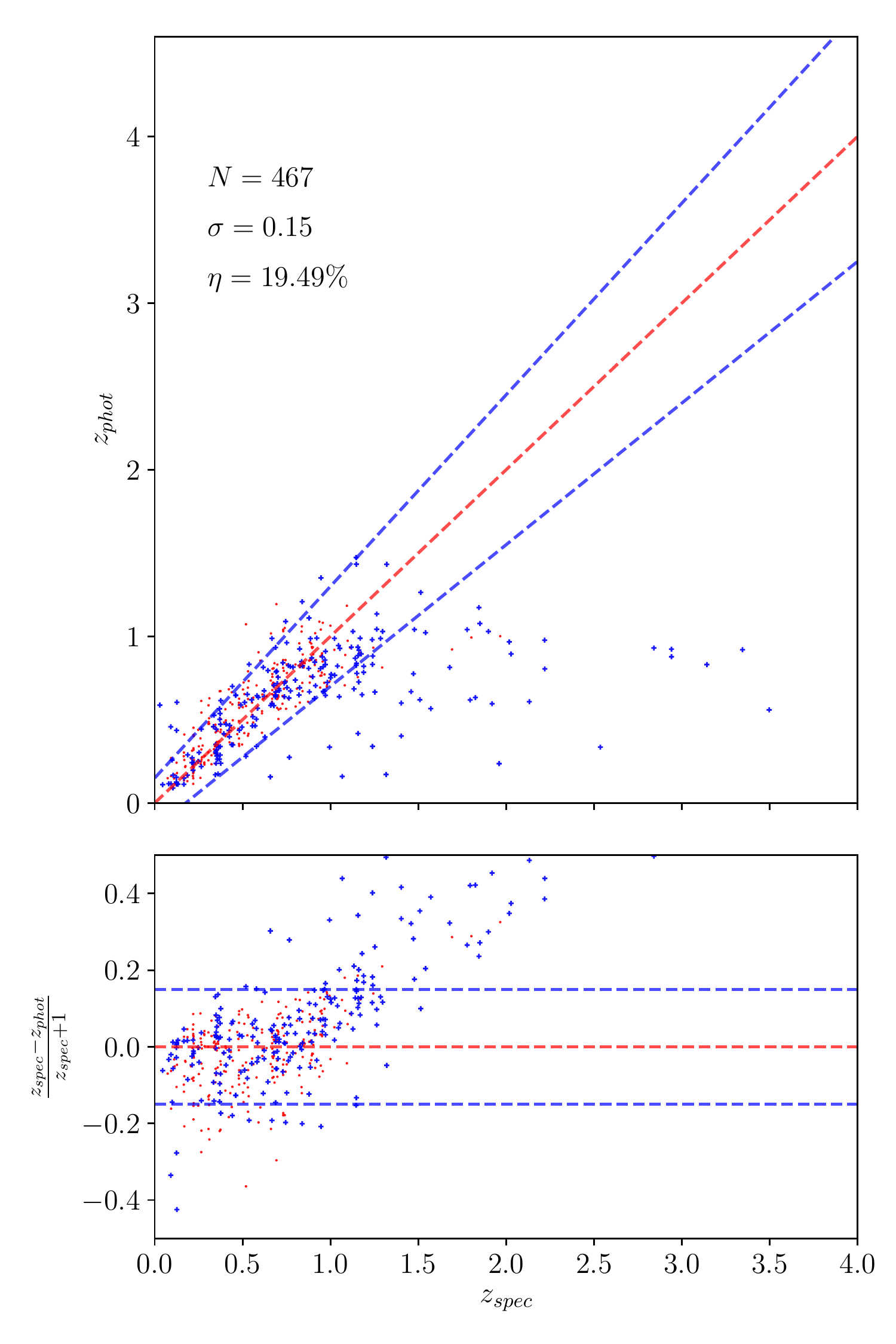} c)
\includegraphics[, scale=0.4]{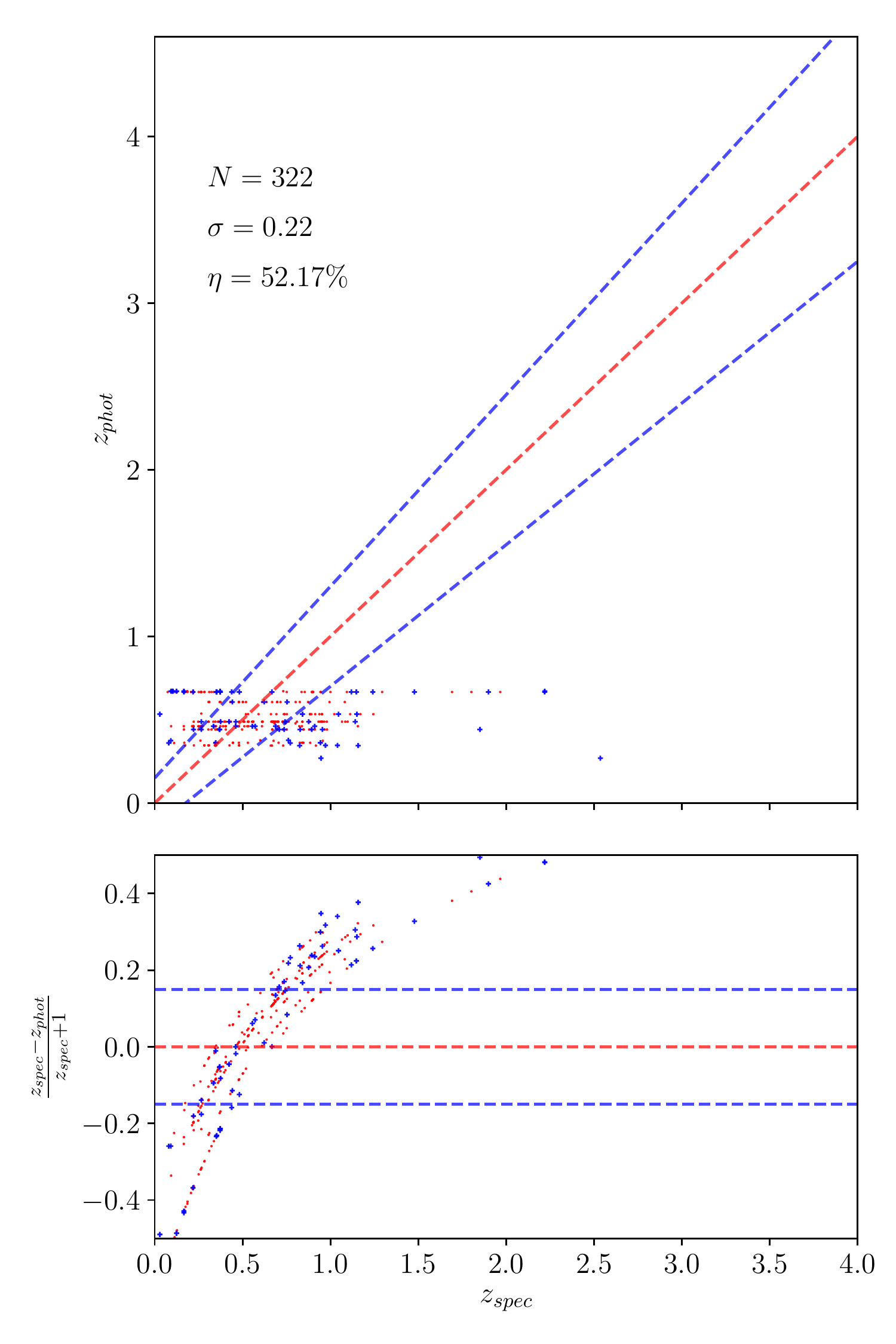} d)
\caption{
Summary of the results obtained in the experiment H2/RSYN with the various methods. 
Panel a): MLPQNA.
Panel b): RF-NA.
Panel c): RF-JHU.
%Panel d): SOM;
Panel d): kNN.
}
\label{FIG:blind_RSYN_shallow}
\end{figure*}

H2 differs from G2 by including radio data. 
\pointless{Figure~\ref{FIG:blind_RSYN_shallow} shows that MLPQNA has a smaller number of systematic outliers than G2, but a slightly lower overall success rate. In addition RF-JHU and RF-NA improve on the outliers and have a slightly hight overall succss rate than G2. In this experiment the large outliers are mainly the AGN that were unidentified at the time of performing the experiment.}

%The results, shown in Figure~\ref{FIG:blind_RSYN_shallow}, have a success rate not significantly changed, confirming that the addition of the radio information, can't 
%of a single radio photometric point does not provide useful information.

%\begin{figure*}\label{FIG:blind_RSYN}
%\includegraphics[, scale=0.20]{h2_brescia_cavuoti_longo_mlpgna_photoz_RSYN_blind_scatterPlot.pdf} a)
%\includegraphics[, scale=0.20]{h2_brescia_cavuoti_longo_rf_photoz_RSYN_blind_scatterPlot.pdf} b)
%\includegraphics[, scale=0.20]{h2_deep_carliles_photoz_RSYN_blind_scatterPlot.pdf} c)
%\includegraphics[, scale=0.20]{h2_deep_geach_photoz_RSYN_blind_scatterPlot.pdf} d)
%%\includegraphics[, scale=0.20]{h2_salvato_zphot_RSYN_blind_scatterPlot.pdf} e)
%\includegraphics[, scale=0.20]{h2_deep_zinn_photoz_RSYN_blind_scatterPlot.pdf} f)
%\caption{
%Summary of the results obtained in the  experiment H2/RSYN with the various methods. 
%Panel a) MLPQNA;
%Panel b)  RF-NA;
%Panel c): RF-JHU;
%Panel d): SOM;
%Panel e) kNN.
%}
%\end{figure*}

%\clearpage

\section{Discussion}
When given all the available data (experiment A2), Le Phare performed better than any of the ML methods, and performed even  better when the X-ray sources were \pointless{treated separately} (C2). Also when the photometry is reduced, its performance are overall superior, but comparable to \pointless{to RF-JHU in terms of accuracy and fraction of outliers}.
%to some of the ML techniques when the data were reduced in sensitivity (E2 and G2).

In the RANDOM-DEEP experiments (A2 to D2), all algorithms performed quite well (with NMAD $\sim$ $0.05$-$0.07$, and $\eta$ $\sim$ $10$-$15$\%, compared to $0.01$-$0.02$ and $3$-$5$\%, respectively, for Le Phare), with RF-NA typically performing rather better than the others, and MLPQNA often best at high redshift. 
%\textcolor{red}{we can't ignore that both MLPQNA and RF-NA DONOT compute redshift for 155 sources!. One way to address the point is the following:} 
However, the high reliability of  both MLPQNA and RF-NA are obtained at the cost of reducing the number of sources for which a photometric redshift is provided, \pointless{resulting in an overall success rate of $\sim 65\%$ for MLPQNA and RF-NA, compared with about 80\% for RF-JHU.} %This is safer than "guessing" a redshift. 
The lack of photoz for about 1/3 of the sources indicate that these sources are not represented by the training sample.
kNN generally performed well at low redshifts, but had a greater tendency than the other techniques to fail at $z \ge 1$, presumably because of the paucity of neighbours at high redshift.

In the RANDOM-SHALLOW experiments (E2 to H2), all algorithms performed less well than with the deeper data (with NMAD $\sim$ $0.1$-$0.2$, $\eta$ $\sim$ $10$-$15$\%, compared to $3$-$5$\% for Le Phare), with RF-NA typically performing rather better than the others. The best performing was RF-JHU and the worst performing was kNN, which essentially failed on these data. \cite{luken18} has shown that kNN is capable of good performance on low sensitivity data, \pointless{and subsequent experiments have shown that that particular implementation of kNN tends to fail on training sets smaller than $\sim$ 100 objects.}

Nevertheless, it is encouraging that most methods give useful results even on the shallow data. For example, of the 70 million EMU sources, about 70\% are likely to have photometry (from SkyMapper, SDSS, VHS, and WISE) comparable to that used in experiments E2 to H2, and RF-JHU was able to provide redshifts for $\sim$ 80\% of these, with an NMAD $\sim$ 0.1. This level of scatter is quite adequate for many of the EMU science goals, and having redshifts for $\geq$ 50\% of the EMU sources will significantly enhance the science from EMU.
%RPN: The following would be interesting, but I think we've run out of time. \textcolor{red}{Could we check whether the sources that are -99 from RF-NA are mostly outliers for RF-JHU ?
%If this is the case, then the two methods should be used together in EMU, because the first tells you how the sample differ from the training sample, while the second provide a  result sufficiently good. So flagging in RF-JHU all the -99 sources of RF-NA provide an additional information.}

Nevertheless, it is likely that even better results can be obtained by (\textit{i}) making full use of the available multi-frequency radio data, and (\textit{ii}) developing the algorithms further to optimise them for the limited data available for most EMU sources. Further work is continuing \citep[e.g.][]{luken18} to achieve this.

\section{Conclusion}
We have tested a number of photometric redshift techniques, including both template-fitting and machine learning techniques, on high-quality photometric data in the COSMOS field, and also explored reducing the quality of the photometric data to match that available from all-sky radio surveys such as EMU. We find that:

\begin{itemize}
\item Given high-quality multi-band photometry such as is available for the COSMOS data, the template-fitting Le Phare technique outperforms the machine learning techniques tested here, especially when X-ray sources are omitted. 
\item When the quality of the photometry is reduced to match that available for the EMU all-sky radio survey, both the template-fitting and the machine learning techniques give comparable results, typically with $\sim$ $20$-$30$\% of sources appearing as outliers, and with NMAD $\sim$ $0.1$-$0.2$.  
\item Most machine learning techniques perform better at $z < 1$ than at higher redshifts, presumably because of the paucity of training data at higher redshifts. 
\item  This level of redshift information from reduced-sensitivity data is still valuable and will result in a significant enhancement to the science from these surveys.
\end{itemize}

In this first set of experiments, we have set a baseline which will no doubt be improved on by further work in this field. Particularly important future directions are to (\textit{i}) obtain better training data for radio sources at high redshifts, and (\textit{ii}) continue developing the algorithms, optimizing them for the lower quality photometry likely to be available, 
\pointless{ 
(\textit{iii}) develop techniques to measure a probability distribution function for the result, rather than a single value, (\textit{iv}) use measured or estimated uncertainties as a weighting function on the input data, and (\textit{v}) use the combination of several different techniques to estimate reliability and to detect possible catastrophic failures.
}

\section*{acknowledgments}

SC acknowledges support from the project ``Quasars at high redshift: physics and cosmology'' financed by the ASI/INAF agreement 2017-14-H.0. MB acknowledges the \textit{INAF PRIN-SKA 2017 program 1.05.01.88.04} and the funding from \textit{MIUR Premiale 2016: MITIC}. VS acknowledges support from the European Union's Seventh Frame-work program under grant agreement 337595 (ERC Starting Grant, ``CoSMass'').

\begin{table*}
\begin{center}

\begin{tabular}{ccccccc}
%\tableline
%%\ray{I need to complete the numbers in this table}\\
Field & X-ray &  UV & OPTICAL& NIR& MIR&Radio\\
&Flux & magnitude &magnitude &magnitude &magnitude & Flux\\
%\multicolumn{1}{c}{$\Theta$\tablenotemark{b}} \\
\hline
& & & & & & \\

COSMOS & XMM,Chandra &GALEX  &22 bands$^a$         & J,H,K            &IRAC band 1 &VLA \\
                 &   $\sim 10^{-14} erg s^{-1}cm^{-2}$              &   25       & $\sim$ 26                                   &$\sim$23.5     &5 $\mu$Jy      &10 $\mu$Jy/bm rms\\
                 & & & & & & \\
% & eROSITA  XXXX & GALEX  XXXX & DES xxxx   &VHS  xxxx            &WISE xxx  & EMU xxxx\\
EMU         &eROSITA          &  GALEX & Skymapper (5 bands)    &VHS (Y,J,H,K)           &WISE W1   &EMU  \\
                 &   $\sim 10^{-14} erg s^{-1}cm^{-2}$               &   20.4     & $\sim$ 21.5            &$\sim$20 &26 $\mu$Jy& 10 $\mu$Jy/bm rms\\

%\tableline
\end{tabular}
\caption{Comparison of multiwavelength coverage for COSMOS and for the EMU all-sky survey.\label{Tab:all sky}}
%% Any table notes must follow the \end{tabular} command.
%$^{a}${from "u" to "i"; 6 broad band and 18 intermediate band filters}
%$^{b}${ 5 bands from "u" to "i"}
%\tablecomments{We can also attach a long-ish paragraph of explanatory material to a table.} 
\end{center}
\label{sensitivity}
\end{table*}

\begin{table*}
\centering
\resizebox{0.8\textwidth}{!}{
\begin{tabular}{lll}
\hline
id   & parameter         & note        \\             
\hline
1     & u\_cfht              & U photometry from CFHT Megaprime       \\ %               & 3   & e\_u\_cfht       & error         \\
2     & B\_subaru        & B photometry from Subaru Suprime-cam       \\ %               & 5    & e\_B\_subaru & error         \\       
3     & V\_subaru        & V photometry from Subaru  Suprime-cam       \\ %              & 7    & e\_V\_subaru  & error        \\
4     & g\_subaru        & G photometry from Subaru  Suprime-cam      \\ %              &  9    & e\_g\_subaru & error        \\   
5   & r\_subaru         &  R photometry from Subaru  Suprime-cam     \\ %              & 11   & e\_r\_subaru  & error       \\       
6   & i\_subaru         &  I photometry from Subaru  Suprime-cam      \\ %              & 13   & e\_i\_subaru   & error       \\          
7   & z\_subaru        &  Z photometry from Subaru   Suprime-cam      \\ %             & 15   & e\_z\_subaru  & error        \\
8  & J\_wfcam         &  J photometry from  UKIRT WFCAM    \\ %              & 17   & e\_J\_wfcam   & error        \\
9   & H\_wircam       &    H photometry from CFHT Wircam            \\ %  &  19       &e\_H\_wircam &       \\         
10   & flamingos\_Ks&  Ks photometry from Gemini FLAMINGOS              \\ %&   21     &e\_flamingos\_Ks&   \\
11   & K\_wircam    &     K photometry from CFHT Wircam             \\ %&  23       & e\_K\_wircam  &      \\         
12   & i\_cfht           &    I photometry from CFHT Megaprime              \\ %&   25      & e\_i\_cfht          &      \\
13   & u\_SDSS       &    u' photometry from SDSS            \\ %&  27       & e\_u\_SDSS      &      \\         
14   & g\_SDSS      &     g' photometry from SDSS              \\ %&  29       & e\_g\_SDSS   &         \\        
15   & r\_SDSS        &    r' photometry from SDSS              \\ %& 31        & e\_r\_SDSS   &           \\
16    & i\_SDSS        &   i' photometry from SDSS              \\ %&  33        & e\_i\_SDSS     &        \\
17    & z\_SDSS       &    z' photometry from SDSS             \\ %&  35        & e\_z\_SDSS   &         \\
18    & f814               &    Subaru Suprime-cam    814nm        \\ %& 37        &  e\_f814         &         \\
19    & IB427           &      Subaru Suprime-cam    427nm     \\ %& 39         &  e\_IB427       &         \\
20    & IB464           &      Subaru Suprime-cam     464nm    \\ %& 41         &  e\_IB464       &         \\
21   & IB484           &       Subaru Suprime-cam      484nm  \\ %& 43         & e\_IB484        &         \\
22    & IB505           &      Subaru Suprime-cam       505nm  \\ %& 45         & e\_IB505        &        \\
23    & IB527           &       Subaru Suprime-cam      527nm  \\ %& 47         & e\_IB527        &        \\
24    & IB574           &       Subaru Suprime-cam      574nm  \\ %& 49         &  e\_IB574       &       \\ 
25    & IB624           &        Subaru Suprime-cam     624nm  \\ %& 51         & e\_IB624       &       \\
26    & IB679           &        Subaru Suprime-cam      679nm \\ %& 53         & e\_IB679       &       \\
27    & IB709           &        Subaru Suprime-cam       709nm\\ %& 55         & e\_IB709       &      \\
28    & IB738           &         Subaru Suprime-cam      738nm\\ %& 57         &  e\_IB738      &      \\
29    & IB767           &        Subaru Suprime-cam       767nm\\ %& 59         & e\_IB767       &      \\
30    & IB827           &         Subaru Suprime-cam     827nm \\ %& 61         & e\_IB827       &      \\
31    & NB711          &        Subaru Suprime-cam       711nm\\ %& 63         & e\_NB711     &     \\
32    & NB816          &       Subaru Suprime-cam        816nm\\ %& 65         &  e\_NB816   &      \\
33    & ch1              &       IRAC band 1 (3.6 $\mu$m)      \\ %& 67         &  e\_ch1        &      \\
34    & ch2              &       IRAC band 2 (4.5 $\mu$m)       \\ %& 69         &  e\_ch2        &     \\
35    & ch3              &       IRAC band 3 (5.8 $\mu$m)        \\ %& 71         &  e\_ch3        &     \\
36    & ch4              &       IRAC band 4 (8.0 $\mu$m)        \\ %& 73         &  e\_ch4        &     \\
37    & galex1500   &     Galex 150 nm           \\ %& 75         &  e\_galex1500 &  \\
38    & galex2500   &    Galex 250nm            \\ %& 77          & e\_galex2500&   \\
39    &  morphology & See http://irsa.ipac.caltech.edu/data/COSMOS/tables/morphology/ \\ %
40	& zspec	&    spectroscopic redshift\\ %
41    & radio\_integrated\_flux  & VLA integrated 20cm radio flux \\ %
42   &   xmm\_soft   &  XMM soft\\%
43	&  xmm\_hard  &  XMM hard      \\%    
44    & ch\_soft       & Chandra soft \\
45    &  ch\_hard     & Chandra hard \\
\end{tabular}}
\caption{Features (observational parameters) used in the tests. Column $2$: feature name, column $3$: explanation. Details of all these filters are given on http://cosmos.astro.caltech.edu/page/filterset. Uncertainties for all measurements were also available in the KB.}
%{\bf This table needs some more work to fill in the blanks}
\label{TAB:parameters}
\end{table*}
%

%\afterpage{\clearpage}
 
%\begin{landscape}

\clearpage

\bibliographystyle{mnras}

\end{document}